%% file: paper_R1_v4_double.tex
\documentclass[10pt,journal,twoside,twocolumn]{IEEEtran}
\input macros

\input labelfig.tex

\newtheorem{proposition}{Proposition}

\newtheorem{lemma}{Lemma}

\newtheorem{problem}{Problem}

\newtheorem{remark}{Remark}

\newcommand{\argmax}{\operatornamewithlimits{argmax}}

\usepackage[dvips]{graphics}
\usepackage[dvips]{graphicx}
\usepackage{amsmath,amssymb,amsfonts,latexsym,verbatim,color,epsfig,psfrag}%epsf,
\usepackage{times,multirow,multicol, array} 
\usepackage{algorithm,algorithmic}
\usepackage{cite}
\usepackage{url}
\usepackage{dblfloatfix}
\usepackage{booktabs}
\usepackage{setspace}
\usepackage{soul} % for strikethrough

\usepackage{caption}
\usepackage[font=small]{caption} % set the size of caption of the figure

\usepackage[letterpaper,hmargin=1in,top=0.85in, bottom=0.97in]{geometry}
\definecolor{gray}{rgb}{0.5,0.5,0.5}

\definecolor{islamicgreen}{rgb}{0.0, 0.56, 0.0}

\begin{document}

\title{\huge Training Sequence Design for Feedback Assisted \\ Hybrid Beamforming in Massive MIMO Systems}

\author{\authorblockN{Song Noh, Michael D. Zoltowski, and David J. Love\thanks{
S. Noh, M. Zoltowski, and D. J. Love are with the School of Electrical and Computer Engineering, Purdue University, West Lafayette, IN 47907, USA (e-mail:songnoh@purdue.edu and \{mikedz,djlove\}@ecn.purdue.edu). 
A preliminary version of this work was presented in \cite{Noh&Zoltowski&Love:14GLOBECOM}.
%\cite{Noh&Zoltowski&Sung&Love:13ASILOMAR}, in which only the MISO
%case is considered. 
%In this paper, the sequential design proposed
%in \cite{Noh&Zoltowski&Sung&Love:13ASILOMAR} is extended to the
%MIMO case, power allocation, and the block-fading case. Extensive
%simulation results with some realistic channel models
%are provided.
}}
%\begin{center}
%EDICS: SPC-CEST, SPC-PERF % SPC-CEST Channel estimation and equalization, SPC-PERF Performance analysis and bounds,  SSP-FILT Filtering < SSP STATISTICAL SIGNAL PROCESSING
%\end{center}
}

%\author{\authorblockN{Song Noh$^\dagger$, Michael D. Zoltowski$^\dagger$, Youngchul Sung$^\ddagger$, and David J. Love$^\dagger$}\\
%\authorblockA{$^\dagger$School of Electrical and Computer Engineering, Purdue University, West Lafayette, IN 47907, USA\\
%Email: songnoh@purdue.edu, mikedz@ecn.purdue.edu, djlove@ecn.purdue.edu}\\
%\authorblockA{$^\ddagger$Department of Electrical Engineering, KAIST, Daejeon, Korea 305-701\\
%Email: ysung@ee.kaist.ac.kr}}

\maketitle

%\vspace{-5.3em}
\begin{abstract} %\vspace{-1.2em}
The use of large-scale antenna systems in future commercial wireless communications is an emerging technology that uses an excess of transmit antennas to realize high spectral efficiency. 
Achieving potential gains with large-scale antenna arrays in practice hinges on sufficient channel estimation accuracy. 
Much prior work focuses on TDD based networks, relying on reciprocity between the uplink and downlink channels. 
However, most currently deployed commercial wireless systems are FDD based, making it difficult to exploit channel reciprocity. 
In massive MIMO FDD systems, the problem of channel estimation becomes even more challenging due to the attendant substantial training resources and feedback requirements which scale with the number of antennas. 
In this paper, we consider the problem of training sequence design that employs a set of training signals and its mapping to the training periods. 
We focus on reduced-dimension training sequence designs, along with transmit precoder designs, aimed at reducing both hardware complexity and power consumption. The resulting designs are extended to hybrid analog-digital beamforming systems, which employ a limited number of active RF chains for transmit precoding, by applying the Toeplitz distribution theorem to large-scale linear antenna systems. 
A practical guideline for training sequence parameter selection is presented along with performance analysis. 
\end{abstract}

%\vspace{-1.5em}
\begin{keywords}%\vspace{-1.4em}
Massive MIMO systems, channel estimation, training sequence design, hybrid beamforming
\end{keywords}
%\vspace{-1.0em}

%%%%%%%%%%%%%%%%%%%%%%%%%%%%%%%%%%%%%%%%%%%%%%%%%%%%%%%%%%%%%%%%%%%%%%%
\section{Introduction} \label{sec:introduction}
Multiple-input multiple-output (MIMO) technology has been demonstrated to be effective in providing reliable wireless links;  the advantages of MIMO communications are widely recognized \cite{Renzo&Haas&Ghrayeb&Sugiura&Hanzo:14IEEE}. 
MIMO systems utilizing a large number of antennas at the base station, referred to as {\em massive MIMO} systems, are emerging as a key technology for the design of high throughput and energy efficient systems for future wireless communications.
Massive MIMO represents a paradigm shift in system configuration, wherein 
the power per antenna is reduced by a factor roughly equal to the number of transmit antennas, and only relatively simple signal processing is performed, e.g., spatial matched filtering \cite{Marzetta:10WCOM}.
This is all enabled by exploiting advantageous assumptions about the propagation environment that arise from asymptotic random matrix analysis
The large size of the transmit antenna array relative to the number of serviced users mitigates thermal noise, fast channel fading, and some forms of interference, all drawing in part at least from the law of large numbers \cite{Marzetta:10WCOM,Rusek&Persson&Lau&Larsson&Edfors&Tufvesson&Marzetta:13SPM}.

However, the potential gains of massive MIMO in practical systems are limited by channel estimation accuracy \cite{Jose&Ashikhmin&Marzetta&Vishwanath:11WCOM}. In contrast to current MIMO systems equipped with a few antennas at each base station, the training signal overhead required for channel estimation in a massive MIMO system can be overwhelming, since the number of time slots for transmitting orthogonal training signals must be at least as large as the number of antennas. % units at the base station.  %to achieve sufficient channel estimation accuracy. 
In addition, an important issue regarding the cost of implementation is that
the number of active RF chains required for channel sounding and transmit precoding is limited relative to the number of antennas \cite{Renzo&Haas&Ghrayeb&Sugiura&Hanzo:14IEEE}. 
%relative to the number of antennas for channel sounding and transmit beamfoming at the transmitter \cite{Renzo&Haas&Ghrayeb&Sugiura&Hanzo:14IEEE}. 
As a result, 
%reliable 
channel estimation schemes that are reliable and require a low training overhead and low-complexity are important in order to efficiently utilize the large antenna array gains.

% TDD massive MIMO
To tackle the challenge of channel estimation, much of the prior work focused on time-division duplex (TDD) operation assuming channel reciprocity 
\cite{Marzetta:10WCOM,Rusek&Persson&Lau&Larsson&Edfors&Tufvesson&Marzetta:13SPM,Hoydis&Brink&Debbah:13JSAC} to acquire channel state information (CSI) at the base station under the assumption of time-invariant channels within the coherence time.
In a TDD mode, the uplink channel sounding enables downlink channel estimation by using channel reciprocity that requires proper calibration of the hardware chains between the terminal uplink and downlink chains \cite{Shepard&Yu&Anand&Li&Marzetta&Yang&Zhong:12MobiCom}.
%When full frequency reuse across neighboring cells is adopted, 
In addition, in a multi-cell environment with a high frequency reuse factor, pilot contamination induced by the use of non-orthogonal uplink training signals in neighboring cells leads to imperfect channel estimation causing severely degraded system performance  \cite{Jose&Ashikhmin&Marzetta&Vishwanath:11WCOM}. %,Yin&Gesbert&Filippou&Liu:13JSAC,Ngo&Larsson:12ICASSP}.
%The simplicity of a TDD approach comes with proper calibration of the hardware chains between the terminal uplink and downlink chains to rely on channel reciprocity.

% FDD massive MIMO
In most wireless systems that employ a frequency-division duplex (FDD) mode, the problem of channel estimation becomes more challenging %because downlink training represents a significant bottleneck 
because downlink channel estimation requires substantial overhead, such as feedback and dedicated times for channel sounding which scales with the number of antennas. 
%
%%%%%%%%%
%There exist several beamforming techniques in MIMO systems by considering hybrid analog digital CSI feedback \cite{Yuan&Ho:10VTC} and simplified analog antenna beamforming design \cite{Gholam&Via&Santamaria:11VT}.
In an FDD mode, it was shown that the overhead for channel estimation does not scale with the number of antennas in conjunction with underlying channel statistics, the spatial sparsity, and the specific antenna arrangement \cite{Noh&Zoltowski&Sung&Love:13ASILOMAR,Noh&Zoltowski&Sung&Love:14STSP,Choi&Chance&Love&Madhow:13COM,Choi&Love&Bidigare:14STSP,Adhikary&Nam&Ahn&Caire:13IT,Adhikary&Caire:14STSP,Lee&Sung:14arXiv,SoKimLeeSung:14SPL}. 
%Note that there is one observation in massive MIMO systems in that antenna correlations in real-world systems are larger than expected under independent and identically distributed (i.i.d.)  channel assumptions; experimental investigations \cite{Hoydis&Hoek&Wild&Brink:12ISWCS,Gao&Edfors&Rusek&Tufvesson:11VTC} and analytical studies \cite{Wagner&Couillet&Debbah&Slock:12IT,Ngo&Larsson&Marzetta:13COMJun} have been presented.
Note that antenna correlations are observed in experimental investigations \cite{Hoydis&Hoek&Wild&Brink:12ISWCS,Gao&Edfors&Rusek&Tufvesson:11VTC} and analytical studies that consider
a very high angular resolution due to its large antenna aperture have been presented \cite{Wagner&Couillet&Debbah&Slock:12IT,Ngo&Larsson&Marzetta:13COMJun}.
%For a massive MIMO cellular system, a concept of joint spatial division and multiplexing (JSDM) was introduced with hybrid analog/digital beamforming so as to efficiently support multiple users in a cell\cite{Adhikary&Nam&Ahn&Caire:13IT}. 
In order to efficiently support multiple users in a massive MIMO cellular system, a technique called joint spatial division and multiplexing (JSDM) was introduced with hybrid analog/digital beamforming  under the assumption that the effective channel rank is known to the system \cite{Adhikary&Nam&Ahn&Caire:13IT}. 
Low complexity algorithms to solve the user scheduling problem were presented in \cite{Adhikary&Caire:14STSP,Lee&Sung:14arXiv}. In addition, there has been work on channel state information feedback based on limited feedback \cite{Choi&Chance&Love&Madhow:13COM,Kudoetal:13JCN} and compressive sensing \cite{Kuo&Kung&Ting:12WCNC,Rao&Lau:14SP} by considering the sparsity features of the channel matrices.
Recently, initial work on channel estimation in massive MIMO systems has been proposed. %  \cite{Noh&Zoltowski&Sung&Love:13ASILOMAR,Noh&Zoltowski&Sung&Love:14STSP,SoKimLeeSung:14SPL,Choi&Love&Bidigare:14STSP}.
One approach is pilot beam pattern design based on channel statistics aimed at minimizing channel mean square error (MSE) \cite{Noh&Zoltowski&Sung&Love:13ASILOMAR,Noh&Zoltowski&Sung&Love:14STSP} and leveraging a received SNR \cite{SoKimLeeSung:14SPL}. The other approach is adaptive codebook selection based on a received SNR or the channel MSE \cite{Choi&Love&Bidigare:14STSP}. 
However, the existing researches generally addressed beamforming design and channel estimation technique separately.  
%Note that the performance of beamforming is closely related to the quality of channel estimation and it is critical to understand how to consider them jointly.
In particular, 
the pilot beam patterns proposed in \cite{Noh&Zoltowski&Sung&Love:14STSP} inherently lie in a high-dimensional space and also make a closed-form performance analysis intractable due to the greedy sequential search of the dominant eigenmodes.

In this paper, we consider the design of a training scheme that properly specifies the training signals and its mapping to the corresponding training period
%have a suitable mapping for the training signal patterns during the training periods 
for downlink channel estimation in FDD massive MIMO systems. 
We refer to this scheme as using a {\em training sequence}. Under a Kalman filtering framework, the proposed training sequence is designed to minimize the steady-state channel mean square error (MSE) to leverage channel estimation performance. 
In addition, we focus on a reduced-dimensionality training sequence and transmit precoding design aimed at reducing the cost of implementation and power consumption \cite{Renzo&Haas&Ghrayeb&Sugiura&Hanzo:14IEEE}. %an energy-efficient system configuration.
%This consideration is to reduce both the cost of implementation and power consumption which are proportional  to the number of active RF chains. 
We then extend the low-dimensional constraint to hybrid analog-digital beamforming scheme that uses a limited number of available RF chains for digital baseband precoding by applying the Toeplitz distribution theorem to an uniformly spaced linear array (ULA) at the base station. 
For performance analysis, we adopt a {\em deterministic equivalent} technique \cite{Hoydis&Brink&Debbah:13JSAC} to handle the case of large antenna arrays and provide a practical guideline for training sequence parameters. 
%We provide a concise framework that extends the result in \cite{Noh&Zoltowski&Sung&Love:14STSP} by accounting the steady-state channel MSE and periodic training signal sequence.

Our main contributions are summarized as follows:
\begin{itemize}
\item We propose a periodic training sequence framework that enables a reduced dimensionality design of the training sequence and transmit precoding. 
We remark that by considering the steady-state channel MSE, the effective channel rank required for transmit precoding can be obtained, even when the channel has continuous power (azimuth) spectrum. 
\item By analyzing the monotonicity property in the steady-state channel MSE, a reduced-complexity suboptimal algorithm that minimizes the maximum steady-state MSE without much loss in performance is proposed. For large-scale linear antenna arrays, the proposed method can extend to a hybrid analog-digital beamforming scheme that requires a limited number of active RF chains for transmit beamforming.
\item We derive a closed-form expression for the steady-state mean square error (MSE) and the SINR under spatial matched filtering, which is close to the exact value obtained from numerical simulations. 
%of the channel estimate as a function of a Gauss-Markov fading process and training signal placements. Under spatial matched filtering, we also derive a closed-form approximation for the SINR, and 
%The approximation is close to the exact value obtained from numerical simulations.  
%We evaluate the influence of the spatio-temporal channel statistics for some scenarios that can represent the downlinks of cellular communication systems.
Our results show that the proposed method yields good performance, even with an imperfect  knowledge of channel statistics.
%\item We derive a closed-form expression for the steady-state mean square error (MSE) of the channel estimate as a function of a Gauss-Markov fading process and training signal placements. Under spatial matched filtering, we also derive a closed-form approximation for the SINR, and the approximation is close to the exact value obtained from numerical simulations.  
%%We evaluate the influence of the spatio-temporal channel statistics for some scenarios that can represent the downlinks of cellular communication systems.
%Our results show that the proposed method yields good performance, even with an imperfect  knowledge of channel statistics.
\end{itemize}

%% ORGANIZATION
%This paper is organized as follows. The system model and
%background are described in Section \ref{sec:systemmodel}. 
%Section \ref{sec:codebookpilotbeampattern} describes
%the proposed training sequence and hybrid analog-digital precoding design.
%In Section \ref{sec:multiuser_massiveMIMO}, an extension to multi-user massive MIMO systems is discussed with performance analysis.
%Numerical results are provided in Section \ref{sec:numericalresults},  followed by conclusions in Section \ref{sec:conclusion}.

\vspace{0.5em}
{\em Notations:} Vectors and matrices are written in boldface with matrices in capitals. 
All vectors are column vectors. For a matrix $\Abf$, $\mathbf{A}^T$, $\mathbf{A}^H$, and  $\text{tr}(\mathbf{A})$ indicate the transpose, Hermitian transpose, and trace of $\mathbf{A}$,
respectively. $\mathbf{A} \odot \mathbf{B}$ denotes the Hadamard product between $\mathbf{A}$ and $\mathbf{B}$. 
%$\Abf_{\Ic}$ is a submatrix of $\Abf$ composed of the column vectors of the index set $\Ic$.
$[\Abf]_{p,q}$ represents the element in the $p$-th row and the $q$-th column of $\Abf$.
%$\Nc(\Abf)$ stands for the null space of $\Abf$.
%at rows $\mathcal{I}$ and columns $\mathcal{J}$ for some index
%sets $\mathcal{I}$ and $\mathcal{J}$, and $i_1:i_2$ denotes the set
%$\{i_1,i_1+1,\cdots,i_2\}$ (without $i_1$ and $i_2$ it denotes the
%entire set). $[\mathbf{A}]_{i,j}$ denotes the element of $\mathbf{A}$ at
%the $i$-th row and $j$-th column, and $[a_{i,j}]$ is the matrix
%composed of element $a_{i,j}$ at the $i$-th row and $j$-th column.
$\mbox{diag}(d_1,\cdots,d_n)$ is the diagonal matrix composed of
elements $d_1,\cdots,d_n$. 
$\mathbf{I}_N$ stands for the identity matrix of size $N$; $\mathbf{1}_{M\times N}$ and $\mathbf{0}_{M\times N}$ denote an $M\times N$ matrix composed of all-ones and all-zeros, respectively.
For a vector $\abf$, $\|\abf\|_p$ represents the $p$-norm.
For a matrix $\Abf$, $\|\Abf\|_F$ denotes the Frobenious norm. % and $\Cc(\Abf)$ denotes the column space of $\Abf$.
%$\mathbf{A}^{\dagger}$, $\|\mathbf{A}\|_F$,
%$\|\mathbf{A}\|_2$, and $\mbox{tr}\{\mathbf{A}\}$ denote the Moore-Penrose
%pseudo-inverse, Frobenius norm, $L_2$-norm, and trace of $\mathbf{A}$,
%respectively. 
%$\mathbf{A} \otimes \mathbf{B}$ denotes the Kronecker product
%and $\mathbf{A} \odot \mathbf{B}$ denotes the Hadamard product between $\mathbf{A}$
%and $\mathbf{B}$. 
$\xbf\sim\mathcal{CN}(\mubf,\Sigmabf)$ means that the random vector $\xbf$ is complex Gaussian distributed with mean $\mubf$ and covariance matrix $\Sigmabf$.
%$\iota=\sqrt{-1}$ is used for the imaginary number so that $i$ and $j$ may be used as indices.
$E\{\cdot\}$ denotes statistical expectation. $\mathbb{N}$ and $\mathbb{C}$ denote the sets of natural numbers and complex numbers, respectively.

%\tableofcontents

%\vspace{-0.4em}
%%%%%%%%%%%%%%%%%%%%%%%%%%%%%%%%%%%%%%%%%%%%%%%%%%%%%%%%%%%%%%%%%%%%%%%
\section{System Model}\label{sec:systemmodel}
%%%%%%%%%%%%%%%%%%%%%%%%%%%%%%%%%%%%%%%%%%%%%%%%%%%%%%%%%%%%%%%%%%%%%%%
%\vspace{-0.5em}
%\vspace{-0.6em}

%\subsection{System Setup}
%\vspace{-0.5em}

\begin{figure}[!t]
\centerline{
\psfrag{(n1)}[c]{\scriptsize $1$} %
\psfrag{(n2)}[c]{\scriptsize {$2$}} %
\psfrag{(nt)}[c]{\scriptsize {$N_t$}} %
\psfrag{(xk)}[c]{\footnotesize {$\xbf_k$}} %
\psfrag{(sk)}[c]{\footnotesize $\sbf_k$} %
\psfrag{(md)}[c]{\scriptsize $M_p< m \le M$} %
\psfrag{(mp)}[c]{\scriptsize $1\le m \le M_p$} %
%\psfrag{(md)}[c]{\scriptsize $(k \notin \mathcal{I}_p)$} %
%\psfrag{(mp)}[c]{\scriptsize $(k \in \mathcal{I}_p)$} %
\psfrag{(nr1)}[c]{\footnotesize $1$} %
\psfrag{(nr)}[c]{\footnotesize {$N_r$}} %
\psfrag{(yk)}[l]{\footnotesize {$y_k$}} %
\psfrag{(ypilot)}[l]{\footnotesize {$y_k\in\mathbb{C}$}} %
\psfrag{(hest)}[l]{\footnotesize {$\hat{\hbf}_{\ell |\ell}$}} %
\psfrag{(h)}[c]{\footnotesize {$\hbf_\ell$}} %
\psfrag{(data)}[c]{\footnotesize Linear} %
\psfrag{(precoder)}[c]{\footnotesize precoder} %
\psfrag{(vk)}[c]{\footnotesize {$\Vbf_k$}} %
\psfrag{(kalman)}[c]{\footnotesize MMSE} %
\psfrag{(filter)}[c]{\footnotesize filter} %
\psfrag{(feedback)}[c]{\scriptsize Feedback channel} %
%\psfrag{(proposed)}[c]{\scriptsize Tracking $\boldsymbol\lambda$} %
%\psfrag{(channel)}[c]{\footnotesize Gauss-Markov channel} %
\includegraphics[scale=0.76]{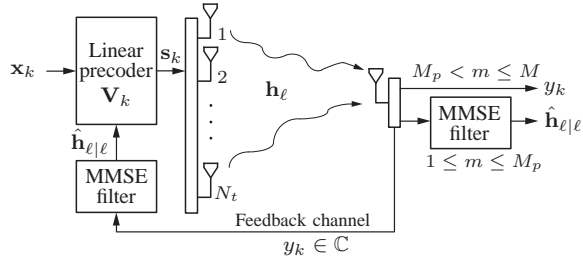}}
\vspace{-0.2em}
\caption{Massive MIMO system model for the symbol time $k=\ell M+m$.} \label{fig:massiveMIMOsystemModel}
\vspace{-2.0em}
\end{figure}

We consider a downlink massive MIMO system with $N_t$ transmit antennas and a single receive antenna operating over flat Rayleigh-fading channels, as shown in Fig. \ref{fig:massiveMIMOsystemModel}. %\tcr{(the extension to the case of wideband frequency-selective channel in OFDM systems follows  in a similar manner)}.
We focus on the single-user case first and then point out the multiple-user case in Section \ref{sec:multiuser_massiveMIMO}.
We assume block transmission with $M$ consecutive symbols for one block composed of
a pilot transmission period of $M_p$ symbols and a data
transmission period of $M_d$ symbols, i.e., $M=M_p+M_d$.
(We will refer to the $M$ consecutive channel transmissions composed of the training period and the data transmission period as a {\em block}.)
The received signal at the $k$-th symbol time is given by
\begin{equation}
y_k  = \hbf_\ell^H\sbf_k +
w_k,~~~\text{ for  } k=\ell M + m,\label{eq:statespacemodel_y1}
\end{equation}
where $\ell=0,1,\ldots$ and $1\le m\le M$ so that $k=1,2, \ldots$.
Here, $\sbf_k\in\mathbb{C}^{N_t}$ is the transmitted symbol vector with power constraint $E\{\|\sbf_k\|_2^2\}=\rho$ and
$\hbf_\ell\in\mathbb{C}^{N_t}$ is the channel vector with additive noise $w_k\sim\mathcal{CN}(0,1)$. 
%and $w_k$ is a zero-mean independent and identically distributed (i.i.d.) complex Gaussian noise with covariance $\sigma_w^2$. 
The transmit vector $\sbf_k$ represents a training signal vector during the training period (i.e., $k=\ell M+m$ where $1\le m \le M_p$). 
On the other hand, during the data transmission period ($M_p< m \le M$), $\sbf_k$ denotes a precoded data vector constructed by mapping the data symbols $\xbf_k=[x_{1,k},\cdots,x_{U,k}]^T\in \mathbb{C}^U$ to the transmit antenna array using the multi-dimensional beamformer $\Vbf_k\in\mathbb{C}^{N_t\times U}$, i.e., $\sbf_k=\Vbf_k \xbf_k$.
We first consider a single data stream transmission ($U=1$) in which a rank-one beamformer is denoted as  $\vbf_k\in\mathbb{C}^{N_t}$, and then extend to the case of $U>1$.

We assume that the channel is block-fading and that the channel
remains constant during the $\ell$-th block. % in \eqref{eq:statespacemodel_y1}.
%, i.e., $\hbf_k=\hbf_{\ell}$ for $k=\ell M+m$, $\ell=0,1,2,\cdots$, and $1\le m \le M$. 
The channel temporal variation across the blocks is modeled using a state-space framework as a first-order stationary
Gauss-Markov process \cite{Tong&Sadler&Dong:04SPM} with
\begin{equation}
\hbf_{\ell+1}  = a \hbf_{\ell} + \sqrt{1-a^2}\bbf_{\ell+1}, \label{eq:statespacemodel_h}
\end{equation}
%that satisfies the Lyapunov equation $\Rbf_\hbf = a^2 \Rbf_\hbf + (1-a^2) \Rbf_\bbf$  
where $a\in(0,1)$ denotes the temporal fading correlation coefficient,\footnote{In Jake's model \cite{Jakes:book}, $a = J_0(2\pi f_D (T_s M))$ where $J_0(\cdot)$ is the zeroth-order Bessel function, $f_D=\frac{vf_c}{c}$ is the maximum Doppler frequency shift, $v$ denotes a mobile speed, $T_s$ is the symbol time interval, and each block is composed of $M$ symbols.} 
$\bbf_{\ell}$ denotes process noise for the $\ell$-th block time index where $\bbf_{\ell}\sim\mathcal{CN}(\mathbf{0},\Rbf_\hbf)$, 
and the channel spatial correlation is given by $\Rbf_\hbf=E\{\hbf_\ell\hbf_\ell^H\}$ for all $\ell$.
%Note that \eqref{eq:statespacemodel_h} satisfies  the Lyapunov equation $\Rbf_\hbf = a^2 \Rbf_\hbf + (1-a^2) \Rbf_\bbf$  where $\Rbf_{\hbf}=E\{\hbf_{\ell}\hbf_{\ell}^H\}=\Rbf_{\bbf}= E\{\bbf_{\ell}\bbf_{\ell}^H\}$ for all $\ell$. %Here, $\gamma$ denotes propagation path loss that will be defined in Section \ref{sec:numericalresults}.
%% and $\bbf_{\ell}$ denotes plant noise for $\ell$-th block time index where $\bbf_{\ell}\sim\mathcal{CN}(\mathbf{0},\Rbf_\bbf)$. %is a zero-mean and temporally independent plant Gaussian vector, and $\ell$ is the block index. 
The channel model can represent a spatially correlated channel 
%a physical parameters of local scatterers or a limited angular spread 
by considering  $\text{rank}(\Rbf_\hbf)=r$ 
where $r \le N_t$. An eigen-decomposition (ED) of $\Rbf_\hbf$ is given by
\begin{equation}
\Rbf_\hbf = \Ubf\Lambdabf\Ubf^H,\label{eq:rankdefChannelCov}
\end{equation}
where $\Ubf=[\ubf_1,\cdots,\ubf_{r}]\in\mathbb{C}^{N_t\times r}$ and $\Lambdabf=\text{diag}(\lambda_1,\cdots,\lambda_r)$ is  composed of the non-zero eigenvalues of $\Rbf_\hbf$ in descending order.
Throughout the paper, we assume that the channel statistics ($a, \Rbf_\hbf$) are known to the system.\footnote{
Wireless channels are usually characterized by local quasi-stationarity \cite{Matz:05WCOM}. Experimental investigations have confirmed that the local quasi-stationarity well describes time-varying channels in an urban macrocell scenario \cite{Ispas&Dorpinghaus&Ascheid&Zemen:13SP}. This means that within local quasi-stationary time periods, the proposed method can operate with the channel statistics of low-mobility users updated from a window-based average.
Please see \cite{Noh&Zoltowski&Sung&Love:14STSP} for a discussion of a practical estimation approach in massive MIMO systems.
}

During the $\ell$-th training period, the received signal in \eqref{eq:statespacemodel_y1} can be rewritten in vector form as
\begin{equation}
\ybf_{\ell,pilot}  = \Sbf_{\ell}^H\hbf_{\ell} + \wbf_{\ell},\label{eq:statespacemodel_y2}
\end{equation}
where $\Sbf_{\ell}=[\sbf_{\ell M+1}\cdots \sbf_{\ell M+M_p}]$ denotes the transmitted training signals subject to an average transmit power constraint $E\{\|\Sbf_\ell\|_F^2\}=\rho M_p$, %with power constraint $\|\sbf_k\|_2^2=\rho$, 
$\ybf_{\ell,pilot}=[y_{\ell M+1},\cdots,y_{\ell M+M_p}]^H$, and $\wbf_\ell$ is similarly defined.
We focus on minimum mean square error (MMSE) channel estimation
based on the current and all previous received training signals given
by $\hat{\hbf}_{\ell |\ell}=E\{\hbf_{\ell}|\ybf_{pilot}^{(\ell)}\}$, where
$\ybf_{pilot}^{(\ell)}=\{\ybf_{\ell^\prime,pilot}|  \ell^\prime \le \ell\}$ denotes all received training signals up to the
$\ell$-th training period. 
%During the training period, a sequence of pilot beam matrix $\{\Sbf_{\ell}\}$ is transmitted for channel estimation.
%A simplified setup for CSI feedback is that we can consider no fading and orthogonal access for its feedback link under the assumption that the SNR on the feedback link is  comparable with the un-faded downlink SNR, i.e., $w_k \sim \mathcal{CN}(0,\beta\sigma_w^2)$ for $\beta\ge 1$ \cite{Caire&Jindal&Kobayashi&Ravindran:07ISIT}.
From \eqref{eq:statespacemodel_h} and \eqref{eq:statespacemodel_y2}, the system can be viewed as a state-space model, % for the channel vector, 
and then optimal channel estimation is given by Kalman filtering, as shown in Table \ref{tab:kalmanfiltering}. %\cite{Kailath:book}. 
Here, $(\Pbf_{\ell|\ell},\Pbf_{\ell|\ell-1})$ are the estimation and prediction error covariance matrices, and $\Kbf_\ell$ denotes the Kalman gain matrix defined as
%\begin{align}
%\Pbf_{\ell |\ell^\prime} &= E\bigl\{\bigl(\hbf_{\ell}-\hat{\hbf}_{\ell |\ell^\prime}\bigr)\bigl(\hbf_{\ell}-\hat{\hbf}_{\ell |\ell^\prime}\bigr)^H | \ybf_{pilot}^{(\ell^\prime)}\bigr\}\nonumber\\%\label{eq:mmsePl}\\
%\Kbf_{\ell}&=\Pbf_{\ell |\ell-1}\Sbf_{\ell}(\Sbf_{\ell}^H\Pbf_{\ell |\ell-1}\Sbf_{\ell}+\Ibf_{M_p})^{-1}.\nonumber
%\end{align}
%\vspace{-0.5em}
\begin{align*}
\Pbf_{\ell |\ell^\prime} &= E\bigl\{\bigl(\hbf_{\ell}-\hat{\hbf}_{\ell |\ell^\prime}\bigr)\bigl(\hbf_{\ell}-\hat{\hbf}_{\ell |\ell^\prime}\bigr)^H | \ybf_{pilot}^{(\ell^\prime)}\bigr\} \\%~~~ \text{and}~~~
\Kbf_{\ell}&=\Pbf_{\ell |\ell-1}\Sbf_{\ell}(\Sbf_{\ell}^H\Pbf_{\ell |\ell-1}\Sbf_{\ell}+\Ibf_{M_p})^{-1}.
%\vspace{-0.5em}
\end{align*}

%\vspace{-1.5em}

%% Requires the booktabs if the memoir class is not being used
%\begin{table}[h]%[tp]
%\normalsize
%\centering
%%\topcaption{Table captions are better up top} % requires the topcapt package
%\begin{tabular}{@{} l @{}} % Column formatting, @{} suppresses leading/trailing space
%\toprule
%{\em Initialization:}\\
%\vbox{\begin{align}
%\hat{\hbf}_{0|-1} ={\mathbf{0}}  
%~\mbox{ and }~  
%\Pbf_{0|-1}=\Rbf_\hbf \label{eq:initialCond}
%\end{align}}\\
%%
%{\bf while } $\ell =0,1,\cdots$ {\bf do}\\
%%
%\quad{\em Measurement update:} \\
%%
%\vbox{\begin{align}
%\hat{\hbf}_{\ell |\ell} &= \hat{\hbf}_{\ell |\ell-1}+\Kbf_{\ell}(\ybf_{\ell,pilot}-\Sbf_{\ell}^H\hat{\hbf}_{\ell |\ell-1}) \label{eq:measurementupdateH}\\
%\Pbf_{\ell |\ell} &= \Pbf_{\ell |\ell-1} - \Kbf_{\ell}\Sbf_{\ell}^H\Pbf_{\ell |\ell-1} \label{eq:measurementupdateP}
%\end{align}}\\
%%
%\quad{\em Time update:} \\
%%
%\vbox{\begin{align}
%\hat{\hbf}_{\ell+1|\ell} &= a\hat{\hbf}_{\ell |\ell} \nonumber \\
%\Pbf_{\ell+1|\ell} &= a^{2}\Pbf_{\ell |\ell}+(1-a^{2})\Rbf_{\hbf}
%\label{eq:timeupdateH}
%\end{align}}\\
%{\bf end while }\\
%\bottomrule
%\end{tabular}
%\vspace{0.1em}
%\caption{Channel estimation based on Kalman filtering \cite{Kailath:book}} \label{tab:kalmanfiltering}
%\vspace{-2.0em}
%\end{table}
\begin{table}[t]%[tp]
%\small
\centering
%\topcaption{Table captions are better up top} % requires the topcapt package
\begin{tabular}{@{} l @{}} % Column formatting, @{} suppresses leading/trailing space
\toprule
{\em Initialization:}\\
\vbox{\vspace{-0.5em}
\begin{equation}
\vspace{-0.5em}
\hat{\hbf}_{0|-1} ={\mathbf{0}}  
~\mbox{ and }~  
\Pbf_{0|-1}=\Rbf_\hbf \label{eq:initialCond}
\end{equation}}\\
{\bf while } $\ell =0,1,\cdots$ {\bf do}\\
\quad{\em Measurement update:} \\
\vbox{\vspace{-0.5em}\begin{align}
\hat{\hbf}_{\ell |\ell} &= \hat{\hbf}_{\ell |\ell-1}+\Kbf_{\ell}(\ybf_{\ell,pilot}-\Sbf_{\ell}^H\hat{\hbf}_{\ell |\ell-1}) \label{eq:measurementupdateH}\\
\Pbf_{\ell |\ell} &= \Pbf_{\ell |\ell-1} - \Kbf_{\ell}\Sbf_{\ell}^H\Pbf_{\ell |\ell-1} \label{eq:measurementupdateP}
\end{align}\vspace{-0.7em}}\\
\vspace{-0.5em}
\quad{\em Time update:} \\
\vbox{%\vspace{0.3em}
\begin{align}
\hat{\hbf}_{\ell+1|\ell} &= a\hat{\hbf}_{\ell |\ell} \nonumber \\
\Pbf_{\ell+1|\ell} &= a^{2}\Pbf_{\ell |\ell}+(1-a^{2})\Rbf_{\hbf} \label{eq:timeupdateH}
\end{align}\vspace{-1.0em}}\\
{\bf end while }\\
\bottomrule
\end{tabular}
\vspace{-0.0em}
\caption{Channel estimation based on Kalman filtering \cite{Kailath:book}} \label{tab:kalmanfiltering}
\vspace{-1.5em}
\end{table}

In this paper, we employ the concept of a {\em training frame}. A training frame is the joint design of the training signals sent over $G$ consecutive blocks. %, where recall that a block is composed of a training period and a data transmission period.
This means for each $i$, $\Sbf_{iG},\Sbf_{iG+1},\ldots,\Sbf_{(i+1)G-1}$ are jointly designed. 
%and we refer to $[\Sbf_{iG},\Sbf_{iG+1},\ldots,\Sbf_{(i+1)G-1}]$ as the $i$-th training frame. 
%In this paper, we consider a training frame composed of $G$ consecutive training periods for joint training signal design. 
We assume $G=2^s$ for $s\in\{0,1,2,\ldots\}$ for simplicity, which will be revisited later.

During the $\ell$-th data transmission period (i.e., channels uses satisfying $k=\ell M+m$ with $M_p<m\le M$), we assume that the data symbol $x_k$ is transmitted with a rank-one beamformer $\vbf_k\in\mathbb{C}^{N_t}$. To realize a low-complexity solution for the beamformer design, we assume the beamformer $\vbf_k$ is restricted to a subspace of dimension $n_d$ which should be optimized to meet the effective channel rank. Then, we can write $\vbf_k$ as a hybrid precoding $\vbf_k=\Fbf\dbf_k$, i.e., $\vbf_k$ lies in column space of $\Fbf\in\mathbb{C}^{N_t\times n_d}$ with its linear combination of $\dbf_k\in\mathbb{C}^{n_d}$ where $n_d\le N_d$. Here, $N_d$ denotes some system constraint with $1\le N_d\le N_t$ (e.g., the number of available RF chains in the case of hybrid analog-digital beamforming in Section \ref{subsec:hybrid_analogdigital}).

In a hybrid beamforming scenario, our goal is to design the pre-beamforming matrix $\Fbf$ that supports a spatial matched filter transmit beamforming (e.g., $\vbf_k=\hat{\hbf}_{\ell |\ell}/\|\hat{\hbf}_{\ell |\ell}\|_2$)\footnote{The results can be straightforwardly generalized for any linear transmit beamformings.} by using channel statistics and training signal design. Here, the pre-beamforming matrix $\Fbf$ is optimized offline and
%to transmit data symbol along the dominant eigenmodes of the channel 
%based on channel statistics by using $n_d$ dominant eigenvectors of $\Rbf_\hbf$ in \eqref{eq:rankdefChannelCov} where the variable $n_d$ should be properly optimized through training signal design. This will be discussed in next sub-section.
%This is because the design of the pre-beamforming is closely connected to those of training signals, which will be discussed in next sub-section. Note that the pre-beamforming is designed by $n_d$ dominant eigenvectors of $\Rbf_\hbf$ in \eqref{eq:rankdefChannelCov} where the variable $n_d$ should be properly optimized to account for the effective channel rank.
%Note that the pre-beamforming is fixed within a time window resulting from a window-based average to update the channel statistics of low-mobility users \cite{Matz:05WCOM,Hoydis&Hoek&Wild&Brink:12ISWCS,Ispas&Dorpinghaus&Ascheid&Zemen:13SP}.
%On the other hand, 
the post-beamforming $\dbf_k$ is determined with respect to (w.r.t.) transmit beamforming schemes by using the current channel estimate. 
%We consider a spatial matched filter transmit beamforming, given by $\vbf_k=\hat{\hbf}_{\ell |\ell}/\|\hat{\hbf}_{\ell |\ell}\|_2$ and focus on an optimization of training signal aimed for enabling the two-stage beamforming (the results can be straightforwardly generalized for any linear beamformers). 

%\vspace{-1.0em}
%%%%%%%%%%%%%%%%%%%%%%%%%%%%%%%%%%%%%%%%%%%%%%%%%%%%%%%%%%%%%%%%%%%%%%%
\subsection{Review of Prior Work}\label{subsec:pilotbeampattern}

We briefly review other work on the sequential design of the pilot beam pattern for channel estimation in massive MIMO systems. % proposed in \cite{Noh&Zoltowski&Sung&Love:13STSP}.
%The scheme exploits knowledge of channel statistics combined with Kalman filtering to design pilot beam pattern for improved channel estimation.
The channel mean square error (MSE) $\text{tr}(\Pbf_{\ell|\ell})$ in \eqref{eq:measurementupdateP} depends on the current training signal $\Sbf_\ell$ and the prediction error covariance $\Pbf_{\ell|\ell-1}$ that is a function of all previous training signals $\Sc_{\ell-1}$ %$\{\Sbf_{\ell^\prime}:\ell^\prime<\ell\}$ 
and the channel statistics $(a,\Rbf_\hbf)$ by the Kalman recursion in \eqref{eq:measurementupdateP} and  \eqref{eq:timeupdateH}.
Thus, given the previous training signal $\Sc_{\ell-1}$, the channel MSE can be minimized by properly designing the pilot beam pattern $\Sbf_\ell$. %$\{\Sbf_\ell\}$.
The following proposition presents a property of pilot beam pattern.

%Note from \eqref{eq:measurementupdateH} and \eqref{eq:measurementupdateP} that the channel MSE at the $\ell$-th training period $\text{tr}(\Pbf_{\ell |\ell})$ should be minimized by properly designed pilot beam pattern sequence $\{\Sbf_{\ell^\prime}:\ell^\prime\le\ell\}$.
%The following proposition presents a property of pilot beam pattern design.

\vspace{0.2em}
\begin{proposition} \label{pro:argminMMSE_mimo_blkfading}\cite{Noh&Zoltowski&Sung&Love:14STSP}
Given all previous pilot signals $\Sbf_{\ell^\prime}$ ($\ell^\prime <\ell$), the pilot beam signal $\Sbf_{\ell}$ at the $\ell$-th training period
minimizing $\text{tr}(\Pbf_{\ell |\ell})$ is given by a properly scaled version of the $M_p$
dominant eigenvectors of the Kalman prediction error covariance matrix $\Pbf_{\ell |\ell-1}$ for the $\ell$-th
training period.
\end{proposition}
\vspace{0.2em}

Proposition \ref{pro:argminMMSE_mimo_blkfading} states that the use of the $M_p$ dominant eigenvectors of $\Pbf_{\ell|\ell-1}$ for training signals minimizes the channel MSE at the $\ell$-th training period. Under this pilot beam pattern design,  all the Kalman matrices $(\Pbf_{\ell|\ell},\Pbf_{\ell|\ell-1})$ and the channel spatial covariance $\Rbf_\hbf$ are {\em simultaneously diagonalizable}, i.e., given the ED of $\Rbf_\hbf$ in \eqref{eq:rankdefChannelCov}, we have $\Pbf_{\ell|\ell}=\Ubf\bar{\Lambdabf}^{(\ell)}\Ubf^H$ and $\Pbf_{\ell|\ell-1}=\Ubf\Lambdabf^{(\ell)}\Ubf^H$ where $\bar{\Lambdabf}^{(\ell)}$ and $\Lambdabf^{(\ell)}$ denote diagonal matrices composed of the eigenvalues of $\Pbf_{\ell|\ell}$ and $\Pbf_{\ell|\ell-1}$, respectively.
%This yields that all the sequentially optimal pilot beam patterns are selected from a finite set that is defined by the eigenvectors of $\Rbf_\hbf$.
This yields that all the eigenvectors of $\Pbf_{\ell|\ell-1}$ over time are selected from the set of eigenvectors of $\Rbf_\hbf$ defined by $\{\ubf_1,\ldots,\ubf_{r}\}$ in \eqref{eq:rankdefChannelCov}.
However, the pilot beams patterns considered in this approach are inherently obtained in a high-dimensional space % requires quite a few kinds of eigenvectors of $\Rbf_\hbf$ over time 
and %also 
makes a closed-form performance analysis intractable due to the greedy %sequential 
search of the dominant eigenvectors of $\Pbf_{\ell|\ell-1}$.

%%%%%%%%%%%%%%%%%%%%%%%%%%%%%%%%%%%%%%%%%%%%%%%%%%%%%%%%%%%%%%%%%%%%%%%
\section{Proposed Training Sequence Framework}\label{sec:codebookpilotbeampattern}
%%%%%%%%%%%%%%%%%%%%%%%%%%%%%%%%%%%%%%%%%%%%%%%%%%%%%%%%%%%%%%%%%%%%%%%
%\vspace{-0.5em}

%Compared to conventional MIMO schemes, it will be of interest to deploy massive MIMO systems where a number of RF chains is less than the number of transmit antennas to reduce implementation complexity and power consumption at the base station \cite{Renzo&Haas&Ghrayeb&Sugiura&Hanzo:14IEEE}. 
%Thus, in massive MIMO systems, channel estimation and beamforming technique should be designed in order to strike a good tradeoff between performance and complexity.
%% improve the without imposing excessive complexity to the transceiver.
%%\tcr{the essential point of} massive MIMO system design is to operate with a limited number of RF chains to reduce implementation complexity and power consumption at the base station \cite{Renzo&Haas&Ghrayeb&Sugiura&Hanzo:14IEEE}. 

In this section, we first focus on the design of a reduced dimensionality training sequence that has a suitable mapping to training signals used during the training periods.
%The proposed training codebook is designed to minimize the steady-state channel MSE by exploiting the channel statistics obtained by uplink training or a feedback-assisted scheme from the receiver. 
We next provide a hybrid analog-digital beamforming method that exploits the limited number of available RF chains relative to the number of antennas.

%\vspace{-0.5em}
%%%%%%%%%%%%%%%%%%%%%%%%%%%%%%%%%%%%%%%%%%%%%%%%%%%%%%%%%%%%%%%%%%%%%%%
\subsection{Motivation for Proposed Scheme}\label{subsec:motivation}

Because of the optimal training signals' properties mentioned in Proposition \ref{pro:argminMMSE_mimo_blkfading}, we assume that each training matrix $\Sbf_\ell$ is a scaled version of $M_p$ eigenvectors of $\Rbf_\hbf$ in \eqref{eq:rankdefChannelCov} to satisfy the power constraint.\footnote{
Note that, 
if each of the training signals 
is %designed by a linearly combined version of a set of eigenvectors of $\Rbf_\hbf$,
a linear combination of the eigenvectors of $\Rbf_\hbf$,
the eigenvectors of $\Pbf_{\ell|\ell-1}$ do not remain the same throughout the training, i.e., they change over time by the Kalman recursion.
This implies that the ED of $\Pbf_{\ell|\ell-1}$ at each training period is required to compute the dominant eigenvectors, and this can be computationally expensive since $N_t$ is assumed to be large for massive MIMO systems.}
%To tackle the optimization problem of the pre-beamforming $\Fbf$ and a variable $n_d$ on the effective channel rank, we assume the columns of the training signal matrix $\Sbf_\ell\in\mathbb{C}^{N_t\times M_p}$ to be scaled versions of $M_p$ eigenvectors of $\Rbf_\hbf$ in \eqref{eq:rankdefChannelCov} to satisfy the power constraint, based on the property of the optimal training signals in Proposition \ref{pro:argminMMSE_mimo_blkfading}. %This means, we can specify each $\Sbf_\ell$ by the indices of the eigenvectors chosen.
From \eqref{eq:measurementupdateH}, the channel estimate at the $\ell$-training period is a {\em linear} combination of all previously used training signals $\Sc_\ell:=\{\Sbf_{\ell^\prime}:\ell^\prime\le\ell\}$ by Kalman recursion in \eqref{eq:measurementupdateP} and \eqref{eq:timeupdateH}, i.e., the channel estimate $\hat{\hbf}_{\ell |\ell}$ lies in the column space of $\Sc_\ell$. 
%This yields that the pre-beamforming matrix $\Fbf$ will span the subspace spanned by the training signal $\Sc_\ell$ for enabling the hybrid beamforming $\vbf_k=\Fbf\dbf_k$. %used for channel estimation. 
That is, in the hybrid beamforming structure of $\vbf_k=\Fbf\dbf_k$, we require that the pre-beamforming matrix $\Fbf$ spans the subspace spanned by the training signal $\Sc_\ell$ 
for subspace sampling of the channel estimate.
%and $\dbf_k$ is given by its linear combination to represent the current channel estimate. 
%Then, we focus on an optimization of the pre-beamformer $\Fbf$ for enabling the two-stage beamforming $\vbf_k=\Fbf\dbf_k$.
%Under training-based channel estimation systems, we focus on an optimization of the pre-beamformer $\Fbf$ aimed at leveraging data rate.
%In the case of spatial matched filter beamforming, the transmit precoding is given by $\vbf_k=\hat{\hbf}_{\ell |\ell}/\|\hat{\hbf}_{\ell |\ell}\|_2$.
%This means that the channel vector should be estimated in the column space of $\Fbf$ for enabling the two-stage beamforming $\vbf_k=\Fbf\dbf_k$.
%Note from \eqref{eq:measurementupdateH}  that the channel estimate at the $\ell$-training period is a function of all previously used training signals $\Sc_\ell:=\{\Sbf_{\ell^\prime}:\ell^\prime\le\ell\}$ by Kalman recursion in \eqref{eq:measurementupdateP} and \eqref{eq:timeupdateH}. 
%This yields, during the downlink data beamforming, the basis vectors $\Fbf$ will span the subspace of the training signal $\Sc_\ell$ used for channel estimation 
%%(i.e., $\Cc(\Fbf)\subseteq\Cc(\Sc_{\ell})$) 
%and $\dbf_k$ is given by its linear combination to represent the current channel estimate. 
Therefore, the training signal $\Sc_\ell$ should be suitably designed to capture the $n_d$ dominant channel eigenmodes under the $n_d\le N_d$ dimensionality constraint where the variable $n_d$ should be properly optimized to account for the effective channel rank.
Note that the pre-beamforming matrix $\Fbf$ is then determined by a set of $n_d$ distinct eigenvectors of $\Rbf_\hbf$ used in the training signals $\Sc_\ell$.

Alternatively, based on applying the Toeplitz distribution theorem to the channels of large-scale linear antenna arrays, the eigenvectors of the correlation matrix $\Rbf_\hbf$ are well approximated by columns of a unitary {\em discrete Fourier transform} (DFT) matrix. 
Then, the pre-beamformer can be designed using some columns of the DFT matrix used in the training signals, as later discussed in Section \ref{subsec:hybrid_analogdigital}.

\begin{figure*}[t]
\centerline{
\psfrag{(G)}[c]{\footnotesize $G$} %
\psfrag{(Nt)}[c]{\footnotesize $N_t$ eigenvectors of $\Rbf_\hbf$} %
\psfrag{(g1)}[l]{\scriptsize $g_1$} %
\psfrag{(g2)}[l]{\scriptsize $g_2=g_4$} %
\psfrag{(g3)}[l]{\scriptsize $g_3$} %
\psfrag{(g5)}[l]{\scriptsize $g_5$} %
\psfrag{(g6)}[l]{\scriptsize $g_6$} %
\psfrag{(m)}[c]{\scriptsize $m=$} %
\psfrag{(1stm)}[c]{\scriptsize $1$} %
\psfrag{(2ndm)}[c]{\scriptsize $2$} %
\psfrag{(Mpthm)}[c]{\scriptsize $3$} %
\psfrag{(1st)}[c]{\scriptsize $1$} %
\psfrag{(2nd)}[c]{\scriptsize $2$} %
\psfrag{(3rd)}[c]{\scriptsize $3$} %
%\psfrag{(Mpth)}[c]{\scriptsize $M_{\hspace{-0.2em}p}$} %
\psfrag{(M)}[c]{\footnotesize $M$} %
\psfrag{(u1)}[c]{\scriptsize $\ubf_1$} %
\psfrag{(u2)}[c]{\scriptsize $\ubf_2$} %
\psfrag{(u3)}[c]{\scriptsize $\ubf_3$} %
\psfrag{(u4)}[c]{\scriptsize $\ubf_4$} %
\psfrag{(u5)}[c]{\scriptsize $\ubf_5$} %
\psfrag{(u6)}[c]{\scriptsize $\ubf_6$} %
\psfrag{(u7)}[c]{\scriptsize $\ubf_7$} %
%\psfrag{(u8)}[c]{\scriptsize $\ubf_8$} %
%\psfrag{(u9)}[c]{\scriptsize $\ubf_9$} %
%\psfrag{(uNrf)}[c]{\scriptsize $\ubf_{\hspace{-0.10em}n_{\hspace{-0.15em}{\tiny d}}}$} %
\psfrag{(uNrf)}[c]{\scriptsize $\ubf_{8}$} %
\psfrag{(ui)}[c]{\scriptsize $\ubf_i$} %
\psfrag{(cdot)}[c]{\scriptsize $\cdots$} %
\psfrag{(figA)}[c]{\footnotesize (a) Allocation of training signal} %
\psfrag{(figB)}[c]{\footnotesize (b) $G$ consecutive blocks } %
\psfrag{(exPilot)}[c]{\footnotesize $(c)$} %
\psfrag{(exData)}[c]{\footnotesize $(d)$} %
\psfrag{(t1)}[l]{\scriptsize $\ell=0$} %
\psfrag{(t2)}[l]{\scriptsize $\ell=1$} %
\psfrag{(tg)}[l]{\scriptsize $\ell=3$} %
\psfrag{(tf)}[c]{\scriptsize Training} %
\psfrag{(tf2)}[c]{\scriptsize frame} %
\psfrag{(codebook1)}[c]{\scriptsize Training} %
\psfrag{(codebook2)}[c]{\scriptsize resource} %
\psfrag{(codebook3)}[c]{\scriptsize block} %
\psfrag{(C)}[c]{\scriptsize ($G\times M_p$)} %
\psfrag{(pilot)}[c]{\scriptsize Training} %
\psfrag{(data)}[c]{\scriptsize Data transmission} %
\includegraphics[scale=1.6]{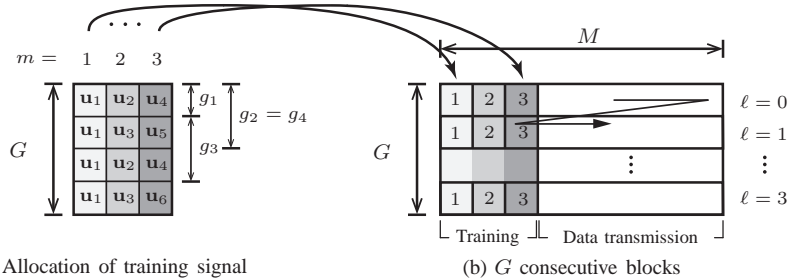}}
\vspace{-0.0em}
\caption{$G$ consecutive blocks where $G=4$, $M_p=3$, $n_d=6$, and symbol time $k=\ell M+m$.
Training signals sent over $G$ consecutive blocks are jointly designed.  
%A training frame denotes a set of channel uses for training over $G$ consecutive blocks. 
%: (c) $\ubf_i$ is used and (d) $\ubf_i$ is not used where $i\in\{1,\cdots,N_t\}$.
} \label{fig:pilotbeam_codebook_design}
\vspace{-1.0em}
\end{figure*}

The $M_p$ columns of each $\Sbf_\ell\in\mathbb{C}^{N_t\times M_p}$ are represented by a $1\times M_p$ index vector where the $i$-th entry of the vector equals to the index of the eigenvector of $\Rbf_\hbf$ that defines the $i$-th column of $\Sbf_\ell$. 
%each entry chosen as an indices of the eigenvectors $\{\ubf_i\}$ of $\Rbf_\hbf$. 
For example, if $\Sbf_{0}=\sqrt{\rho}[\ubf_1,\ubf_2,\ubf_4]$ and $\Sbf_{1}=\sqrt{\rho}[\ubf_1,\ubf_3,\ubf_5]$ are given for $M_p=3$, those training signals are characterized by the index vectors of  $[1, 2, 4]$ and $[1, 3, 5]$, respectively.
By collecting $G$ consecutive training periods to define the training signal $\Sc_{G-1}=\{\Sbf_\ell: 0\le \ell <G\}$,  we can then succinctly define the training signal $\Sc_{G-1}$ by an index matrix $\Cbf\in\mathbb{C}^{G\times M_p}$ with each row representing the eigenvector indices used during the corresponding training period.
For example, Fig.  \ref{fig:pilotbeam_codebook_design}(b) shows $G$ consecutive blocks starting from $\ell=0$
%a $G\times M_p$ training resource block 
where the training signal matrix $\Sbf_\ell$ will be  transmitted at the $\ell$-th training period where $0\le \ell < G$.
An example of the index matrix with $G=4$ and $M_p=3$ in Fig. \ref{fig:pilotbeam_codebook_design} is given by
{%\small
\begin{align}
\Cbf&=
\left[\begin{array}{cccc}
1 & 1 & 1 & 1 \\
2 & 3 & 2 & 3 \\
4 & 5 & 4 & 6 
\end{array}\right]^T, \label{eq:trainingcodebookex}
\end{align}}\noindent
then we have the  $\ell$-th training signal matrix $\Sbf_{\ell}=\sqrt{\rho}[\ubf_{[\Cbf]_{\ell+1,1}},\ubf_{[\Cbf]_{\ell+1,2}},\cdots,\ubf_{[\Cbf]_{\ell+1,M_p}}]$. % where $0\le \ell < G$.

We have discussed how the index matrix $\Cbf$ defines the training signals $\{\Sbf_\ell:0\le \ell <G\}$ where we will refer to $\Cbf$ as the {\em training sequence} (index) matrix. 
In the next subsections, we present a systematic approach to the training sequence optimization.

%\vspace{-0.5em}
%%%%%%%%%%%%%%%%%%%%%%%%%%%%%%%%%%%%%%%%%%%%%%%%%%%%%%%%%%%%%%%%%%%%%%%
\subsection{Problem Formulation}\label{subsec:problemFormulation}

To construct the training sequence $\Cbf$, we need to define how the selected $n_d$ eigenvectors should be allocated in the $G$ consecutive training periods.
%For simplicity, we restrict the value of $G$ to be $2^s$ where $s\in\{0,1,2,\ldots\}$. 
We assume that each of the selected eigenvectors will not be transmitted more than once within each of the training periods in order to allocate the training signals across $G$ consecutive training periods. Note that in this case each training signal matrix $\Sbf_\ell$ is composed of $M_p$ distinct eigenvectors.
Therefore, for each $\ell$, $\Sbf_\ell^H\Sbf_\ell=\Ibf_{M_p}$.
%That is, some dominant eigenvectors can be used more frequently than others, e.g., $\ubf_1$ and $\ubf_2$ are constantly transmitted at symbol time $\ell M+1$ and $\ell M+2$ where $0\le \ell <G$ and others are transmitted with some block time-wise intervals in Fig. \ref{fig:pilotbeam_codebook_design}(a).  
%
%To define pilot placement in the multiple training phases, 
%
%For $1\le i\le n_d$, denote by $g_i$ the block time-wise interval 
%%for the use of $\ubf_i$ as a training signal where $0\le g_i\le G$ and $1\le i \le N_t$. That is, 
%where if $\ubf_i$ is firstly used as a training signal at symbol time $\tilde{\ell} M+m$ for $0\le \tilde{\ell}<G$, $1\le m\le M_p$, and $0<g_i\le G$, the training signal on $\ubf_i$ will be re-transmitted at symbol time $(\tilde{\ell} + g_i\cdot(\ell)_{G/g_i})M + m$ for $0\le \ell <G$. Here, $(\cdot)_{b}$ denotes the integer modulo $b$. 
%For $n_d<i\le N_t$, we set $g_i=0$ because $\ubf_i$ is not used as a training signal. 
For $1\le i\le n_d$ and $0<g_i\le G$, denote the block time-wise interval used for the training signal $\ubf_i$ allocation across $G$ blocks by $g_i$, 
%for the use of $\ubf_i$ as a training signal where $0\le g_i\le G$ and $1\le i \le N_t$. That is, 
where if $\ubf_i$ is firstly used as a training signal at symbol time $\ell^\prime M+m$ for some $0\le \ell^\prime<G$ and $1\le m\le M_p$, the training signal corresponding to $\ubf_i$ will be re-transmitted at symbol time $\{(\ell^\prime+ q\cdot g_i)M+m: \text{ for } 1\le q <G/g_i\}$. (For the example of $g_3=2$ in Fig. \ref{fig:pilotbeam_codebook_design}, $\ubf_3$ is firstly transmitted at symbol time $M+2$, and re-transmitted at symbol time  $3M+2$.)
%$(\tilde{\ell} + g_i\cdot(\ell)_{G/g_i})M + m$ for $0\le \ell <G$. Here, $(\cdot)_{b}$ denotes the integer modulo $b$. 
For $n_d<i\le N_t$, we set $g_i=0$ because $\ubf_i$ is not used as a training signal.
An example with $g_1=1$, $g_2=g_3=g_4=2$, $g_5=g_6=4$, and $n_d=6$ %, and $g_{n_{RF}+1}=\cdots=g_{N_t}=0$ 
is shown in Fig. \ref{fig:pilotbeam_codebook_design}(a), where we omit the illustration for $g_5$ and $g_6$ for the brevity.

The dimension of the offline-designed precoder, denoted as $n_d$, is clearly constrained by the system.  
First, the training signal construction allows the transmitter to sound {\em at least} $M_p$ different subspace dimensions.  
Therefore, $n_d$ should be restricted to be at least $M_p$ in order to span as large of a subspace as possible. In the same way, the transmitter structure (e.g., hybrid precoder structure) means that the precoder must satisfy $n_d\leq N_d$ and $n_d\le r$ where $\text{rank}(\Rbf_\hbf)=r$ in \eqref{eq:rankdefChannelCov}.  
Second, %a training frame can sound {\em at most} $GM_p$ different dimensions.  
there are {\em at most} $GM_p$ different dimensions in the training sequence $\Cbf$. Thus, $n_d \leq G M_p$.
Therefore, we notice that the number of distinct eigenvectors $(n_d)$ used in the training sequence should satisfy $M_p\le n_d\le \min\{GM_p,N_d,r\}$. %, where recall that $\text{rank}(\Rbf_\hbf)=r$ in \eqref{eq:rankdefChannelCov}. 
%Here, we can use up to $r$ eigenvectors for designing $\Cbf$ to account for the dominant channel directions because using non-dominant eigenvectors for the training signal may lead to a loss in effective received power. 
The unknown parameter $n_d$ should be jointly optimized with $\{g_i\}$, which defines the training sequence $\Cbf$.
We then consider the following condition for the design of training sequence.

\vspace{0.5em}
\noindent{\em Condition (C.1)}: For each $i$ ($1\le i\le n_d$), the block time-wise interval $g_i$ is a divisor of $G$, i.e., $g_i\in \Ic_{G}:=\{d_j| (G)_{d_j}=0 \text{ for } 1\le d_j \le G \text{ and } d_j< d_{j+1}\}$, then we have an integer of $G/g_i$.
\vspace{0.5em}

\noindent
Here, $(\cdot)_{b}$ denotes the integer modulo $b$. 
%Here, $(\cdot)_{g_i}$ denotes the integer modulo $g_i$.
Condition {\em (C.1)} guarantees that once the training sequence $\Cbf\in\mathbb{N}^{G\times M_p}$ is designed with $n_d$ and $\{g_i\}$, 
the training sequence can be used for all subsequent training frames due to the periodic pilot allocation patterns,
%the training sequence can be used periodically across $G$ successive block transmissions due to periodic pilot allocation patterns, %with the block time-wise intervals $\{g_i\}$, 
i.e., we transmit the training signals $\Sbf_{\ell}=\Sbf_{(\ell)_G}$ where $\ell=0,1,\ldots$.
Note that, when $G=2^s$ for some nonnegative integer $s$, the set of divisors of $G$ is given by 
{%\small
\begin{equation}
\Ic_G=\{1,2,\ldots,2^{s-1},2^s\}.\label{eq:divisorset}
\end{equation}}\noindent
 % with $|\Ic_G|=s+1$.
%Then, the training codebook $\Cbf\in\mathbb{N}^{G\times M_p}$ can specify the training signals $\{\Sbf_{\ell},\Sbf_{\ell+1},\cdots,\Sbf_{\ell+G-1}\}$ periodically over $G$ contingent transmission. 
%transmitted during G 
%$\{\Sbf_{\ell},\Sbf_{\ell+1},\cdots,\Sbf_{\ell+G-1}\}$ for $\ell=0,1,\ldots$.
%$\Sc_{\ell=nG}$ where $n\in\mathbb{N}$.

Given the block time-wise vector $\gbf=[g_1,g_2,\cdots,g_{n_d}]^T$ satisfying {\em (C.1)}, the training signal vectors $\{\ubf_i:1\le i\le n_d\}$ are interspersed corresponding to the block time-wise intervals $\{g_i\}$ across $G$ consecutive training periods. 
%Thus, when the training signal of $\ubf_i$ is not used during $g_i-1$ inter-block symbol times, the channel MSE along the direction of $\ubf_i$ monotonically increases by \eqref{eq:timeupdateH} where $1\le i\le n_d$. 
Then, we evaluate the channel estimation performance by deriving the minimum steady-state channel MSE for $1\le i \le n_d$ given by
{%\small 
\begin{align}
\lambda_{i,g_i}^{(\underline{\infty})}
&:=  \lim_{\ell\rightarrow \infty}  \bar{\lambda}_i^{(\ell)} \nonumber\\%\label{eq:minSSMSE_v1}\\
%:= \min\lim_{\ell\rightarrow \infty}  \lambda_i^{(\ell)} \label{eq:minSSMSE_v1}\\
&=  \frac{\bigl(a^{2g_i}\lambda_{i,g_i}^{(\underline{\infty})} + (1-a^{2g_i})\lambda_i\bigr)}{\rho\bigl(a^{2g_i}\lambda_{i,g_i}^{(\underline{\infty})} + (1-a^{2g_i})\lambda_i\bigr) + 1} \nonumber\\%\\
%&=  -\frac{(1-a^{2g_i})(1+\rho\lambda_i)}{2a^{2g_i}\rho} \nonumber\\
%&~~~ + \sqrt{\left(\frac{(1-a^{2g_i})(1+\rho\lambda_i)}{2a^{2g_i}\rho}\right)^2 + \frac{(1-a^{2g_i})\lambda_i}{a^{2g_i}\rho}}, \label{eq:minSSMSE_v2}
%&
&=\frac{\lambda_i}
{\bigl(\frac{1}{2}(1+\lambda_i\rho)\bigr) + \sqrt{\bigl(\frac{1}{2}(1+\lambda_i\rho)\bigr)^2+\frac{a^{2g_i}}{1-a^{2g_i}}\lambda_i\rho}},\label{eq:minSSMSE_v2}
\end{align}}\noindent
from the Riccati equation\cite{Dong&Tong&Sadler:04SP}
{%\small
\begin{align}
\bar{\lambda}_i^{(\ell)}
&= 
\frac{\bigl(a^{2g_i}\bar{\lambda}_{i}^{(\ell-g_i)} + (1-a^{2g_i})\lambda_i\bigr)}{\rho\bigl(a^{2g_i}\bar{\lambda}_{i}^{(\ell-g_i)} + (1-a^{2g_i})\lambda_i\bigr) + 1},
\end{align}}\noindent
where $\bar{\lambdabf}^{(\ell)}=[\bar{\lambda}_1^{(\ell)},\cdots,\bar{\lambda}_{r}^{(\ell)}]^T:=\text{diag}(\bar{\Lambdabf}^{(\ell)})$  and $\lambdabf=[\lambda_1,\cdots,\lambda_{r}]^T:=\text{diag}(\Lambdabf)$ denote the eigenvalues of $\Pbf_{\ell|\ell}$ and $\Rbf_\hbf$, respectively.
%Recall from Section \ref{subsec:pilotbeampattern} that the EDs of $\Pbf_{\ell|\ell-1}$ and $\Rbf_\hbf$ are given by $\Pbf_{\ell |\ell-1}=\Ubf\Lambdabf^{(\ell)}\Ubf^H$ and $\Rbf_\hbf=\Ubf\Lambdabf\Ubf^H$ with the initial conditions \eqref{eq:rankdefChannelCov} and \eqref{eq:initialCond}.
Note that $\lambda_{i,g_i}^{(\underline{\infty})}$ represents the minimum steady-state channel MSE because it is the steady-state response obtained from the measurement step in \eqref{eq:measurementupdateP} when the training signal corresponding to $\ubf_i$ is transmitted according to the block time-wise interval $g_i$ at each training period.
%Here, the minimum steady-state channel MSE of $\lambda_{i,g_i}^{(\underline{\infty})}$ is obtained when the training signal corresponding to $\ubf_i$ is  transmitted according to the block time-wise interval $g_i$ at training period.
A closed-form expression for the right-hand side of (11) can be derived by algebraic manipulation of the Riccati equation in a steady-state condition which is a linear second-order equation in $\lambda_{i,g_i}^{(\underline{\infty})}$.

%The minimum steady-state channel MSE $\lambda_{i,g_i}^{(\underline{\infty})}$ is obtained when the index $\ell_p$ matches with the block index time (with modulo operation) at which the training vector $\ubf_i$ is transmitted as a training signal. %, i.e., matches with the gray bin in Fig. \ref{fig:pilotbeam_codebook_design}(a).
Once the training signal corresponding to $\ubf_i$ is transmitted, the training signal is not transmitted during the next $g_i-1$ successive blocks where the channel MSE along the direction of $\ubf_i$ monotonically increases by \eqref{eq:timeupdateH} for $1\le i\le n_d$. 
Note that the (vector) prediction step in \eqref{eq:timeupdateH} can be decomposed into a set of scalar equations by its projection to the eigenvectors $\Ubf$ (i.e., $\Lambdabf^{(\ell+1)}=a^2\bar{\Lambdabf}^{(\ell)}+(1-a^2)\Lambdabf$), where we use $\Pbf_{\ell|\ell}=\Ubf\bar{\Lambdabf}^{(\ell)}\Ubf^H$, $\Pbf_{\ell|\ell-1}=\Ubf\Lambdabf^{(\ell)}\Ubf^H$, and $\Rbf_\hbf=\Ubf\Lambdabf\Ubf^H$.
Thus, during the $g_i-1$ blocks, the minimum steady-state channel MSE $\lambda_{i,g_i}^{(\underline{\infty})}$ grows according to \eqref{eq:timeupdateH} because the Kalman filter predicts the channel state along the direction of the training signal. 
In this case, the maximum steady-state MSE of the training signal (denoted as $\lambda_{i,g_i}^{(\overline{\infty})}$) is reached after the $g_i-1$ blocks and is given by
%when $\ubf_i$ is not transmitted for a training signal during the $g_i-1$ inter-blocks, the minimum steady-state channel MSE $\lambda_{i,g_i}^{(\underline{\infty})}$ grows up by \eqref{eq:timeupdateH}. 
%Then, the maximum steady-state channel MSE (denoted as $\lambda_{i,g_i}^{(\overline{\infty})}$) corresponding to $\ubf_i$ is given by
\begin{align}
\lambda_{i,g_i}^{(\overline{\infty})} 
%&:= \max_{\ell_p\in\{1,\cdots,g_i\}} \lim_{\substack{\ell\rightarrow \infty\\\text{s.t. }(\ell)_{g_i}+1=\ell_p}}  \lambda_i^{(\ell)} \label{eq:maxSSMSE_v1}\\
&:=  a^{2(g_i-1)}\lambda_{i,g_i}^{(\underline{\infty})}   + (1-a^{2(g_i-1)})\lambda_i, \label{eq:maxSSMSE_v2}
\end{align}
where $1\le i\le n_d$.
Note that the steady-state channel MSEs of the unused eigenvectors in the training sequence
%corresponding to the unused eigenvectors for training signals 
remain constant over time, i.e., $\lim_{\ell\rightarrow \infty} \lambda_i^{(\ell)} = \lambda_{i}$ for $n_d< i \le r$.

\vspace{0.5em}
\begin{remark}\label{rem:boundSSMSE}
From \eqref{eq:minSSMSE_v2} and \eqref{eq:maxSSMSE_v2}, the steady-state channel MSE is bounded by
\begin{align}
%&\sum_{i=1}^{N_p}\lambda_i^{(\underline{\infty})}
%\le \lim_{\ell\rightarrow \infty} \text{tr}(\Pbf_{lM+m| lM+m}) 
%\le \sum_{i=1}^{N_p}\lambda_i^{(\overline{\infty})}\label{eq:boundSSMSE}
%&\|\lambdabf_{\gbf}^{(\underline{\infty})}\|_1 
%\le \lim_{\ell\rightarrow \infty} \text{tr}(\Pbf_{\ell M+m| \ell M+m}) 
%\le \|\lambdabf_{\gbf}^{(\overline{\infty})}\|_1,\label{eq:boundSSMSE}
&\text{diag}\bigl(\lambdabf_{\gbf}^{(\underline{\infty})}\bigr)
\preceq \lim_{\ell\rightarrow \infty} \Pbf_{\ell |\ell}
\preceq \text{diag}\bigl(\lambdabf_{\gbf}^{(\overline{\infty})}\bigr), \label{eq:boundSSMSE}
\end{align}
where %$c:=\sum_{i=N_p+1}^{N_t} \lambda_i^{(0)}$, 
$\lambdabf_{\gbf}^{(\underline{\infty})} = [\lambda_{1,g_1}^{(\underline{\infty})},\cdots,\lambda_{n_d,g_{n_d}}^{(\underline{\infty})}, \lambda_{n_d+1},\cdots,\lambda_{r}]^T$ and $\lambdabf_{\gbf}^{(\overline{\infty})}$ is similarly defined. $\Abf\succeq 0$ denotes a positive semidefinite matrix.
The gap between upper and lower bounds for the steady-state channel MSE $\text{tr}(\Pbf_{\ell|\ell})$ is given by 
%\begin{align}
$\sum_{i=1}^{n_d}(1-a^{2(g_i-1)})\bigl(\lambda_i-\lambda_{i,g_i}^{(\underline{\infty})}\bigr).$
%\end{align}
\end{remark}
\vspace{0.5em}

%We propose an optimization problem to design training codebook by minimizing an upper bound on the steady-state channel MSE.
Note that the trace of the steady-state channel MSE is bounded by $\|\lambdabf_{\gbf}^{(\overline{\infty})}\|_1$ in \eqref{eq:boundSSMSE}, then we formulate the optimization problem that designs the training sequence by minimizing an upper bound on the steady-state channel MSE, which is formally stated as follows.%\footnote{\tcgry{XXXXX Gap between upper and lower bound is given by: $(1-a^{2(g_i-1)})\left(\lambda_i^{(0)}-\lambda_{i,g_i}^{(\underline{\infty})}\right)$}}

\vspace{0.5em}
\begin{problem}[MMSE upper bound minimization]\label{prob:limitedRF}
Given the parameters $(G, M_p, N_d, r)$, 
solve for $\gbf^*$ and $n_d^*$ such that
%the optimal $\gbf^*$ and $n_d^*$ are chosen such that
%design $\gbf^*\in\mathbb{N}^{n_d}$ and $n_d^*$ such that they are the minimizers of 
\begin{align}
\min_{\mathbf{g},n_d} &~~~ \|\lambdabf_{\gbf}^{(\overline{\infty})}\|_1 \label{eq:objfunc} \\
\text{subject to} &~~~ \text{\em (C.1)} ~ \text{ and } ~ M_p\le n_d\le \min\{GM_p, N_d, r\},\nonumber \\
&~~~ 
\sum_{i=1}^{n_d}{1}/{g_i}=M_p. \label{eq:nonlinearConst}
\end{align}
\end{problem}
\vspace{0.5em}

The nonlinear constraint \eqref{eq:nonlinearConst} yields a total training resource constraint in the $G\times M_p$ training sequence $\Cbf$. % in the $B\times M_p$ pilot signal block. 
That is, for $1\le i\le n_d$, each training signal $\ubf_i$ is transmitted according to the block time-wise interval $g_i$ across the $G$ consecutive training periods where the training signal on $\ubf_i$ is transmitted $G/g_i$ times during the training periods. Then the total number of channel uses for sounding the $n_d$ training signals should be equal to $GM_p$, i.e.,  $\sum_{i=1}^{n_d}G/g_i = GM_p$.
Thus, the constraint \eqref{eq:nonlinearConst} guarantees that all the entries of $\Cbf$ can be constructed with a proper allocation scheme, which will be discussed in Proposition \ref{cor:sufficientCondForTrainingAllocation} of the next subsection. 
The nonlinear inequality \eqref{eq:nonlinearConst} and the periodicity condition of {\em (C.1)} make solving Problem \ref{prob:limitedRF}  extremely difficult, particularly because the optimization variables $\gbf$ and $n_d$ are interconnected.

%\vspace{-0.5em}
%%%%%%%%%%%%%%%%%%%%%%%%%%%%%%%%%%%%%%%%%%%%%%%%%%%%%%%%%%%%%%%%%%%%%%%
\subsection{Training Sequence Design}\label{subsec:trainingCodebookDesign}

To tackle the challenge of this problem, we first consider an exhaustive search in a finite search space that arises from the integer constraint of Problem \ref{prob:limitedRF}. 
In this case, it is important to reduce the computational complexity of the exhaustive search, and thus we derive a property of the objective function in \eqref{eq:objfunc}.
\vspace{0.5em}
\begin{proposition}\label{lem:monotonicitySSMSE}
$\lambda_{i,g_i}^{(\overline{\infty})}$ is a monotonic increasing function of $g_i$, i.e., $\lambda_{i,g_i^\prime}^{(\overline{\infty})} \le \lambda_{i,g_i}^{(\overline{\infty})}$ if $1\le g_i^\prime \le g_i$.
%\tcb{\% Monotonicity of the objective function in Problem 1 with respect to $g_i$!}
\end{proposition}

%\vspace{0.5em}
{\em Proof:} See Appendix \ref{subsec:monotonicitySSMSE}.
\vspace{0.5em}

Proposition \ref{lem:monotonicitySSMSE} shows that the maximum steady-state channel MSE of $\lambda_{i,g_i}^{(\overline{\infty})}$ is monotonically reduced with decreasing $g_i$ (i.e., MSE is reduced by transmitting the training signal corresponding to $\ubf_i$ more frequently).
Thus, we assume that the block time-wise interval $\gbf$ is arranged in ascending order such that $g_i\le g_j$ for $i\le j$ in order to effectively minimize the dominant channel MSE of $\lambdabf_{\gbf}^{(\overline{\infty})}$ that corresponds to the dominant channel directions.
It is numerically confirmed that the assumption of the ordered $\gbf$ is consistent with the result of the exhaustive search and 
yields a much simpler implementation compared to the initial number of trials for the exhaustive search $\Oc(|\Ic_G|^{\min\{GM_p,N_d,r\}})$.
Furthermore, by exploiting the monotonicity in Proposition \ref{lem:monotonicitySSMSE}, we propose an efficient algorithm for training sequence design that sequentially minimizes the maximum upper bound of the steady-state channel MSE. 
The corresponding algorithm is summarized in Algorithm \ref{alg:minmaxCodebookDesign}, which requires substantially less computational complexity $\Oc(GM_p)$ while achieving most of the performance gain compared to the exhaustive search.

%Simulation results will be provided in Section \ref{sec:numericalresults}.
\begin{algorithm} [h]                       % enter the algorithm environment
\caption{Min-Max Training Sequence Design}          % give the algorithm a caption
\label{alg:minmaxCodebookDesign}                 % and a label for \ref{} commands later in the document
\begin{algorithmic}[1]                          % enter the algorithmic environment
\REQUIRE Perform the ED of $\Rbf_\hbf=\Ubf\Lambdabf\Ubf^H$.
Store $\lambdabf=\text{diag}(\Lambdabf)$.

\STATE Set $\gbf=(G+1)\mathbf{1}_{N_t\times 1}$, $\qbf=\mathbf{0}_{N_t\times 1}$, $\lambdabf_{\gbf}^{(\overline{\infty})}=\lambdabf$, $\Nc_d=\{1,\cdots,N_d\}$, and $N_{blk}=GM_p$.
%\STATE $\lambdabf=\lambdabf^{(1)}$ and partition
%$\lambdabf=[\lambdabf_1^T,\cdots,\lambdabf_{N_t}^T]^T$

\WHILE{$N_{blk}>0$}
\STATE $\left(\begin{array}{rl}
\hspace{-0.2em}i^\prime =&\hspace{-0.5em} \textstyle\argmax_{ i \in \Nc_d} \lambda_{i,g_i}^{(\overline{\infty})}\\
\hspace{-0.2em}d^* =&\hspace{-0.5em} \textstyle\max_{j: d_j<  g_{i^\prime}} d_j  \text{ where } d_j \in \Ic_G  
\end{array}\right.$
%\vspace{-1.5em}
%\begin{align*} 
%\hspace{-1.5em}
%\left(\begin{array}{rl}
%\hspace{-0.2em}i^\prime =&\hspace{-0.5em} \textstyle\argmax_{ i \in \Nc_d} \lambda_{i,g_i}^{(\overline{\infty})}\\
%\hspace{-0.2em}d^* =&\hspace{-0.5em} \textstyle\max_{j: d_j<  g_{i^\prime}} d_j  \text{ where } d_j \in \Ic_G  
%\end{array}\right. ~~~\text{(Step $\dagger$)}
%\end{align*} %\vspace{-1.0em}
%\STATE $i^\prime=\argmax_{ i \in \Nc_d} \lambda_{i,g_i}^{(\overline{\infty})}$ 
%\STATE $d^* = \max_{j: d_j<  g_{i^\prime}} d_j $ where $d_j \in \Ic_G$ ~~~(Step $\dagger$)

\IF{($N_{blk}+ q_{i^\prime}\cdot G/g_{i^\prime}) \ge G/d^* $}
\STATE $N_{blk}\leftarrow (N_{blk} +  q_{i^\prime}\cdot G/g_{i^\prime}) - G/d^*$
\STATE $g_{i^\prime}\leftarrow d^*$ and $q_{i^\prime}=1$
%\STATE $q_{i^\prime}=1$
\STATE Compute $\lambda_{{i^\prime},g_{i^\prime}}^{(\underline{\infty})}$~~ (See \eqref{eq:minSSMSE_v2})
\STATE $\lambda_{i^\prime,g_{i^\prime}}^{(\overline{\infty})}=a^{2(g_{i^\prime}-1)}\lambda_{{i^\prime},g_{i^\prime}}^{(\underline{\infty})}   + (1-a^{2(g_{i^\prime}-1)})\lambda_{i^\prime}$ (See \eqref{eq:maxSSMSE_v2}) %$\hfill \text{(See \eqref{eq:maxSSMSE_v2})}$

\ELSE
\STATE $\Nc_d\leftarrow \Nc_d \setminus \{i^\prime\}$

\ENDIF
%\ENDIF

\ENDWHILE
\STATE $\gbf=\gbf(\qbf)$ %\hspace{13em}(Step $\ddagger$)
\end{algorithmic}
%(Here, Step $\dagger$ is to sequentially minimize the maximum steady-state channel MSE by reducing the interval $g_i$ and 
%Step $\ddagger$  is to select some entries of $\gbf$ optimized in this algorithm.) 
%such that $1\le g_i\le G$.)
%(Here, $\delta_{\Ic_G}(g_{i^*})$ denotes the delta function, i.e., $\delta_{\Ic_G}(g_{i^*})=1$ if $g_{i^*}\in \Ic_G$ $./$ denotes the element-wise division and $\lambda_{ij}$
%is the $j$-th element of $\lambdabf_i$. Step * incorporates the
%measurement update step \eqref{eq:measurementUpdateReductionMIMO}
%and Step ** incorporates the prediction step
%\eqref{eq:datammseMIMO}.)
\end{algorithm} 
\vspace{-0.5em}
%After computing $\lambdabf_{\gbf}^{(\overline{\infty})}$ from \eqref{eq:minSSMSE_v2} and \eqref{eq:maxSSMSE_v2}, 
In Step 3 of Algorithm \ref{alg:minmaxCodebookDesign}, we select the eigenvector index $i^\prime$ corresponding to the largest maximum steady-state MSE and choose the largest value $d^*$ among the subset of elements of $\Ic_G$ that are smaller than the pre-defined block time-wise interval $g_{i^\prime}$ (where $g_{i^\prime}$ is a design variable updated during iteration). % at each iteration  by Algorithm \ref{alg:minmaxCodebookDesign} and is initially set up to $g_{i^\prime}=G+1$ and updated in each iteration
In this case, $\lambda_{i^\prime,g_{i^\prime}}^{(\overline{\infty})}$ can be reduced by replacing $g_{i^\prime}$ by $d^*$ because $\lambda_{i^\prime,d^*} ^{(\overline{\infty})}\le \lambda_{i^\prime,g_{i^\prime}}^{(\overline{\infty})}$ if $d^*\le g_{i^\prime}$ from Proposition \ref{lem:monotonicitySSMSE}. 
Thus, the idea of Algorithm \ref{alg:minmaxCodebookDesign} is to sequentially reduce the largest maximum steady-state MSE among $\{ \lambda_{i,g_i}^{(\overline{\infty})}\}$ by allocating a small block time-wise interval.  %in a step-wise manner.
We then check whether it is possible to allocate the index $i^\prime$ corresponding to $d^*$ in the available resources (i.e., $N_{blk}+q_{i^\prime}\cdot G/g_{i^\prime}$) in Step 4.
After that, if the condition in Step 4 is satisfied, $\lambda_{i^\prime,g_{i^\prime}}^{(\overline{\infty})}$ is updated through Steps 5-8. Otherwise, the index $i^\prime$ is excluded from the set $\Nc_d$.
Step 13 is to select some entries of $\gbf$ corresponding to the non-zero elements of $\qbf$ (i.e., the block time-wise intervals obtained by Algorithm \ref{alg:minmaxCodebookDesign}).
 
%Here, Step $\dagger$ is to sequentially minimize the maximum steady-state channel MSE by reducing the interval $g_i$ and Step $\ddagger$  is to select some entries of $\gbf$ optimized in this algorithm.

%Proposition \ref{prop:increasingSSMSE} shows that the initial number of trials for the exhaustive search $\Oc(|\Ic_G|^{\min\{GM_p,N_d,r\}})$ %$\sum_{j=M_p}^{\min\{GM_p,\bar{N}_p\}} |\Ic_G|^j=\frac{|\Ic_G|^{\min\{GM_p,\bar{N}_p\}+1}-|\Ic_G|^{M_p}}{|\Ic_G|-1}$ 
%can be reduced by considering only the case of $\gbf$ in \tcr{ascending order, i.e., $g_i\le g_j$ if $\lambda_i \ge \lambda_j$ where $\lambda_i$ and $\lambda_j$ denote ordered singular values of $\Rbf_\hbf$ in \eqref{eq:rankdefChannelCov}.} %without loss of performance.
%Furthermore, by exploiting the monotonicity of the maximum steady-state channel MSE w.r.t. the inter-block interval $\{g_i\}$ in Lemma \ref{lem:monotonicitySSMSE}, we propose an efficient algorithm for training sequence design that sequentially minimizes the maximum upper bound of the steady-state channel MSE. 
%The corresponding algorithm is summarized in Algorithm \ref{alg:minmaxCodebookDesign}, which require substantially less computational complexity $\Oc(GM_p)$ \tcr{while achieving most of the performance gain compared to the exhaustive search.} %without much loss in performance. 
%Simulation results will be provided in Section \ref{subsec:practicalGuideline} and Section \ref{sec:numericalresults}. 

Since the design of the block time-wise interval $\gbf$ and $n_d$ is complete, we finish this subsection by explaining the construction of $\Cbf$ from the optimized intervals $\{g_i: g_i=2^{k_i}\in \Ic_G \text{ in } \eqref{eq:divisorset},  1\le i \le n_d\}$ which uses the assumption that $G=2^s$.
Note that, given a set of $\{g_1,\ldots,g_{n_d}\}$, there can be several (row-wise and column-wise) permutated versions of a training signal allocation in the training sequence $\Cbf$. However, they have the same steady-state performance, with only minor changes during the transient phase. Thus, we present an efficient method for construction of $\Cbf$ which is done iteratively.

A new variable $U_j$ represents the number of undetermined entries of the $j$-th column of the matrix $\Cbf$, which is initially set to $U_j=G$ for $1\le j \le M_p$. Denote by $q$ the row index of the matrix $\Cbf$ for $1\le q\le G$. 
First, we set the initial values as $q=1$, and $I_q=1$, and then
%so that we allocate the eigenvector index of $i_{q,j}:=I_q+(j-j^\prime)$ at the $j$-th column with a row-wise mapping $g_{i_{q,j}}$. Mathematically, this means 
%\begin{equation}
%[\Cbf]_{q,j}=[\Cbf]_{g_{i_{q,j}}+q,j}=\cdots=[\Cbf]_{(G/g_{i_{q,j}}-1)g_{i_{q,j}}+q,j}=i_{q,j},\label{eq:step1_v1}
%\end{equation}
%where $j^\prime\le j\le M_p$.
%Note that $G/g_{i_{q,j}}$ entries of the $j$-th column are determined using a row-wise allocation of $i_{q,j}$ while leaving its $U_j - G/g_{i_{q,j}}$ undetermined entries. 
we loop over all row indices $q$ in order to determine the entries of $\Cbf$ as follows.

%We define the row index $q$ of the matrix $\Cbf$ for $1\le q\le G$, and a column index $j^\prime\in\{1,\ldots,M_p\}$ of $\Cbf$ such that $[\Cbf]_{q,j^\prime}$ is not determined while $\{[\Cbf]_{q,1},\ldots,[\Cbf]_{q,j^\prime-1}\}$ are all determined by some eigenvector indices in the previous step. 
%Denote by  $I_q\in\{1,\ldots,n_d\}$ an eigenvector index indicating the unused index to be allocated to $[\Cbf]_{q,j^\prime}$.
%(At the beginning, we set the initial values as $q=1$, $j^\prime=1$, and $I_q=1$.) 

%\noindent {\em Step (1)} 
%\tcr{We denote by $q$ the row index of the matrix $\Cbf$ for $1\le q\le G$ and by $j^\prime\in\{1,\ldots,M_p\}$ a column index of $\Cbf$ such that 
%% indicate a undetermined entry at the first time in the $q$-th row of $\Cbf$, respectively. That is, 
%$[\Cbf]_{q,j^\prime}$ is not determined while $\{[\Cbf]_{q,1},\ldots,[\Cbf]_{q,j^\prime-1}\}$ are all determined by some eigenvector indices in the previous step. 
%We define an index $I_q\in\{1,\ldots,n_d\}$, which indicates the eigenvector index to be allocated to $[\Cbf]_{q,j^\prime}$.
%(For example, we set the initial values as $q=1$, $j^\prime=1$, and $I_q=1$.) 
%}

\vspace{0.5em}
%\noindent {\em Step (1)}~ \tcr{
%At the beginning, we set the initial values as $q=1$, $j^\prime=1$, and $I_q=1$ so that we allocate the eigenvector index of $i_{q,j}:=I_q+(j-j^\prime)$ at the $j$-th column with a row-wise mapping $g_{i_{q,j}}$. Mathematically, this means 
%\begin{equation}
%[\Cbf]_{q,j}=[\Cbf]_{g_{i_{q,j}}+q,j}=\cdots=[\Cbf]_{(G/g_{i_{q,j}}-1)g_{i_{q,j}}+q,j}=i_{q,j},\label{eq:step1_v1}
%\end{equation}
%where $j^\prime\le j\le M_p$.
%Note that $G/g_{i_{q,j}}$ entries of the $j$-th column are determined using a row-wise allocation of $i_{q,j}$ while leaving its $U_j - G/g_{i_{q,j}}$ undetermined entries. We then update the variables by 
%\begin{align}
%I_{q+1}&=I_q+M_p-j^\prime +1, &U_j=U_j - G/g_{i_{q,j}}, \text{ and } \nonumber\\
%q&= q+1, \label{eq:step1_v2}
%\end{align}
%for the next step.}

\noindent {\em Step (1)}~
Given $q$, $I_q$, and $U_j$, select a column index $j^\prime\in\{1,\ldots, M_p\}$ of $\Cbf$ such that $[\Cbf]_{q,j^\prime}$ is not determined while $\{[\Cbf]_{q,1},\ldots,[\Cbf]_{q,j^\prime-1}\}$ are all determined by some eigenvector indices in the previous step (e.g., $j^\prime=1$ at the first iteration). 
%so that we allocate the eigenvector index of $i_{q,j}:=I_q+(j-j^\prime)$ at the $j$-th column with a row-wise mapping $g_{i_{q,j}}$. Mathematically, this means 
%\begin{equation}
%[\Cbf]_{q,j}=[\Cbf]_{g_{i_{q,j}}+q,j}=\cdots=[\Cbf]_{(G/g_{i_{q,j}}-1)g_{i_{q,j}}+q,j}=i_{q,j},\label{eq:step1_v1}
%\end{equation}
%where $j^\prime\le j\le M_p$.
%Note that $G/g_{i_{q,j}}$ entries of the $j$-th column are determined using a row-wise allocation of $i_{q,j}$ while leaving its $U_j - G/g_{i_{q,j}}$ undetermined entries. 
We then allocate the eigenvector index of $i_{q,j}:=I_q+(j-j^\prime)$ at the $j$-th column with a row-wise mapping $g_{i_{q,j}}$. Mathematically, this means 
\begin{equation}
[\Cbf]_{q,j}=[\Cbf]_{g_{i_{q,j}}+q,j}=\cdots=[\Cbf]_{(G/g_{i_{q,j}}-1)g_{i_{q,j}}+q,j}=i_{q,j},\label{eq:step1_v1}
\end{equation}
where $j^\prime\le j\le M_p$.
Note that $G/g_{i_{q,j}}$ entries of the $j$-th column are determined using a row-wise allocation of $i_{q,j}$ while leaving its $U_j - G/g_{i_{q,j}}$ undetermined entries.\footnote{If all entries of the $q$-th row of $\Cbf$ are determined, we complete {\em Step (1)} by updating $I_{q+1}=I_q$ and $q=q+1$ with the same $U_j$.}  
%As shown in \eqref{eq:step1_v1}, we then allocate the eigenvector index of $i_{q,j}$ with a row-wise allocation. % into the unused entries of the $j$-th column.
For the next step, we update the variables given by
%\begin{align}
%I_{q+1}&=I_q+M_p-j^\prime +1, &U_j=U_j - G/g_{i_{q,j}}, \text{ and } \nonumber\\
%q&= q+1. \label{eq:step1_v2}
%\end{align} 
\begin{align}
I_{q+1}&=I_q+M_p-j^\prime +1,~U_j=U_j - G/g_{i_{q,j}},\text{ and } \nonumber\\
q&= q+1. \label{eq:step1_v2}
\end{align}
%by $I_{q+1}=I_q+M_p-j^\prime +1$, $U_j=U_j - G/g_{i_{q,j}}$, and $q\leftarrow q+1$ for the next step.

\noindent {\em Step (2)}~ 
Repeat {\em Step (1)} until all entries of $\Cbf$ are determined, i.e., $U_j=0$ for all $j$.

\vspace{0.5em}
\begin{proposition}\label{cor:sufficientCondForTrainingAllocation}
Given the constraints of Problem \ref{prob:limitedRF} and the arranged block time-wise interval $\gbf$ in ascending order, 
if $G=p^s$ for some prime number $p$ and nonnegative number $s$, 
a training sequence $\Cbf$ can be constructed.
\end{proposition}

%\vspace{0.5em}
{\em Proof:} See Appendix \ref{subsec:sufficientCondForTrainingAllocation}.
\vspace{0.5em}

\subsection{Hybrid Analog-Digital Beamforming}\label{subsec:hybrid_analogdigital}

%We assume that the antenna arrays are uniformly placed  in a one-dimensional or two-dimensional space at the base station, 
%where the channel covariance matrix $\Rbf_\hbf$ becomes Toeplitz under a far-field assumption.
We assume that the antennas are uniformly spaced in a one-dimensional or two-dimensional grid at the base station. 
In this case, the channel covariance matrix $\Rbf_\hbf$ can be well approximated by a Toeplitz matrix under the virtual channel representation and a far-field assumption \cite{Sayeed:02SP}.\footnote{
%For a physically large scale MIMO system, the near-field effect on antenna arrays may not negligible \cite{Wuetal:14JSAC}. 
The focus of the paper is not on near-field analysis in physically large antenna arrays but training sequence and beamforming design for the antenna array located in a finite area.
When the antennas are spaced over a rectangular aperture for the same number antennas of the linear aperture, 
a far-field approximation is still appropriate because the largest dimension of the rectangular aperture will be significantly less than the length of the linear aperture, thereby lowering the far-field distance (Fraunhofer distance). 
%a far-field approximation is still appropriate due to a reduced (largest) dimension of the anteena
%%because the far-field distance (Fraunhofer distance) depends on the largest dimension of the aperture and the wavelength.
%due to a reduced largest dimension of the antenna array,  where the far-field distance (Fraunhofer distance) depends on the size of the aperture and the wavelength.
%\tcb{a} large number of antenna elements are placed in a finite area such as a two-dimensional planar array, a far-field approximation \tcb{for a physically small antenna array}  
% is still appropriate by lowering Fraunhofer distance that classifies the far and near regional boundaries.
%(e.g., the macro-cellular tower-mounted base station). 
A theoretical channel model based on a physically large scale MIMO is available in \cite{Wuetal:14JSAC}. 
%we consider a practical limitation that the antenna array is located in a finite area due to the limited room on the macro-cellular (tower-mounted) base station as addressed in \cite{Nametal:13COMMAG}.
} %and the one-ring model \cite{Adhikary&Nam&Ahn&Caire:13IT}. 
It is known that when the size of a Toeplitz matrix is large, the Toeplitz matrix can be decomposed by a DFT matrix, referred to as the Toeplitz distribution theorem (TDT) \cite{Grenander&Szego:book,Sungetal:09IT}, i.e., 
\begin{equation}
%\Rbf_\hbf=E\{\hbf_\ell \hbf_\ell^H\}\approx \Fbf_{\Ic}{\Lambdabf}\Fbf_{\Ic}^H, \label{eq:ToeplitzApproxTheorem}
\Rbf_\hbf=E\{\hbf_\ell \hbf_\ell^H\} \approx \tilde{\Fbf}{\Lambdabf}\tilde{\Fbf}^H, \label{eq:ToeplitzApproxTheorem}
\end{equation} 
where $\tilde{\Fbf}=[\tilde{\fbf}_1,\cdots,\tilde{\fbf}_r]\in\mathbb{C}^{N_t\times r}$ denotes a matrix of distinct columns of the $N_t$-point DFT matrix and 
${\Lambdabf}=\text{diag}({\lambda}_{1},\cdots, {\lambda}_{r})$ is a matrix of the non-zero eigenvalues of $\Rbf_\hbf$ in descending order.
%where $\Fbf_{\Ic}\in\mathbb{C}^{N_t\times r}$ denotes a submatrix of a DFT matrix $\Fbf=[\fbf_1,\cdots,\fbf_{N_t}]\in\mathbb{C}^{N_t\times N_t}$ with the index set $\Ic:=\{i_1,\cdots,i_r\}$ of the selected columns 
%and ${\Lambdabf}=\text{diag}({\lambda}_{i_1},\cdots, {\lambda}_{i_r})$ is comprised of all approximated non-zero eigenvalues of $\Rbf_\hbf$ in descending order.
%the eigenvalues corresponding to the approximated eigenvectors $\Fbf_{\Ic}$.
% of the selected columns for all approximated non-zero eigenvalues of $\Rbf_\hbf$ and
%$\Dbf\in\mathbb{R}^{\tilde{R}\times \tilde{R}}$ is a diagonal matrix composed of the eigenvalues corresponding to the approximated eigenvectors $\Fbf_{\Ic}$.
%%all approximated non-zero eigenvalues of $\Rbf_\hbf$ in descending order.
%%in which the eigenvalues are ordered corresponding to the approximated eigenvectors $\Fbf$.
%
%
%\tcgry{For example, in the case of  a {\em one-ring} channel model defined by the random local scattering around the receiver \cite{Shiu&Foschini&Gans&Kahn:00COM}, the diagonal entries of ${\Lambdabf}$ are non-zero only in the angle spectrum $(\theta-\Delta,~\theta+\Delta)$ where recall that angle-of-arrival (AoA: $\theta$) and angle spread (AS:$\Delta$) of the receiver \cite{Adhikary&Nam&Ahn&Caire:13IT,Noh&Zoltowski&Sung&Love:13STSP}.
%Thus, given knowledge of $\theta$ and $\Delta$,  the spatial correlation matrix $\Rbf_\hbf$ can be characterized by the columns of DFT matrix  $\tilde{\Fbf}$ and its angular power spectrum of ${\Lambdabf}$ in \eqref{eq:ToeplitzApproxTheorem}.}
Note that the $k$-th column of $\tilde{\Fbf}$ is given by
$[1, e^{j 1 \psi_k 2\pi/N_t},  e^{j 2 \psi_k 2\pi/N_t}, \cdots, e^{j (N_t-1) \psi_k 2\pi/N_t}]^H/\sqrt{N_t}$. 
%\begin{equation} %\label{eq:onebeamdirection}
%\frac{1}{\sqrt{N_t}}[1, e^{j 1 \psi_k 2\pi/N_t},  e^{j 2 \psi_k 2\pi/N_t}, \cdots, e^{j (N_t-1) \psi_k 2\pi/N_t}]^H. \nonumber
%\end{equation}
This is simply the transmit steering vector for the physical angle $\theta_k = \sin^{-1}(\psi_k\lambda/d)$,\footnote{The virtual angle $\psi_k$ is related to the physical angle $\theta_k$ by $\psi = \frac{d}{\lambda}\sin(\theta_k)$, i.e.,  if $d/\lambda=1/2$, $-\frac{\pi}{2} \le \theta_k \le \frac{\pi}{2}$ corresponds to $-\frac{1}{2} \le \psi_k \le \frac{1}{2}$, where $d$ and $\lambda$ denote the antenna spacing and the carrier wavelength, respectively \cite{Noh&Zoltowski&Sung&Love:14STSP}.}
and thus the diagonal matrix $\Lambdabf$ can be viewed as the channel power spectral density corresponding to the virtual angular domain.

%In this case, 
Then, the channel dynamics in \eqref{eq:statespacemodel_h} can be rewritten by a parametric channel model \cite{Molisch:04SP,Forenza&Love&Heath:07VT}
\begin{align}
\hbf_{\ell+1} &= a\hbf_\ell + \sqrt{1-a^2} \tilde{\Fbf}{\Lambdabf}^{1/2}\tilde{\bbf}_{\ell+1}, \label{eq:statespacemodel_h_parametric}
\end{align}
where the entries of $\tilde{\bbf}_\ell$ are i.i.d., %independent and identically distributed (i.i.d.), 
i.e., $\tilde{\bbf}_\ell \sim\mathcal{CN}(\mathbf{0},\Ibf_r)$.
This yields that the channel vector is characterized by a random linear combination of the columns of $\tilde{\Fbf}$ and channel estimation can be viewed as estimation of the linear combination coefficients corresponding to the set of basis vectors $\tilde{\Fbf}$. 
Thus, we use the DFT-based training signal (i.e., $\Sbf_\ell \subset \{\tilde{\fbf}_i\}$) during the training period because the columns of DFT are approximated eigenvectors of $\Rbf_\hbf$ in \eqref{eq:ToeplitzApproxTheorem}.
%the channel spatial correlation $\Rbf_\hbf$ in \eqref{eq:ToeplitzApproxTheorem}.
% by transmitting the DFT-based training signals $\Fbf$ during the training phase, i.e., $\Sbf_\ell \subset \{\fbf_i\}$.

%\begin{figure}[!t]
%\centerline{
%\psfrag{(x1)}[l]{\small $x_{1,k}$} %
%\psfrag{(xu)}[l]{\small $x_{U,k}$} %
%\psfrag{(vk)}[c]{\small $\Vbf_k$} %
%\psfrag{(1st)}[c]{\footnotesize $1$} %
%\psfrag{(Npnd)}[c]{\footnotesize $N_d$} %
%\psfrag{(analog)}[c]{\footnotesize Analog} %
%\psfrag{(digital)}[c]{\footnotesize Digital} %
%\psfrag{(precoder)}[c]{\footnotesize precoder} %
%\psfrag{(dk)}[c]{\footnotesize {$\Dbf_k$}} %
%\psfrag{(a1)}[c]{\footnotesize {$\tilde{\fbf}_{i_1}$}} %
%\psfrag{(aNp)}[c]{\footnotesize {$\tilde{\fbf}_{i_{N_d}}$}} %
%\psfrag{(RF)}[c]{\footnotesize RF} %
%\psfrag{(chain)}[c]{\footnotesize chain} %
%\includegraphics[scale=1.4]{figures/hybridprecoding_v3.eps}}
%%\includegraphics[scale=0.9]{figures/systemmodel_miso.eps}}
%\vspace{-0.5em}
%\caption{Hybrid beamforming architecture with digital precoding $\Dbf_k\in\mathbb{C}^{n_d\times U}$ and analog precoding $\{\tilde{\fbf}_i\}$ for $U$ data streams ($U\le n_d\le N_d$)} 
%\label{fig:hybridPrecodingModel}
%\vspace{-2.0em}
%\end{figure}

During the $\ell$-th data transmission period (i.e., $k=\ell M+m$ and $M_p<m\le M$), we assume that the data symbol sequence $\{x_{u,k}\}$ is transmitted with the multi-dimensional transmit beamformer $\Vbf_k\in\mathbb{C}^{N_t\times U}$. %, as shown in Fig. \ref{fig:hybridPrecodingModel}. 
Here, we assume that the base station is equipped with $1\le N_D\le N_t$ available RF chains for digital baseband precoding of $\Dbf_k\in\mathbb{C}^{n_d\times U}$, given by $\Vbf_k=\Fbf\Dbf_k$ where
a pre-beamforming matrix $\Fbf\in\mathbb{C}^{N_t\times n_d}$ is implemented by analog beamforming techniques (e.g., use analog phase shifters with constant magnitude entries) \cite{Ayach&Rajagopal&AduSurra&Pi&Heath:14WC}. % and its linear combination $\Dbf_k$ is implemented by digital baseband beamforming that uses $n_d$ active RF chains.
%where $\Fbf_k\in\mathbb{C}^{N_t\times n_d}$ and $\Dbf_k\in\mathbb{C}^{n_d\times U}$ denote a set of basis vectors and a linear combination of those vectors, respectively. 
%
The variable $n_d\le N_d$ denotes the number of used RF chains to have a low-dimensional solution while capturing the effective channel rank
 %specify effective number of analog beams 
%that accounts for the effectively dominant channel directions 
aimed at enabling low-complexity and energy-efficient system implementation. %\footnote{Note that the base station can turn off the unused $(N_d-n_d)$ RF chains including filters and mixers, which saves additional power.}
%The basis vectors $\Fbf_k$ are implemented by using analog phase shifters with constant magnitude entries (i.e., analog beamforming) and a linear combination of those vectors $\Dbf_k$ is implemented by digital baseband precoding by using $n_d$ active RF chains. 

As explained in Section \ref{sec:systemmodel}, the $\ell$-th channel estimate $\hat{\hbf}_{\ell|\ell}$ lies in the column space of all the used training signal $\Sc_\ell$ composed of the DFT columns. 
%In the MRT beamforming, 
Then, the pre-beaforming $\Fbf$ will span the subspace of the training signal $\Sc_\ell$, i.e., the pre-beamforming matrix is determined by the distinct DFT columns used in the construction of the training sequence that specifies the training signals.
%(i.e., the subspace spanned by the distinct DFT columns used in the construction of the training codebook).  
Therefore, we focus on the design of the DFT-based training sequence under the constraint of the $N_d$ available RF chains.
To meet the constraint on the number of RF chains, we restrict the number of distinct DFT columns $(n_d)$ used in the training sequence to be less than or equal to $N_d$.
Such a design is obtained from the proposed methods of Problem \ref{prob:limitedRF} by substituting the eigenvalues ${\Lambdabf}$ of \eqref{eq:ToeplitzApproxTheorem} into \eqref{eq:minSSMSE_v2} and \eqref{eq:maxSSMSE_v2}. 
Simulations will be presented in Section \ref{sec:numericalresults} where the training signals are approximated by DFT vectors without much loss in performance.

\section{Extension to Multiuser Massive MIMO Systems}\label{sec:multiuser_massiveMIMO}
%%%%%%%%%%%%%%%%%%%%%%%%%%%%%%%%%%%%%%%%%%%%%%%%%%%%%%%%%%%%%%%%%%%%%%%

%\vspace{-0.5em}
%%%%%%%%%%%%%%%%%%%%%%%%%%%%%%%%%%%%%%%%%%%%%%%%%%%%%%%%%%%%%%%%%%%%%%%
\subsection{System Set-Up}\label{subsec:systemsetup}

Consider the downlink of a cellular system serving $U$ single antenna users.
Let $\hbf_{u,\ell}\in\mathbb{C}^{N_t}$ be the channel vector of user $u$ at the $\ell$-th block symbol time where $1\le u\le U$.
In line with \eqref{eq:statespacemodel_h}, we consider a state-space model for $\hbf_{u,\ell}$ with the channel covariance matrix $\Rbf_{\hbf_u}=E\{\hbf_{u,\ell}\hbf_{u,\ell}^H\}$ such that $\text{rank}(\Rbf_{\hbf_u})=r_u$.
Define $\Hbf_{\ell}=[\hbf_{1,\ell},\cdots,\hbf_{U,\ell}]\in\mathbb{C}^{N_t\times U}$ as the combined channel matrix for a block length of $M$ channel uses.
Denote by $\xbf_{k}=[x_{1,k},\cdots,x_{U,k}]^T\in\mathbb{C}^{U}$ be the data symbols at the symbol time $k=\ell M+m$ to service $U$ user terminals with the same average transmit power of $\rho$ so that $E\{|x_{u,k}|^2\}=\rho$.
We assume that the base station uses the multi-dimensional transmit beamformer $\Vbf_{k}=[\vbf_{1,k},\cdots,\vbf_{U,k}]\in\mathbb{C}^{N_t\times U}$ to map $\xbf_{k}$ to the transmit antennas, i.e., $\sbf_k=\Vbf_k\xbf_k$.
Denote by $n_{d,u}$ the number of disjoint training signals used in the construction of the training sequence to service the user $u$. %such that $M_p\le \sum_{u=1}^U n_{d,u}\le \min\{GM_p, N_d,\sum_{u=1}^U r_u\}$.

%% OMITTED for CLEAR STORY DEVELOPMENT of PAPER
%\tcgry{From \eqref{eq:statespacemodel_y2}, 
%%and by replacing $\Vbf_k$ with the pilot beam patterns $\{\sbf_{u,k}\}$ for $U$ users in \eqref{eq:multiuser_systemEq}, 
%the received signal for user $u$ at the $\ell$-th training period ($k=\ell M + m$  and $1\le m\le M_p$) is given by 
%\begin{align}
%y_{u,k} 
%&= \sbf_{u,k}^H\hbf_{u,\ell}   +   \sum_{u^\prime=1:u^\prime\neq u}^U   \sbf_{u^\prime,k}^H \hbf_{u,\ell} + w_{u,k} \nonumber \\
%&= \sum_{u^\prime\notin \Jc_{u,k}} \sbf_{u^\prime,k}^H \hbf_{u,\ell}  
%  + \sum_{u^\prime\in \Jc_{u,k}} \sbf_{u^\prime,k}^H \hbf_{u,\ell} + w_{u,k}  \nonumber \\
%&= \sum_{u^\prime\notin \Jc_{u,k}} \sbf_{u^\prime,k}^H \hbf_{u,\ell}  +  w_{u,k}, \label{eq:multiuser_eq_training} 
%%&= \frac{1}{\sqrt{U}}\sbf_{u,k}^H\hbf_{u,\ell}   +  \bar{w}_{u,k}, \nonumber
%\end{align}
%where $\Jc_{u,k}:=\{u^\prime: \sbf_{u^\prime,k}\in\Nc(\Rbf_{\hbf_u}), 1\le u^\prime \le U \}$ denotes the user index set having pilot beam patterns spanning the orthogonal subspace of $\Rbf_{\hbf_u}$ and the transmit power constraint $\|\sbf_{k,u^\prime}\|_2^2=\rho_p/U$.  % and $w_{u,k}\sim\mathcal{CN}(0,\sigma_w^2)$.
%By transmitting back the received training signal $y_{u,k}$ as explained in Section \ref{subsec:channelestimation}, the channel vector of $\hbf_{u,\ell}$ is estimated with Kalman filtering with prior knowledge of $\{\sbf_{u^\prime,k}: u^\prime \in \Jc_{u,k}\}$ at the base station.}

For channel estimation, a training sequence for each user is designed by the proposed algorithm in Section III, where the obtained training sequences among users are generally different.
During the training period, we assume that the training sequences of different users are sounded over non-overlapping time intervals, i.e., $UM_p$ out of over $M$ channel uses is allocated for training. 
Then, a quantized (or analog) version of the received signal $y_{u,k}\in\mathbb{C}$ is fed back over some sort of control channel 
to enable channel estimation at the base station. %through some control channel.
On the other hand, we assume that the beamformed data signals are sent simultaneously to analyze the effect of the downlink channel estimation error and the inter-user interference.

The collection of received symbols for all $U$ users at the $\ell$-th data transmission period (i.e., channels uses satisfying $k=\ell M+m$ with $UM_p< m\le M$) is denoted as 
{%\small
\begin{align}
\ybf_{k} 
&= \Hbf_{\ell}^H\Vbf_{k}\xbf_k + \wbf_k \nonumber\\%\label{eq:multiuser_systemEq}\\% ~~\text{for } k=\ell M + m, ~1\le m\le M
&= 
\left[\begin{array}{cccc}
\hbf_{1,\ell}^H\vbf_{1,k} & \hbf_{1,\ell}^H\vbf_{2,k} & \cdots & \hbf_{1,\ell}^H\vbf_{U,k} \\
\hbf_{2,\ell}^H\vbf_{1,k} & \hbf_{2,\ell}^H\vbf_{2,k} & \cdots & \hbf_{2,\ell}^H\vbf_{U,k} \\
\vdots & \vdots & \cdots & \vdots \\
\hbf_{U,\ell}^H\vbf_{1,k} & \hbf_{U,\ell}^H\vbf_{2,k} & \cdots & \hbf_{U,\ell}^H\vbf_{U,k} \\
\end{array}\right]\xbf_k + \wbf_k, %\nonumber
\end{align}}\noindent
where $\ybf_{k}=[y_{1,k},\cdots,y_{U,k}]^T$ and $\wbf_k\sim \mathcal{CN}(0,\Ibf_U)$ is the additive white Gaussian noise vector. 
%, and $\alpha$ is the power normalization to satisfy the average transmit power constraint so that $\alpha =1/\text{tr}(\Vbf_k^H\Vbf_k)$.
Focusing only on the received signal for user $u$, we have
{%\small
\begin{align}
y_{u,k} % \nonumber\\
&= \hbf_{u,\ell}^H\vbf_{u,k}x_{u,k}  +  \sum_{u^\prime\neq u}  \hbf_{u,\ell}^H\vbf_{u^\prime,k}x_{u^\prime,k} + w_{u,k} \nonumber\\
&\stackrel{(a)}{=} 
\alpha_u\hbf_{u,\ell}^H\hat{\hbf}_{u,\ell |\ell}x_{u,k}   \nonumber\\
&~~~
+  \sum_{u^\prime\neq u} \alpha_{u^\prime} \hbf_{u,\ell}^H\hat{\hbf}_{u^\prime,\ell |\ell}x_{u^\prime,k} + w_{u,k} \nonumber\\
&\stackrel{(b)}{=} 
\alpha_{u}\hat{\hbf}_{u,\ell |\ell}^H\hat{\hbf}_{u,\ell |\ell} x_{u,k}  +  \alpha_u \tilde{\hbf}_{u,\ell}^H\hat{\hbf}_{u,\ell |\ell}x_{u,k}  \nonumber\\
&~~~ 
+   \sum_{u^\prime\neq u} \alpha_{u^\prime} \hbf_{u,\ell}^H\hat{\hbf}_{u^\prime,\ell |\ell}x_{u^\prime,k} + w_{u,k},\label{eq:multiuser_eq_data}
%&\stackrel{(b)}{=} 
%\sqrt{\alpha}\hat{\hbf}_{u,\ell |\ell}^H\hat{\hbf}_{u,\ell |\ell} x_{u,k}  +  \sqrt{\alpha}\sum_{u^\prime=1:u^\prime\neq u}^U  \hat{\hbf}_{u,\ell}^H\hat{\hbf}_{u^\prime,\ell |\ell}x_{u^\prime,k} \nonumber\\
%&~~~ +  \sqrt{\alpha}\sum_{u^\prime=1}^U  \tilde{\hbf}_{u,\ell}^H\hat{\hbf}_{u^\prime,\ell |\ell}x_{u^\prime,k} + w_{u,k},\label{eq:multiuser_eq_data}
%&y_{u,k} \nonumber\\
%&= \sqrt{\alpha}\hbf_{u,\ell}^H\vbf_{u,k}x_{u,k}  +  \sqrt{\alpha}\sum_{u^\prime=1:u^\prime\neq u}^U  \hbf_{u,\ell}^H\vbf_{u^\prime,k}x_{u^\prime,k} + w_{u,k} \nonumber\\
%&\stackrel{(a)}{=} 
%\sqrt{\alpha}\hbf_{u,\ell}^H\hat{\hbf}_{u,\ell |\ell}x_{u,k}  +  \sqrt{\alpha}\sum_{u^\prime=1:u^\prime\neq u}^U  \hbf_{u,\ell}^H\hat{\hbf}_{u^\prime,\ell |\ell}x_{u^\prime,k} + w_{u,k} \nonumber\\
%&\stackrel{(b)}{=} 
%\sqrt{\alpha}\hat{\hbf}_{u,\ell |\ell}^H\hat{\hbf}_{u,\ell |\ell} x_{u,k}  +  \sqrt{\alpha} \tilde{\hbf}_{u,\ell}^H\hat{\hbf}_{u,\ell |\ell}x_{u,k}  \nonumber\\
%&~~~ +  \sqrt{\alpha}\sum_{u^\prime=1:u^\prime\neq u}^U  \hbf_{u,\ell}^H\hat{\hbf}_{u^\prime,\ell |\ell}x_{u^\prime,k} + w_{u,k},\label{eq:multiuser_eq_data}
%%&\stackrel{(b)}{=} 
%%\sqrt{\alpha}\hat{\hbf}_{u,\ell |\ell}^H\hat{\hbf}_{u,\ell |\ell} x_{u,k}  +  \sqrt{\alpha}\sum_{u^\prime=1:u^\prime\neq u}^U  \hat{\hbf}_{u,\ell}^H\hat{\hbf}_{u^\prime,\ell |\ell}x_{u^\prime,k} \nonumber\\
%%&~~~ +  \sqrt{\alpha}\sum_{u^\prime=1}^U  \tilde{\hbf}_{u,\ell}^H\hat{\hbf}_{u^\prime,\ell |\ell}x_{u^\prime,k} + w_{u,k},\label{eq:multiuser_eq_data}
\end{align}}\noindent
where $(a)$ follows the matched filtering precoder $\vbf_{u,k}=\alpha_u\hat{\hbf}_{u,\ell |\ell}$ and $(b)$ holds by $\tilde{\hbf}_{u,\ell}:=\hbf_{u,\ell}-\hat{\hbf}_{u,\ell |\ell}$.
Here, $\alpha_u$ denotes the power normalization per user such that $\text{tr}(\Vbf_k^H\Vbf_k)=1$, defined as $\alpha_u=1/(\|\hat{\hbf}_{u,\ell|\ell}\|_2\sqrt{U})$ for $1\le u \le U$.

Applying the method in \cite{Hassibi&Hochwald:03IT}, a lower bound on the training-based capacity is obtained by considering the worst-case uncorrelated additive noise.
From \eqref{eq:multiuser_eq_data}, we have %the SINR of user $u$ as
\begin{align}
SINR_{u,\ell}:= \frac{\eta_{u,\ell}}{\sigma^2_{u,\ell}}, \label{eq:sinr_ul}
\end{align}
where the desired signal power (scaled by $\frac{1}{\alpha_u^2\rho}$) is given by
$\eta_{u,\ell} = |\hat{\hbf}_{u,\ell |\ell}^H\hat{\hbf}_{u,\ell |\ell}|^2$,
%\begin{align*}
%\eta_{u,\ell} &= |\hat{\hbf}_{u,\ell |\ell}^H\hat{\hbf}_{u,\ell |\ell}|^2,
%\end{align*}
and the interference plus noise power is given by
$\sigma^2_{u,\ell} 
= \frac{1}{\alpha_u^2\rho}
+ |\tilde{\hbf}_{u,\ell}^H\hat{\hbf}_{u,\ell |\ell}|^2 
+ \sum_{u^\prime=1:u^\prime\neq u}^U (\alpha_{u^\prime}^2/\alpha_u^2) |\hbf_{u,\ell }^H\hat{\hbf}_{u^\prime,\ell |\ell}|^2$.
%\begin{align*}
%\sigma^2_{u,\ell} 
%&= \frac{1}{\alpha\rho}
%+ |\tilde{\hbf}_{u,\ell}^H\hat{\hbf}_{u^\prime,\ell |\ell}|^2 
%+ \sum_{u^\prime=1:u^\prime\neq u}^U  |\hbf_{u,\ell |\ell}^H\hat{\hbf}_{u^\prime,\ell |\ell}|^2.
%\end{align*}

%\tcr{The notation $SINR_{u,\ell}$ in (22) represents the instantaneous received SNR for the user $u$ when the proposed method is applied to the multi-user case,
%where we consider the effects of imperfect channel estimation and the inter-user interference plus noise.}

%\vspace{-0.5em}
%%%%%%%%%%%%%%%%%%%%%%%%%%%%%%%%%%%%%%%%%%%%%%%%%%%%%%%%%%%%%%%%%%%%%%%
\subsection{Performance Analysis}\label{subsec:performance_analysis}

In order to find a convenient expression for the SINR in \eqref{eq:sinr_ul}, 
we focus on the asymptotic results when $N_t\rightarrow \infty$ in \cite{Hoydis&Brink&Debbah:13JSAC}.
For simplicity, we assume a symmetric scenario with the same number $n_{d,u}=n_d$ for users. 
The results proposed here, however, extends immediately to the general case.
%\tcr{Add more for several $N_{p}$, e.g., allocate another frequency-time resource.}
The following proposition provides a closed-form expression of the SINR. % using the asymptotic deterministic equivalents.

\vspace{0.5em}
\begin{proposition}\label{prop:deterministicSINR}
Under a Kalman filtering framework and spatial matched filtering, the deterministic equivalent SINR of \eqref{eq:sinr_ul}  % and $\gbf\in\mathbb{C}^{N_p}$ generated by Proposition \ref{prop:increasingSSMSE} and Lemma \ref{lem:monotonicitySSMSE}
is given by
\begin{align}
SINR_{u,\ell} - \overline{SINR}_{u,\ell}
~~\substack{a.s.\\ \overrightarrow{N_t\rightarrow \infty}} ~~0
%\overline{SINR}_{u,\ell} %\nonumber\\
%%&= 
%%\frac{|\text{tr}(\Rbf_{\hbf_u}-\Pbf_{u,\ell |\ell})|^2}
%%{\sum_{u^\prime=1:u^\prime\neq u}^U   \text{tr}\left((\Rbf_{\hbf_u}-\Pbf_{u,\ell |\ell})(\Rbf_{\hbf_{u^\prime}}-\Pbf_{u^\prime,\ell |\ell})\right) +
%% \sum_{u^\prime=1}^U  \text{tr}\left(\Pbf_{u,\ell |\ell}(\Rbf_{\hbf_{u^\prime}}-\Pbf_{u^\prime,\ell |\ell})\right) + \frac{\sigma_w^2}{\alpha\rho_p}} \nonumber \\
%%&= 
%%|\text{tr}(\Rbf_{\hbf_u}-\Pbf_{u,\ell |\ell})|^2
%%\left(\frac{1}{\alpha\rho} + \text{tr}\left(\Pbf_{u,\ell |\ell}(\Rbf_{\hbf_{u}}-\Pbf_{u,\ell |\ell})\right)  \right.\nonumber\\
%%&~~~\left. + \sum_{u^\prime\neq u}^U   \text{tr}\left(\Rbf_{\hbf_u}(\Rbf_{\hbf_{u^\prime}}-\Pbf_{u^\prime,\ell |\ell})\right)\right)^{-1} \label{eq:sinr_ul_v2}\\
%&~\substack{\\=\\{N_t\rightarrow \infty}}~
%\frac{A_{u,\ell}}{\frac{1}{\alpha\rho}+B_{u,\ell}+C_{u,\ell}}
\label{eq:sinr_ul_v3},
%\bigl|\text{tr}(\Lambdabf_{u} - \bar{\Lambdabf}_{u}^{(\ell)} )\bigr|^2
%\left(\frac{1}{\alpha\rho} +  \text{tr}\bigl(  \bar{\Lambdabf}_{u}^{(\ell)} (\Lambdabf_{u} - \bar{\Lambdabf}_{u}^{(\ell)} )  \bigr)  \right.\nonumber\\
%&~~~\left. +  \sum_{u^\prime\neq u}^U   \text{tr}\bigl(\Lambdabf_u\Ubf_u^H\Ubf_{u^\prime} (\Lambdabf_{u^\prime}-\bar{\Lambdabf}_{u^\prime}^{(\ell)} )\Ubf_{u^\prime}^H\Ubf_u\bigr) \right)^{-1}, \label{eq:sinr_ul_v3}
\end{align}
where $\overline{SINR}_{u,\ell}$ is given by $\overline{SINR}_{u,\ell} ={A_{u,\ell}}/\left({{1}/{(\alpha_u^2\rho)}+B_{u,\ell}+C_{u,\ell}}\right)$ with 
%\begin{align}
%A_{u,\ell} &= \bigl|\text{tr}(\Lambdabf_{u} - \bar{\Lambdabf}_{u}^{(\ell)} )\bigr|^2 \label{eq:A}\\
%%B_{u,\ell} &= \frac{1}{\alpha\rho}  \label{eq:B}\\
%B_{u,\ell} &= \text{tr}\bigl(  \bar{\Lambdabf}_{u}^{(\ell)} (\Lambdabf_{u} - \bar{\Lambdabf}_{u}^{(\ell)} )  \bigr) \label{eq:B}\\
%C_{u,\ell} &=\sum_{u^\prime=1:u^\prime\neq u}^U   \text{tr}\bigl(\Lambdabf_u\Ubf_u^H\Ubf_{u^\prime} (\Lambdabf_{u^\prime}-\bar{\Lambdabf}_{u^\prime}^{(\ell)} )\Ubf_{u^\prime}^H\Ubf_u\bigr). \label{eq:C}
%\end{align}
{%\small
\begin{align}
%\left.\begin{array}{l}
A_{u,\ell} &= \bigl|\text{tr}(\Lambdabf_{u} - \bar{\Lambdabf}_{u}^{(\ell)} )\bigr|^2 \nonumber\\%,~~~ %\nonumber\\
%B_{u,\ell} &= \frac{1}{\alpha\rho}  \label{eq:B}\\
B_{u,\ell} &= \text{tr}\bigl(  \bar{\Lambdabf}_{u}^{(\ell)} (\Lambdabf_{u} - \bar{\Lambdabf}_{u}^{(\ell)} )  \bigr), ~~~\text{and} \nonumber\\
C_{u,\ell}&=\sum_{\footnotesize\shortstack{$u^\prime=1$\\$:u^\prime\neq u$}}^U  \frac{\alpha_{u^\prime}^2}{\alpha_u^2}  
\text{tr}\bigl(\Lambdabf_u\Ubf_u^H\Ubf_{u^\prime} (\Lambdabf_{u^\prime}-\bar{\Lambdabf}_{u^\prime}^{(\ell)} )\Ubf_{u^\prime}^H\Ubf_u\bigr).
%\end{array}\right.  
\label{eq:C}
\end{align}}\noindent
%\tcr{Here, $\substack{a.s.\\ \longrightarrow}$ denotes the almost sure convergence.}
Given the ED of $\Rbf_{\hbf_u}=\Ubf_u\Lambdabf_u\Ubf_u^H$ where $\Ubf_u\in\mathbb{C}^{N_t\times r_u}$ and $\Lambdabf_u=\text{diag}(\lambda_{u,1},\cdots,\lambda_{u,r_u})$ composed of the non-zero eigenvalues in descending order, the estimation error covariance matrix $\Pbf_{u,\ell |\ell}$ is eigen-decomposed by  $\Pbf_{u,\ell |\ell}=\Ubf_u\bar{\Lambdabf}_u^{(\ell)}\Ubf_u^H$.
\end{proposition}

%\vspace{0.5em}
{\em Proof:} See Appendix \ref{subsec:deterministicSINR}.
\vspace{0.5em}

The three terms in the denominator of \eqref{eq:sinr_ul_v3} characterize the following effects:  $1/(\alpha\rho)$ for the post-processed (average) transmit signal-to-noise power ratio, $B_{u,\ell}$ for the imperfect channel estimation, and $C_{u,\ell}$ for the inter-user interference from the other users sharing the same time-frequency slot.
%The second term in the denominator of \eqref{eq:sinr_ul_v3} represents the effect of imperfect channel estimation and the third term in the denominator denotes those of inter-user interference.
Proposition \ref{prop:deterministicSINR} provides some intuition about how the SINR can be analyzed in training-based channel estimation. 
First, in order to maximize $A_{u,\ell}$ of \eqref{eq:C}, the diagonal entries of $\bar{\Lambdabf}_{u}^{(\ell)}$ should be minimized according to the absolute values of the diagonal entries of $\Lambdabf_{u}$. Note that the proposed training sequence design can be leveraged to increase $A_{u,\ell}$ because the training sequence $\Cbf$ reduces the $n_d^*$ dominant eigenvalues of $\bar{\Lambdabf}_{u}^{(\ell)}$ by using its $n_d^*$ dominant eigenvectors of $\Rbf_{\hbf_u}$ as training signals corresponding to the block time-wise interval $\gbf^*$. Here, $\gbf^*$ and $n_d^*$ denote the minimizers of Problem \ref{prob:limitedRF} obtained by the proposed algorithm.
On the one hand, $B_{u,\ell}$ of \eqref{eq:C} can be viewed as a weighed version of $A_{u,\ell}$ where we can constrain $B_{u,\ell}$ through the training sequence design by reducing the dominant entries of $\bar{\Lambdabf}_{u}^{(\ell)}$. 
Second, $C_{u,\ell}$ of \eqref{eq:C} can be reduced when the users serviced simultaneously on the same time-frequency are scheduled so that the dominant eigenvectors of the users are orthogonal to each other. Here, user scheduling techniques can be applied w.r.t. the angle-of-arrival range and angle spread \cite{Adhikary&Caire:14STSP,Lee&Sung:14arXiv}.\footnote{
%By considering 
For the case of the macro cellular (tower-mounted) base station, there can be scatterers surrounding the mobile terminals without significant scattering around the base station \cite{Shiu&Foschini&Gans&Kahn:00COM}. In this case, we can jointly service the angular-separated users %in distinct annular regions 
\cite{Adhikary&Nam&Ahn&Caire:13IT}.}

%For example, in the case of  a {\em one-ring} channel model defined by the random local scattering around the receiver \cite{Shiu&Foschini&Gans&Kahn:00COM}, the diagonal entries of ${\Lambdabf}$ are non-zero only in the angle spectrum $(\theta-\Delta,~\theta+\Delta)$ where recall that angle-of-arrival (AoA: $\theta$) and angle spread (AS:$\Delta$) of the receiver \cite{Adhikary&Nam&Ahn&Caire:13IT,Noh&Zoltowski&Sung&Love:13STSP}.
%Thus, given knowledge of $\theta$ and $\Delta$,  the spatial correlation matrix $\Rbf_\hbf$ can be characterized by the columns of DFT matrix  $\tilde{\Fbf}$ and its angular power spectrum of ${\Lambdabf}$ in \eqref{eq:ToeplitzApproxTheorem}.

For performance metric analysis, the expression of \eqref{eq:sinr_ul_v2} can be precomputed before the Kalman filter is run and 
the %(deterministic) 
achievable throughput for user $u$ at the $\ell$-th block is given by \cite{Hoydis&Brink&Debbah:13JSAC}
{%\small
\begin{align}
R_{u,\ell} &= \left(1-\frac{UM_p}{M}\right)\cdot \log(1+\overline{SINR}_{u,\ell}), \label{eq:spectralefficiency}
\end{align}}\noindent
%where the pre-log factor $(1-M_p/M)$ comes with the consideration of real data transmission ratio in the block transmission length  $M$.
where the pre-log factor $(1-UM_p/M)$ is needed because $UM_p$ out of over $M$ channel uses is allocated for training.
%For a steady-state throughput analysis, the following corollary provides a lower bound on the SINR.
Based on the closed-form expressions for the steady-state channel MSE in \eqref{eq:minSSMSE_v2} and \eqref{eq:maxSSMSE_v2}, we further derive a closed-form lower bound of \eqref{eq:sinr_ul_v3}, given by
{%\small
\begin{align}
%\hspace{-1.2em}
&\overline{SINR}_{u} %\nonumber\\
%&
= \lim_{\ell\rightarrow\infty} \overline{SINR}_{u,\ell}  \nonumber\\
&\ge
\bigl\|\lambdabf_u - {\lambdabf}_{\gbf_u}^{(\overline{\infty})}\bigr\|_1^2 %\nonumber\\
%&~~~
\left(\frac{1}{\alpha_u^2\rho} %\right.  \nonumber\\
%&~~~\left.
+ \bigl\|{\lambdabf}_{\gbf_u}^{(\overline{\infty})}\odot(\lambdabf_{u} - {\lambdabf}_{\gbf_u}^{(\underline{\infty})})\bigr\|_1 \right.\nonumber\\
&\left.
+ \sum_{u^\prime \neq u}  \frac{\alpha_{u^\prime}^2}{\alpha_u^2}
\text{tr}\bigl({\Lambdabf}_u\Ubf_{u}^H\Ubf_{u^\prime}({\Lambdabf}_{u^\prime} - {\Lambdabf}_{\gbf_{u^\prime}}^{(\underline{\infty})}) \Ubf_{u^\prime}^H\Ubf_{u}\bigr)\right)^{-1} . \label{eq:sinr_ul_steadystate_v2}
\end{align}}\noindent 
%{\small\begin{align}
%\hspace{-1.2em}
%\overline{SINR}_{u} 
%%&
%= \lim_{\ell\rightarrow\infty} \overline{SINR}_{u,\ell} %\nonumber\\
%&\ge
%\bigl\|\lambdabf_u - {\lambdabf}_{\gbf_u}^{(\overline{\infty})}\bigr\|_1^2 %\nonumber\\
%%&~~~
%\left( \frac{1}{\alpha_u^2\rho} 
%+ \bigl\|{\lambdabf}_{\gbf_u}^{(\overline{\infty})}\odot(\lambdabf_{u} - {\lambdabf}_{\gbf_u}^{(\underline{\infty})})\bigr\|_1   \right.\nonumber\\
%&\left. 
%~~~+ \sum_{u^\prime \neq u}  (\alpha_{u^\prime}^2/\alpha_u^2)
%\text{tr}\bigl({\Lambdabf}_u\Ubf_{u}^H\Ubf_{u^\prime}({\Lambdabf}_{u^\prime} - {\Lambdabf}_{\gbf_{u^\prime}}^{(\underline{\infty})}) \Ubf_{u^\prime}^H\Ubf_{u}\bigr) \right)^{-1}. \label{eq:sinr_ul_steadystate_v2}
%\end{align}}\noindent 
A complete derivation of \eqref{eq:sinr_ul_steadystate_v2} is available in Appendix \ref{subsec:lowerbound_steadystate_SINR}.
\section{Numerical Results}\label{sec:numericalresults}
%%%%%%%%%%%%%%%%%%%%%%%%%%%%%%%%%%%%%%%%%%%%%%%%%%%%%%%%%%%%%%%%%%%%%%%
%\vspace{-0.5em}

In this section, we provide numerical results to evaluate the performance of the proposed algorithms.  
We consider two different base station antenna arrays,
%antenna arrays located at the base station, 
i.e., a ULA with $N_t=32$ antenna elements and a $15\times 25$ UPA with $N_t=375$ antenna elements. 
%We considered $N_t\in\{32,375\}$ transmit ULA to serve multiple single antenna users. 
For the time-varying channel model in \eqref{eq:statespacemodel_h}, 
we set $f_c=2.5$GHz carrier frequency and $T_s = 100\mu\text{s}$ for each symbol duration corresponding to %a typical mobile speed range from $v=3\text{km/h}$ to $v=30\text{km/h}$
a mobile speed $v=3\text{km/h}$.
%\footnote{In Jake's model \cite{Jakes:book}, $a = J_0(2\pi f_D (T_s M))$ where $J_0(\cdot)$ is the zeroth-order Bessel function, $f_D=\frac{vf_c}{c}$ is the maximum Doppler frequency shift, and each block is composed of $M$ symbols.}
%we adopted $f_c=2.5\text{GHz}$ carrier frequency and $T_s=0.5\text{ms}$ symbol duration with a typical mobile speed range
%from $v=3\text{km/h}$ $(a=0.9999)$  to $30\text{km/h}$ $(a=0.9881)$.\footnote{In Jake's model \cite{Jakes:book}, $a=J_0(2\pi f_D T_s)$ where $J_0(\cdot)$ is the zeroth-order Bessel function and $f_D=\frac{vf_c}{c}$ is the maximum Doppler frequency shift.} 
%For all considered channel sounding methods, we used Kalman filtering and prediction for the channel estimator. 
We consider channel estimation performance using the normalized mean square error (NMSE), given by $\text{NMSE}=\text{tr}(\Pbf_{\ell |\ell})/\text{tr}(\Rbf_{\hbf})$. 
%Given the unit noise variance normalization $(\sigma_w^2=1)$, the SNR is defined as $SNR=\rho_p$. 
The channel estimation performance for each of the considered methods was averaged over $500$ Monte Carlo runs.

% FIGURE
\begin{figure}[!t]
\centerline{\includegraphics[scale=0.39]{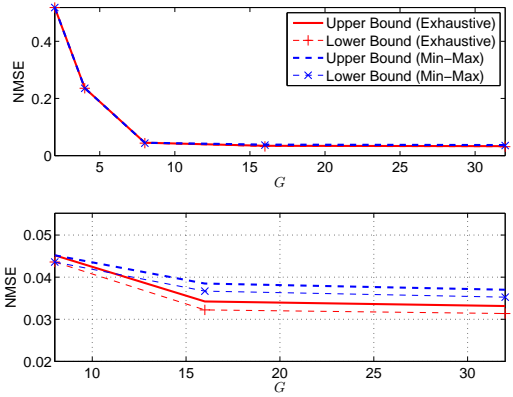}}
\vspace{-0.5em} \caption{NMSE versus the training sequence size $G$ where $N_t=375$, $M=5$, $M_p=2$, $N_d=GM_p$, $\rho=10$, $d_s=100\text{m}$, AS$=16.7^\circ$, and $v=3km/h$.}
\label{fig:nmse_planar_snr10db_3kmh_upperlower}  \vspace{-1.3em}
\end{figure}
% FIGURE
\begin{figure}[!t]
\centerline{\includegraphics[scale=0.39]{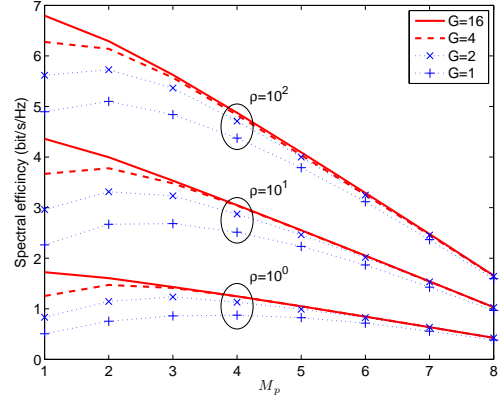}}
\vspace{-0.5em} \caption{Spectral efficiency versus training period length $M_p$ where $N_t=375$, $M=10$, $N_d=GM_p$, $d_s=100\text{m}$, AS$=16.7^\circ$, and $v=3km/h$.}
\label{fig:spectralEfficiency_planar_3kmh} \vspace{-1.3em}
\end{figure}
%% MINIPAGE
%\begin{figure*}[t]
%\begin{minipage}[b]{0.45\linewidth}
%%\SetLabels
%%\L(0.50*0.47)  \footnotesize (a)\\
%%\L(0.50*-0.01) \footnotesize (b)\\
%%\endSetLabels
%%\leavevmode
%%%\ShowGrid
%\strut\AffixLabels{
%\hspace{0.4em}
%\includegraphics[scale=0.33]{figures/fig_Nt375Nr1_Planar_a99988.0848_snr10_M5_pilot12_h60_radius30_s100_NtV15_NtH25_Nrf64_wrtG_v1.eps}
%}
%\vspace{-1.2em} \caption{NMSE versus the training sequence size $G$ where $N_t=375$, $M=5$, $M_p=2$, $N_d=GM_p$, $\rho=10$, $d_s=100\text{m}$, AS$=16.7^\circ$, and $v=3km/h$.}
%\label{fig:nmse_planar_snr10db_3kmh_upperlower} % \vspace{-1.3em}
%\end{minipage}
%\quad \hspace{0.5em}
%\begin{minipage}[b]{0.45\linewidth}
%
%\strut\AffixLabels{
%\hspace{0.4em}
%\includegraphics[scale=0.33]{figures/fig_Nt375Nr1_Planar_velocity3_M10_h60_radius30_s100_NtV15_NtH25_Nrf24_snr020_B116_specEffi_r1.eps}}
%\vspace{-1.2em} \caption{Spectral efficiency versus training period length $M_p$ where $N_t=375$, $M=10$, $N_d=GM_p$, $d_s=100\text{m}$, AS$=16.7^\circ$, and $v=3km/h$.}
%\label{fig:spectralEfficiency_planar_3kmh} %\vspace{1.3em}
%\end{minipage}
%\vspace{-2.0em}
%\end{figure*}

%To generate a channel model for simulation, we adopt the {\em one-ring} channel model that well models typical cellular systems 
We adopt the {\em one-ring} channel model to generate each channel realization during simulation \cite{Shiu&Foschini&Gans&Kahn:00COM,Adhikary&Nam&Ahn&Caire:13IT}. The channel spatial correlation is characterized by angle spread (AS: $\Delta$), angle-of-arrival (AoA: $\theta$), and antenna geometry. 
Based on the one-ring channel model, we considered a $15\times 25$ uniform planar array at the base station.
Then, the channel covariance matrix $\Rbf_\hbf$ is given by $\Rbf_\hbf=\Rbf_H\otimes\Rbf_V$ where $\Rbf_H\in\mathbb{C}^{N_H\times N_H}$ and $\Rbf_V\in\mathbb{C}^{N_V\times N_V}$ denote the horizontal and vertical covariance matrices, respectively. Each of the spatial correlation matrices is defined by % in \eqref{eq:channelCovOneRing}.
{%\small
\begin{align}
[\Rbf_t]_{p,q} &= \frac{\gamma}{2\Delta}
\int^{\theta+\Delta}_{\theta-\Delta} e^{-j \pi (p-q)
\sin(\xi)} d\xi, \label{eq:channelCovOneRing}
\end{align}}\noindent
%The AoA ($\theta_s$) and AS ($\Delta_s$) depend on the physical environment around transmit antenna and users.
where $t\in\{H,V\}$ and $\gamma$ denotes propagation path loss between the transmitter and the receiver
%, defined in Section \ref{sec:numericalresults}.
%The propagation path loss between the transmitter and the receiver is  
given by $\gamma = (1+(\frac{d_s}{d_0})^{\alpha_{0}})^{-1}$, where the path loss exponent is set to  $\alpha_0=3.8$, $d_s$ is the distance from the transmitter in meters, and $d_0$ is the reference distance set to $d_0=30\text{m}$ \cite{Adhikary&Nam&Ahn&Caire:13IT}. 
We assume that the transmit antenna is located at an elevation of $h=60\text{m}$ and the local scattering ring around the user has radius $d_r=30\text{m}$. %, and the distance from the transmitter is $s$ (m). 
Then, the parameters for the channel covariance matrices  $\Rbf_{V}$ and $\Rbf_{H}$ are given by
$\Delta_V=\frac{1}{2}\left(\arctan(\frac{d_s+d_r}{h})-\arctan(\frac{d_s-d_r}{h})\right)$,
$\theta_V=\frac{1}{2}\left(\arctan(\frac{d_s+d_r}{h})+\arctan(\frac{d_s-d_r}{h})\right)$,
$\Delta_H=\arctan(\frac{d_r}{d_s})$, and $\theta_H\in(-\frac{\pi}{3},\frac{\pi}{3})$ for a sector in a cell.

\subsection{Practical Guidelines for Training Sequence}\label{subsec:practicalGuideline}

In this subsection, a practical guideline for training sequence parameters is developed with quantitative analysis.

First, we can improve channel estimation performance by choosing the row length of $\Cbf$ large enough to incorporate more dominant eigen-directions of the channel in the $G\times M_p$ training sequence.
Intuitively, the channel MSE of the $nG\times M_p$ training sequence ($n\in\mathbb{N}$) is {\em at least} equal to those of the $G\times M_p$ training sequence by $n$ times repetition of the shorter version of the training sequence.
%We consider channel estimation performance by the normalized mean square error (NMSE), i.e., $\text{NMSE}=\frac{1}{\text{tr}(\Rbf_{\hbf})}\text{tr}(\Pbf_{\ell |\ell})$. 
Fig. \ref{fig:nmse_planar_snr10db_3kmh_upperlower} shows the closed-form expressions of the upper and lower bounds in \eqref{eq:minSSMSE_v2} and \eqref{eq:maxSSMSE_v2}.
It is seen that increasing $G$ is indeed beneficial in terms of the channel MSE, but the effect becomes marginal when $G$ is too large. 
That is, the proposed training sequence can operate in a finite $G$ regime and achieve reasonably good channel estimation performance, 
which implies that increasing the training period length $M_p$ also has similar effect on the channel MSE due to the increased training sequence size.
%which also implicitly yields that increasing $M_p$ can leverage channel estimation performance.
%(See Section \ref{sec:numericalresults} for a detailed description of the parameter setting.)

%one can improve the channel MSE by choosing $G$ large enough but the effect becomes marginal when $G$ is too large.
%Since the gap of our lower/upper bound on the channel MSE is small as shown in Fig. \ref{fig:nmse_planar_snr10db_3kmh_upperlower}, one can use these closed-form analytical expressions as performance metrics in advance.

Second, though the increased $M_p$ enables large beamforming gain by leveraging channel estimation performance, 
an increment of $M_p$ can degrade achievable data rate because the remaining $M-M_p$ channel uses are only available for downlink data transmission.
%Second, note that increasing the training period length $M_p$ enables large beamfoming gain by transmitting orthogonal training signals that span the $M_p$ dominant channel directions.
%On the other hand, the increased $M_p$ may degrade achievable data rate because the remaining $M-M_p$ symbols are available for downlink data transmission.
%Note that the value of $M_p$ is related to achievable data rate because the remaining $M-M_p$ symbols are available for downlink data transmission.
Therefore, we examine the trade-off of spectral efficiency in \eqref{eq:spectralefficiency} corresponding to the value of $M_p$, which was obtained by using Algorithm \ref{alg:minmaxCodebookDesign} for simplicity.
In Fig. \ref{fig:spectralEfficiency_planar_3kmh}, when the value of $G$ is small, the spectral efficiency benefits from the slightly increased $M_p$ since increasing $M_p$ enables the $G\times M_p$ training sequence to incorporate more dominant directions of the channel for channel estimation accuracy. 
However, increasing $M_p$ over some threshold limits the spectral efficiency due to the shorter length of data transmission period, as expected from the pre-log factor in \eqref{eq:spectralefficiency}.
The tension between channel estimation accuracy and achievable data rate yields that the value of $M_p$ should be properly selected under given system parameters.
Instead of this nontrivial choice, we can again increase the (vertical) sequence size $G$ for channel estimation accuracy without affecting the pre-log term.
%Fig. \ref{fig:spectralEfficiency_planar_3kmh} shows that increasing $G$ makes the design of $M_p$ less dependent on the spectral efficiency.
Fig. \ref{fig:spectralEfficiency_planar_3kmh} shows that the increased $G$ makes the spectral efficiency quite insensitive w.r.t. $M_p$ for the practical range of the value of $G$.

% FIGURE
\begin{figure}[!t]
%\hspace{-1.5em}
\centerline{\includegraphics[scale=0.39]{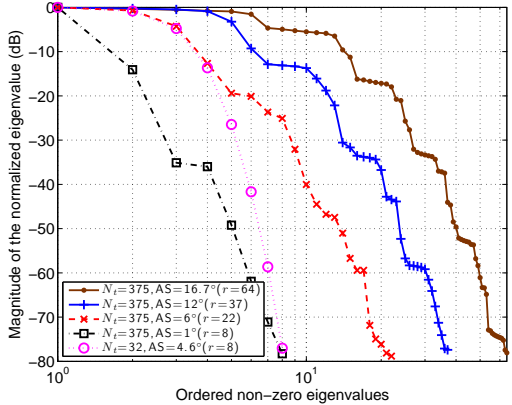}}
\vspace{-0.5em} \caption{The magnitude of the normalized non-zero eigenvalues of $\Rbf_\hbf$ where $\text{rank}(\Rbf_\hbf)=r$.}
\label{fig:eigenvaluDistributionofSpatialCorrelation} \vspace{-1.3em}
\end{figure}
% FIGURE
\begin{figure}[!t]
%\hspace{-1.5em}
\centerline{\includegraphics[scale=0.39]{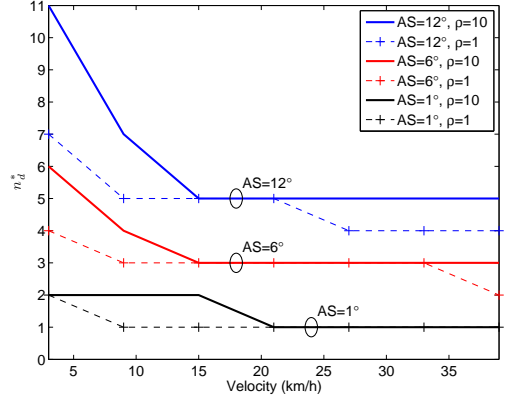}}
\vspace{-0.5em} \caption{The optimized value of $n_d^*$ for training sequence and transmit precoding where $N_t=375$, $M=5$, $M_p=1$, $G=32$, $N_d=GM_p$, and $d_s=150\text{m}$.}
\label{fig:optimal_N_p_distance150} \vspace{-1.3em}
\end{figure}
%% MINIPAGE
%\begin{figure}[t]
%\begin{minipage}[b]{0.45\linewidth}
%%\SetLabels
%%\L(0.50*0.47)  \footnotesize (a)\\
%%\L(0.50*-0.01) \footnotesize (b)\\
%%\endSetLabels
%%\leavevmode
%%%\ShowGrid
%\strut\AffixLabels{
%\hspace{0.4em}
%\includegraphics[scale=0.33]{figures/eigValDist.eps}
%}
%\vspace{-1.2em} \caption{The magnitude of the normalized non-zero eigenvalues of $\Rbf_\hbf$ where $\text{rank}(\Rbf_\hbf)=r$.}
%\label{fig:eigenvaluDistributionofSpatialCorrelation}% \vspace{-1.3em}
%\vspace{2.0em}
%%\vspace{-2.0em}
%%\caption{NMSE versus SNR for different $T$ and $\rho$ in a SISO case where $N_s=300$: (a) equal power delay profile and (b) exponential power delay profile} 
%%\label{fig:single_wrt_tap_off}%\vspace{-3.0em}
%\end{minipage}
%\quad \hspace{0.5em}
%\begin{minipage}[b]{0.43\linewidth}
%\strut\AffixLabels{
%\hspace{0.4em}
%\includegraphics[scale=0.33]{figures/fig_minNp_Nt375Nr1_Planar_r0_velocity339_snr-1515_AS130_h60_s150_NtV15_NtH25_N_rf375_Mp1_B32.eps}}
%\vspace{-1.2em} \caption{The optimized $n_d^*$ for training sequence and transmit precoding where $N_t=375$, $M=5$, $M_p=1$, $G=32$, $N_d=GM_p$, and $d_s=150\text{m}$.}
%\label{fig:optimal_N_p_distance150} \vspace{1.05em}
%%\vspace{-2.0em}
%%\caption{NMSE versus the block size $N_s$ in the SISO case where SNR $40$ dB (assuming known phase ambiguity)$^4$}\label{fig:nmse_single_analysis}
%\end{minipage}
%\vspace{-3.7em}
%\end{figure}

Furthermore, we focus on the optimal number of dimensionality variable $n_d^*$ (or the number of active RF chains in the case of hybrid precoding) obtained from the proposed method.
The reduced dimensionality $n_d^*$ used for training sequence and transmit beamforming design
%That is, the number of unique training beam patterns in the reduced dimensionality $n_d^*$ 
provides insight into the (effective) dominant channel rank considered for transmitting multiple data streams or the beamforming gain.  
%that depends on system parameters. 
%Based on the proposed method, Fig. \ref{fig:optimal_N_p_distance150} shows the optimal number of fixed dimensions $n_d^*$ under several conditions (e.g., the number of active RF chains $n_d^*$ in the case of hybrid precoding).
Fig. \ref{fig:eigenvaluDistributionofSpatialCorrelation} shows the magnitude of the eigenvalues of $\Rbf_\hbf$, which is rank-deficient due to
insufficient scatterers around a tower-mounted base station and a high angular resolution due to its large aperture. The rank of $\Rbf_\hbf$ is determined by a few dominant eigenvalues and the number of less significant eigenvalues.
Fig. \ref{fig:optimal_N_p_distance150} shows that, at high SNR, 
%more RF chains are activated to incorporate channel gains by subspace sampling in a sufficient broad space. 
more training beam patterns are used to incorporate sufficient channel gains by subspace sampling in a sufficient broad space. 
On the other hand, 
%smaller RF chains are required to focus on a few dominant eigen-directions of the channel.
a small number of training beamforming vectors are required to account for the most dominant eigen-directions of the channel, in the low-SNR regime.
In addition, the user's mobility also affects the value of $n_d^*$  because the estimated channel is more likely outdated in the fast-mobility case.
Thus, one can mitigate the channel aging effect on the most dominant eigen-directions by properly reducing the dimension of the sampling subspace of the channel, i.e., properly reduce the number of unique training beam patterns $n_d$.
%This yields that the number of training beam patterns $n_d$ should be reduced to leverage channel estimation accuracy within the reduced dimensional space.
This result indicates the influence of the various system parameters such as channel spatial correlation, angular spread, transmit power, and user terminals' mobility on the effective channel rank based on the proposed method.
%considered for transmit beamforming. %, in conjunction of the proposed method.

%\vspace{-0.5em}
%%%%%%%%%%%%%%%%%%%%%%%%%%%%%%%%%%%%%%%%%%%%%%%%%%%%%%%%%%%%%%%%%%%%%%%
\subsection{Performance Evaluation of Training Techniques}\label{subsec:practicalGuideline}

% FIGURE
\begin{figure*}[!t]
%\vspace{-3em}
\centerline{ \SetLabels %\vspace{.3cm}
\L(0.090*-0.02) \footnotesize (a) Channel estimation (same legend as in (b)) \\
\L(0.695*-0.02) \footnotesize (b) Received SNR \\
\endSetLabels
\leavevmode
%\ShowGrid
\strut\AffixLabels{
\hspace{-1em}
\hspace{0.4em}
\includegraphics[scale=0.39]{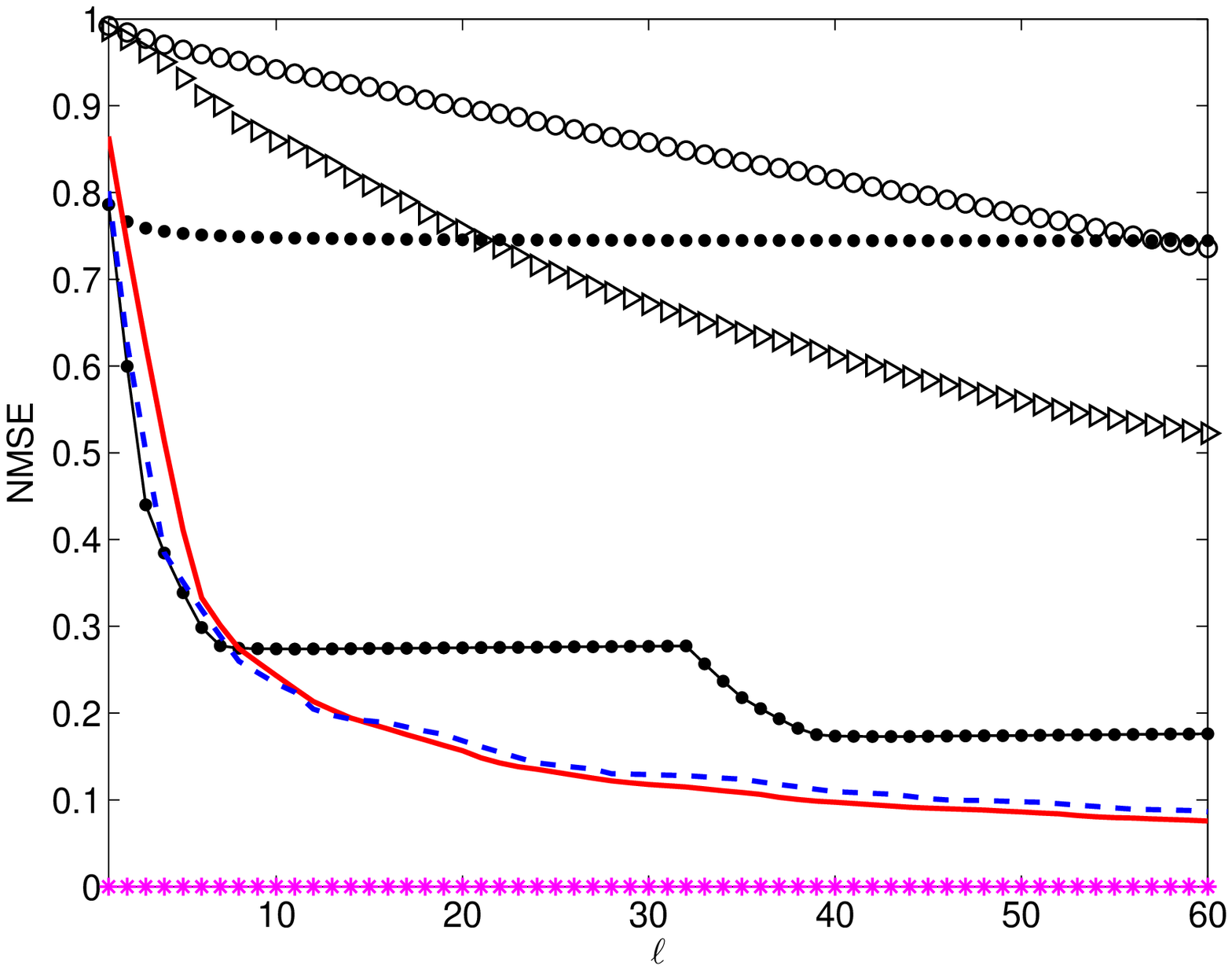}\hspace{0.7em}
\includegraphics[scale=0.39]{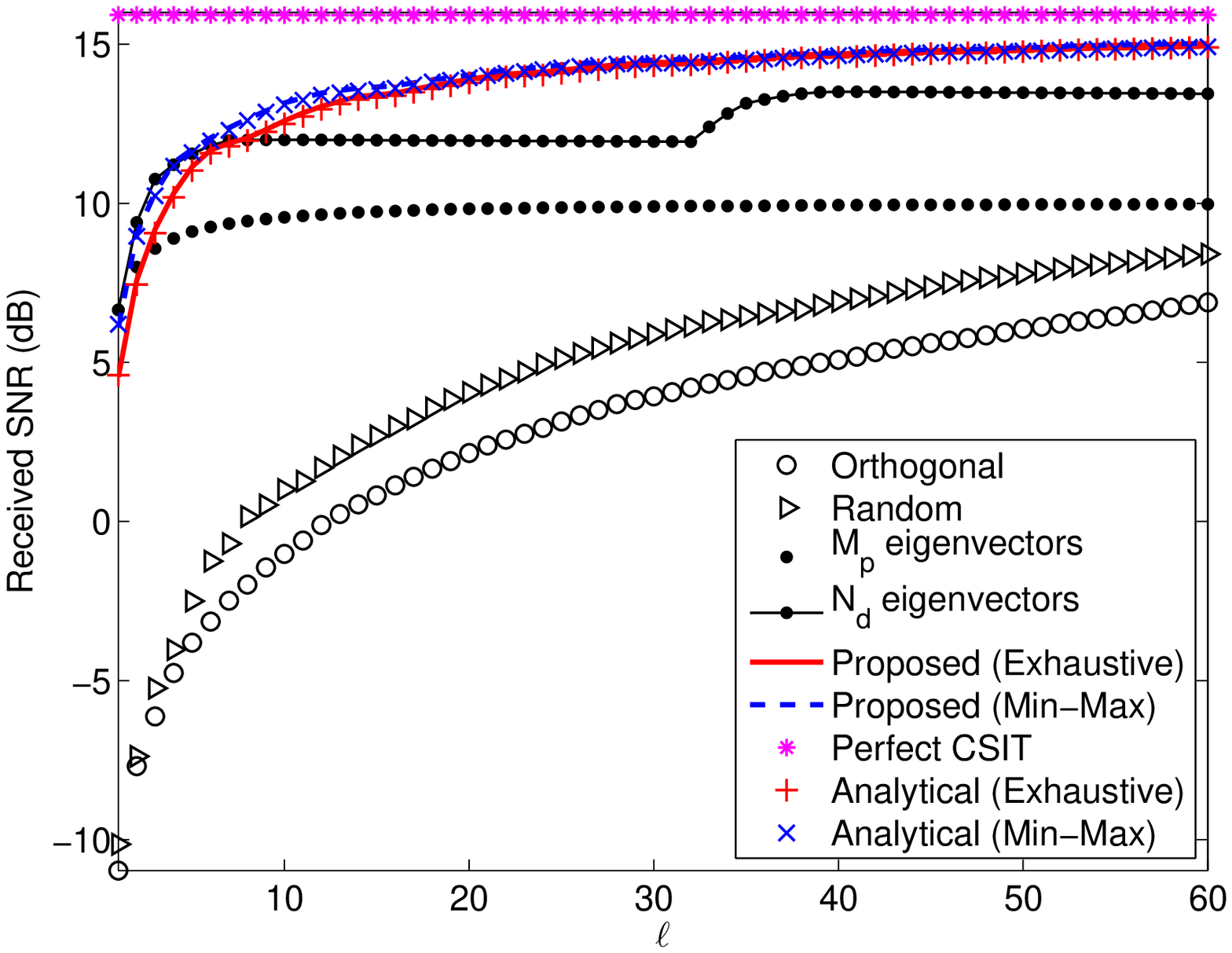}}}
\vspace{0.5em}
%%%%%%%%%%%%%%
%\centerline{ \SetLabels %\vspace{.3cm}
%\L(0.115*-0.03) \footnotesize (c) Steady-state tracking (channel estimation) \\
%\L(0.635*-0.03) \footnotesize (d) Steady-state tracking (received SNR) \\
%\endSetLabels
%\leavevmode
%%\ShowGrid
%\strut\AffixLabels{
%\includegraphics[scale=0.46]{figures/fig_Nt375Nr1_Planar_iter500_slot2062_a99988.0848_r0_snr10_M5_pilot12_h60_radius30_s100_NtV15_NtH25_B32_Bp24_Nrf375_v2_nmsess.eps}\hspace{1em}
%\includegraphics[scale=0.46]{figures/fig_Nt375Nr1_Planar_iter500_slot2062_a99988.0848_r0_snr10_M5_pilot12_h60_radius30_s100_NtV15_NtH25_B32_Bp24_Nrf375_v2_snrss.eps}}}
%\vspace{0.5em}
%%%%%%%%%%%%%%
%\centerline{ \SetLabels 
%\L(0.115*-0.03) \footnotesize (c$^\prime$) Steady-state tracking (channel estimation) \tcr{(Enlarged)} \\
%\L(0.635*-0.03) \footnotesize (d$^\prime$) Steady-state tracking (received SNR) \tcr{(Enlarged)} \\
%\endSetLabels
%\leavevmode
%%\ShowGrid
%\strut\AffixLabels{
%\includegraphics[scale=0.46]{figures/fig_Nt375Nr1_Planar_iter500_slot2062_a99988.0848_r0_snr10_M5_pilot12_h60_radius30_s100_NtV15_NtH25_B32_Bp24_Nrf375_v2_nmsess_enlarged.eps}\hspace{1em}
%\includegraphics[scale=0.46]{figures/fig_Nt375Nr1_Planar_iter500_slot2062_a99988.0848_r0_snr10_M5_pilot12_h60_radius30_s100_NtV15_NtH25_B32_Bp24_Nrf375_v2_snrss_enlarged.eps}}}
\vspace{-0.5em} 
\caption{NMSE and received SNR versus block time index $\ell$ where $N_t=375$, $M=5$, $M_p=2$, $G=32$, $N_d=64$, $\rho=10$, AS$=16.7^\circ$, and $v=3\text{km/h}$.}
\label{fig:nmsesnr_planar_snr10db_3kmh_comparison_w_others} 
\vspace{-1.3em}
\end{figure*}

We compare the performance of the proposed methods to those of several downlink training techniques \cite{Santipach&Honig:10IT,Kaltenberger&Kountouris&Gesbert&Knopp:09WCOM,Kotecha&Sayeed:04SP}.
For all considered channel sounding methods, we use Kalman filtering for channel estimation. 
Fig. \ref{fig:nmsesnr_planar_snr10db_3kmh_comparison_w_others} shows the performance comparison with several training signal design methods 
\cite{Santipach&Honig:10IT,Kaltenberger&Kountouris&Gesbert&Knopp:09WCOM,Kotecha&Sayeed:04SP} with $N_t=375~(N_V=15, N_H= 25)$, $N_d=64$, $\theta_H=\frac{\pi}{6}$, and $d_s=100\text{m}$.  
Orthogonal and random training signals are chosen at the beginning of simulation and used in a round-robin manner. 
These methods are ineffective in terms of the amount of training duration for achieving reasonable channel estimation accuracy since such training signal patterns cannot effectively capture the dominant channel directions over all the $N_t$-dimensional space at each training period. 
The training signal composed of the fixed $M_p$ dominant eigenvectors of $\Rbf_\hbf$ can only minimize the channel MSE in the limited subspace spanned by the fixed $M_p$ training vectors.
Thus, the fixed training signal approach saturates quickly. We also consider the modified scheme that initially selects the $N_d$ dominant eigenvectors of $\Rbf_\hbf$ and transmits $M_p$ training signals among the chosen $N_d$ training signal patterns across $G$ consecutive training periods where $N_d=GM_p$.
The $N_d$ fixed training scheme shows the best performance up to the initial 7 blocks and  becomes inefficient for the remaining duration.
This result indicates that about 14 eigen-directions contain the most dominant channel gain which is not known a priori.

The proposed methods with the optimal number of training signal patterns $n_d^*=24$ substantially reduce the training duration necessary to achieve good channel estimation accuracy. 
This yields that the proper use of less dominant eigen-directions of the channel indeed leverages channel estimation performance.
%
%By using up to the  $\bar{N}_p$ active RF chains,  the proposed algorithm of the structured training codebook also tracks the channel state fast 
%and the performance gap between the sequential pilot beam pattern is negligible, except the very beginning of algorithm.
%Since the steady-state performance of Kalman filtering depends the transient prediction and estimation error,
%the training codebook minimizing the dominant channel uncertainty in the steady-state sense also leverage the transient performance of channel estimation.
Within the first few blocks, the min-max approach in Algorithm \ref{alg:minmaxCodebookDesign} shows better performance than the exhaustive approach in Fig. \ref{fig:nmsesnr_planar_snr10db_3kmh_comparison_w_others}(a).
This is because the min-max training sequence is designed to sequentially minimize the dominant steady-state channel MSE, thus this approach shows a slightly steeper initial slope on the channel MSE. 
As a matter of fact, the exhaustive approach will eventually provide the best channel estimation performance, but only a marginal performance difference is observed in comparison with the min-max approach as shown in Table \ref{tab:steadystate_comparison}. %, which shows the steady-state performance of training techniques.
The proposed methods outperform other methods over almost all of transmission periods in terms of the channel MSE and the received SNR.\footnote{Note that the performance gain of the proposed method is due to a well-designed training sequence and transmit precoder by exploiting all available redundancy in space and time (i.e., spatio-temporal correlation).
For the case of idealized independent identically distributed (i.i.d.) channel coefficients, the performance gain can be reduced since it is difficult to estimate the long channel vector within a constrained training time.}
%based on exhaustive search and min-max approach yield a good performance over all transmission period in the channel MSE and the received SNR with reduced $\bar{N}_p$ dimensional transmit beamforming compared other methods.
Our simulation results also matches the analytic result of \eqref{eq:sinr_ul_v2} very well in Fig. \ref{fig:nmsesnr_planar_snr10db_3kmh_comparison_w_others}(b).

\begin{figure}[!t]
\vspace{-1.0em}
\centerline{\includegraphics[scale=0.39]{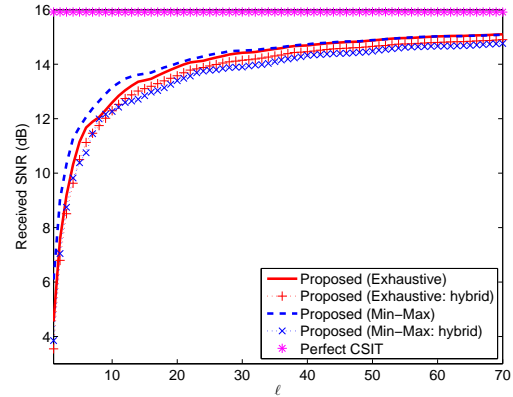}}
\vspace{-0.5em} \caption{Received SNR versus block time index $\ell$ where $N_t=375$, $M=5$, $M_p=2$, $G=32$, $\rho=10$, AS$=16.7^\circ$, and $v=3\text{km/h}$.}
\label{fig:specEffic_planar_snr10db_3kmh_comparison_w_DFT} \vspace{-0.5em}
\end{figure}
% TABLE
% Requires the booktabs if the memoir class is not being used
\begin{table}[t]%[htbp]
%\footnotesize
\small
\centering
%\topcaption{Table captions are better up top} % requires the topcapt package
\begin{tabular}{@{} lcc @{}} % Column formatting, @{} suppresses leading/trailing space
	\toprule
%      	\multicolumn{2}{c}{Item} \\
   	\cmidrule(r){2-3} % Partial rule. (r) trims the line a little bit on the right; (l) & (lr) also possible
      	Method    & NMSE & Received SNR (dB)\\
      	\midrule
      	Orthogonal				& 0.13 	& 13.8 \\
        Random					& 0.13	& 13.8 \\
      	$M_p$ eigenvectors			& 0.74	& 9.3 \\
      	$N_d$ eigenvectors			& 0.05	& 14.9 \\
%      	Sequential design			& 0.04	& 15.4 \\
    	Proposed (Exhaustive)		& 0.03 	& 15.3 \\
      	Proposed (Min-Max)			& 0.04	& 15.3 \\
     	Proposed (Exhaustive: hybrid)	& 0.04 	& 15.2 \\
      	Proposed (Min-Max: hybrid)	& 0.05	& 15.2 \\
	Perfect CSIT				& $0.00$	& 15.8 \\      	\bottomrule
\end{tabular}
%\vspace{-0.0em}
\caption{Steady-state performance: Comparison of several methods}\label{tab:steadystate_comparison}
\vspace{-1.5em}
\end{table}

Fig. \ref{fig:specEffic_planar_snr10db_3kmh_comparison_w_DFT} shows the performance of the proposed hybrid precoding design, 
where the DFT-based training sequence is used by exploiting the approximated channel spatial correlation $\Rbf_\hbf$ in \eqref{eq:ToeplitzApproxTheorem}.
%For the training codebook using the DFT matrix in Problem \ref{prob:limitedRF_analog}, we considered the approximated eigenvalues of $\Rbf_\hbf$ in \eqref{eq:ToeplitzApproxTheorem}.
%It is seen that the DFT/TDT-based method yields almost the same performance as the proposed algorithm with perfectly known $\Rbf_\hbf$, in Fig. \ref{fig:specEffic_planar_snr10db_3kmh_comparison_w_DFT}.
It is seen that the proposed hybrid precoding method that uses imperfect channel correlation knowledge yields almost the same performance as the method with perfectly known $\Rbf_\hbf$ during the transient phase in Fig. \ref{fig:specEffic_planar_snr10db_3kmh_comparison_w_DFT}, and also shows a negligible performance difference in the steady-state phase as shown in Table \ref{tab:steadystate_comparison}.
%
%Table \ref{tab:steadystate_comparison} shows that the proposed method with hybrid design reasonably guarantees the performance of the channel MSE and the received SNR.
An observation of practical importance is that the proposed hybrid precoding method based on a rough estimation of $\Rbf_\hbf$ by using the DFT vectors seems to work well in FDD massive MIMO systems even with a limited number of RF chains for transmit beamforming.
Due to space limitations, simulation results for a ray-based channel model are not provided, but our simulations using ray-based channel models have also yielded good channel estimation performance. 
%This result indicates that the simple practical estimation of $\Rbf_\hbf$ based on the DFT-based training codebook seems to work well in FDD massive MIMO systems that employ analog precoding  and a limited number of RF chains for transmit beamforming.

% FIGURE
\begin{figure}[!t]
%\hspace{-1.5em}
\centerline{\includegraphics[scale=0.39]{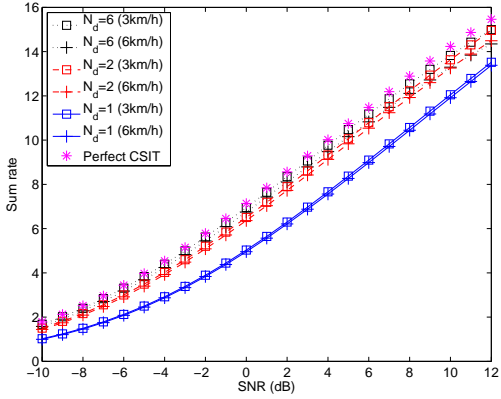}}
\vspace{-0.5em} \caption{A lower bound on sum spectral efficiency versus SNR (dB) where $N_t=32$, $M=10$, $M_p=1$, AS$=4.6^\circ$, and $G=32$ in the multi-user case.}%where $G=32$, $M=20$, $M_p=2$, $\Delta=10^{\circ}$, and $s=150m$}
\label{fig:specEffic_multiuser_ULA-1D} \vspace{-1.3em}
\end{figure}

Finally, we evaluated the proposed method in the multiple-user situation with the ULA ($N_t=32$) at the base station to service $U$ users in a sector of a cell for the same setup as before in the one-ring model.
%We considered the ULA with $N_t=128$ at the base station to service $U$ users in a sector of a cell for the same setup as before.
We assume that users are uniformly distributed in a sector $\{\theta_u\in(-\pi/2,\pi/2):1\le u \le U\}$ with %$\theta_u=-\frac{\pi}{3} + \Delta +(u-1)\frac{2\pi}{3}\frac{1}{U}$ (angle of arrival of user $u$) and $\Delta=\arctan(\frac{d_r}{d_s})$ (angular spread) for
 $d_r=8\text{m}$, $d_s=100\text{m}$, and $U=5$.
Here, the SNR is defined as $\gamma\rho$ to account for the signal transmit power and the propagation path loss in \eqref{eq:channelCovOneRing}.
Fig. \ref{fig:specEffic_multiuser_ULA-1D} shows that the performance of the lower bound on sum spectral efficiency in \eqref{eq:sinr_ul_steadystate_v2} under the several parameters of the dimensionality constraints $N_d$ and the terminal velocity. 
The performance of perfect CSIT case is shown as the performance reference.
%The case of perfect CSIT is also shown for comparison purposes.
In Fig.  \ref{fig:specEffic_multiuser_ULA-1D}, the proposed method achieves close performance of full CSIT with the reasonably increased dimensionality constraint.

%% TABLE
%% Requires the booktabs if the memoir class is not being used
%\begin{table}[t]%[htbp]
%\small
%\centering
%%\topcaption{Table captions are better up top} % requires the topcapt package
%\begin{tabular}{@{} lcc @{}} % Column formatting, @{} suppresses leading/trailing space
%	\toprule
%%      	\multicolumn{2}{c}{Item} \\
%   	\cmidrule(r){2-3} % Partial rule. (r) trims the line a little bit on the right; (l) & (lr) also possible
%      	Item    & NMSE & Received SNR (dB)\\
%      	\midrule
%     	Proposed (Exhaustive)			& 0.03 	& 15.4 \\
%     	Proposed (Exhaustive-hybrid)		& 0.04 	& 15.3 \\
%      	Proposed (MinMax)				& 0.04	& 15.4 \\
%      	Proposed (MinMax-hybrid)		& 0.05	& 15.3 \\
%	Perfect CSIT					& $-$	& 15.8 \\
%      	\bottomrule
%\end{tabular}
%\vspace{1.0em}
%\caption{\hspace{-1em}Steady-state performance of proposed methods}\label{tab:steadystate_proposed}
%\vspace{-2em}
%\end{table}

%\tcr{Continue...}\footnote{
%\tcb{\%\% To be included.... in Numerical Results \%\%}\\
%\tcgry{- Include figures for performance w.r.t. the codebook size $G$ and the number of RF chains$\bar{N}_p$}\\
%\tcgry{- Include figures for {achievable data} rate and \st{BER}?} \\
%\tcgry{\st{- Include figures for the use of received data feedback}} \\
%\tcgry{- Include figures for multi-user massive MIMO} \\
%}

%%%%%%%%%%%%%%%%%%%%%%%%%%%%%%%%%%%%%%%%%%%%%%%%%%%%%%%%%%%%%%%%%%%%%%%
\section{Conclusion}\label{sec:conclusion}
%%%%%%%%%%%%%%%%%%%%%%%%%%%%%%%%%%%%%%%%%%%%%%%%%%%%%%%%%%%%%%%%%%%%%%%
%\vspace{-0.5em}

We considered a reduced dimensionality training sequence and transmit precoder design aimed at enabling low-complexity and energy-efficient system implementation.
We proposed a new method for training sequence design that leverages steady-state channel estimation performance in conjunction with Kalman filtering. 
The low-dimensionality constraint on training sequence and transmit precoding extends to a hybrid analog-digital precoding scheme that uses a limited number of active RF chains for transmit precoding by applying the Toeplitz distribution theorem with specific antenna configurations.
We derived some necessary conditions for the optimal solution and  provide a practical guideline for selecting the training sequence parameters along with performance analysis. 
The proposed method can provide a way to realize energy-efficient large-scale antenna systems. %\vfill
\appendix 
\subsection{Proof of Proposition \ref{lem:monotonicitySSMSE}}\label{subsec:monotonicitySSMSE}
%\vspace{-0.2em}

Given $g_i \ge g_i^\prime=g_i-c$ for some $0\le c \le g_i-1$, we have
{%\small
\begin{align}
\frac{a^{2g_i}}{1-a^{2g_i}} \le \frac{a^{2g_i^\prime}}{1-a^{2g_i^\prime}}
&= 
\frac{a^{2g_i}}{1-a^{2g_i}}\gamma_c,\label{eq:increasingg_i}
\end{align}}\noindent
where $\gamma_c:=\frac{1-a^{2g_i}}{a^{2c}-a^{2g_i}} \ge 1$.
From \eqref{eq:minSSMSE_v2} and \eqref{eq:increasingg_i}, the channel MSE $\lambda_{i,g_i}^{\underline{\infty}}$ is increasing on $g_i$ as
{%\small
\begin{align*}
\lambda_{i,g_i}^{(\underline{\infty})} %\nonumber\\
&= \frac{\lambda_i}
{\bigl(\frac{1}{2}(1+\lambda_i\rho)\bigr) + \sqrt{\bigl(\frac{1}{2}(1+\lambda_i\rho)\bigr)^2+\frac{a^{2g_i}}{1-a^{2g_i}}\lambda_i\rho}} \nonumber\\
&
\ge \frac{\lambda_i}
{\bigl(\frac{1}{2}(1+\lambda_i\rho)\bigr) + \sqrt{\bigl(\frac{1}{2}(1+\lambda_i\rho)\bigr)^2+\frac{a^{2g_i}}{1-a^{2g_i}}\gamma_c\lambda_i\rho}} \nonumber\\
%&= \frac{\lambda_i}
%{\bigl(\frac{1}{2}(1+\lambda_i\frac{\rho_p}{\sigma_w^2})\bigr) + \sqrt{\bigl(\frac{1}{2}(1+\lambda_i\frac{\rho_p}{\sigma_w^2})\bigr)^2+\frac{a^{2g_i}}{1-a^{2g_i}}\lambda_i\frac{\rho_p}{\sigma_w^2}}} \nonumber\\
%&\ge \frac{\lambda_i}
%{\bigl(\frac{1}{2}(1+\lambda_i\frac{\rho_p}{\sigma_w^2})\bigr) + \sqrt{\bigl(\frac{1}{2}(1+\lambda_i\frac{\rho_p}{\sigma_w^2})\bigr)^2+\frac{a^{2g_i}}{1-a^{2g_i}}\gamma_c\lambda_i\frac{\rho_p}{\sigma_w^2}}} \nonumber\\
&
= \lambda_{i,g_i^\prime}^{(\underline{\infty})} %\label{eq:increasingSSMSE}
\end{align*}}\noindent
In \eqref{eq:maxSSMSE_v2}, $\lambda_{i,g_i}^{(\overline{\infty})}$ is a convex combination of $\lambda_{i,g_i}^{(\underline{\infty})}$ and $\lambda_i$ satisfying $\lambda_{i,g_i}^{(\underline{\infty})}<\lambda_i$ and thereby an increasing function as $g_i$ increases. 
Since the composite of increasing functions is increasing:
{%\small
\begin{align*}
\lambda_{i,g_i^\prime}^{(\overline{\infty})} 
&
=  a^{2(g_i^\prime-1)}\lambda_{i,g_i^\prime}^{(\underline{\infty})}   + (1-a^{2(g_i^\prime-1)})\lambda_i \\
&\le a^{2(g_i-1)}\lambda_{i,g_i^\prime}^{(\underline{\infty})}   + (1-a^{2(g_i-1)})\lambda_i \\
&\le a^{2(g_i-1)}\lambda_{i,g_i}^{(\underline{\infty})}   + (1-a^{2(g_i-1)})\lambda_i = \lambda_{i,g_i}^{(\overline{\infty})},
\end{align*}}\noindent
 we have the claim. $\hfill\blacksquare$ %$\lambda_{i,g_i}^{(\overline{\infty})}$ is increasing of $g_i$. 

\vspace{-0.65em}
%%%%%%%%%%%%%%%%%%%%%%%%%%%%%%%%%%%%%%%%%%%%%%%%%%%%%%%%%%%%%%%%%%%%%%%
\subsection{Proof of Proposition \ref{cor:sufficientCondForTrainingAllocation}}\label{subsec:sufficientCondForTrainingAllocation}
\vspace{-0.2em}

For the proof of Proposition \ref{cor:sufficientCondForTrainingAllocation}, it suffices to show that the matrix $\Cbf$ of size $G\times M_p$ is constructed (i.e., $U_j=0$ for $1\le j\le M_p$) by allocating the $n_d$ eigenvector indices corresponding to the block time-wise interval $\{g_1,\ldots,g_{n_d}\}$ since $\sum_{i=1}^{n_d}G/g_i = GM_p$ in \eqref{eq:nonlinearConst}. 

The proof is by induction, where the notations follow those of {\em Step (1)} and {\em Step (2)}.
When $q=1$, let $U_j=G=p^s$ be the initial value for $1\le j\le M_p$ where given $\{g_1,\ldots,g_{n_d}\} \in \Ic_G=\{1,p,\ldots,p^s\}$ as in \eqref{eq:divisorset}. % by the proposed algorithm.
%For $q=1$, after inserting the index $i_{q,j}$ at the $j$-th column as in step (1), $U_j$ is updated as $U_j=n_j \cdot 2^{s-g_{i_{q,j}}}$ where $n_j=2^{i_{q,j}}-1$ and $g_{i_{q,j}}=2^{k_{i_{q,j}}}\in\Ic_G$.
%This means that an index allocation w.r.t. the block time-wise interval $g_{i_{q,j}}$ is possible with $n_j$ times, starting at different row indices.
For any $1\le q\le G$, if $U_j\neq 0$, suppose that the unused $U_j$ entries at the $j$-th column can be described by the $n_j$ disjoint sets of equi-spacing $g_{i_{q,j}}=p^{k_{i_{q,j}}}\in\Ic_G$, i.e., $U_j=n_j\cdot p^{s-k_{i_{q,j}}}$ for some nonnegative integers $n_j$ and $k_{i_{q,j}}$. 
This means that  the unused $U_j$ entries can be viewed as a collection of $n_j$ disjoint sets where the entries of each set are equi-spaced with $g_{i_{q,j}}=p^{k_{i_{q,j}}}$.
%{\em at most} $n_j$ disjoint indices can be allocated using a row-wise mapping $g_{i_{q,j}}$ at the unused entries of the $j$-th column with starting at different row indices. 
%an index allocation w.r.t. the block time-wise interval $g_{i_{q,j}}$ into the unused entries of the $j$-th column is possible with $n_j$ times, starting at different row indices.
Thus, after inserting the index of $i_{q,j}$ with a row-wise allocation at {\em Step (1)}, $U_j$ is updated as $U_j=(n_j-1) \cdot p^{s-k_{i_{q,j}}}$.

In the subsequent iteration, if $U_j\neq 0$, there exist some row index $q^\prime>q$  such that we need to allocate the index of $i_{q^\prime,j}$ using a row-wise mapping $g_{i_{q^\prime,j}}$ starting from $[\Cbf]_{q^\prime,j}$ as shown in \eqref{eq:step1_v1}. 
%Note that, by Proposition \ref{prop:increasingSSMSE}, it follows that $g_{i_{q^\prime,j}}\ge g_{i_{q,j}}$.
Note that, by the assumption, it follows that $g_{i_{q^\prime,j}}\ge g_{i_{q,j}}$.
Since each set of equi-spacing $g_{i_q,j}=p^{k_{i_{q,j}}}$ at the preceding step can be separated by the $g_{i_{q^\prime,j}}/g_{i_q,j}=p^{k_{i_{q^\prime,j}}-k_{i_{q,j}}}$ disjoint subsets of equi-spacing $g_{i_{q^\prime,j}}=p^{k_{i_{q^\prime,j}}}$,
the remaining $U_j$ entries can be viewed as the $(n_j-1) p^{k_{i_{q^\prime,j}}-k_{i_{q,j}}}$ disjoint sets of equi-spacing $g_{i_{q^\prime,j}}$, i.e., 
$U_j=((n_j-1) p^{k_{i_{q^\prime,j}}-k_{i_{q,j}}}) \cdot p^{s-k_{i_{q^\prime,j}}}$. % where $n_j\leftarrow (n_j-1) 2^{k_{i_{q^\prime,j}}-k_{i_{q,j}}}$.
Therefore, it is possible to allocate the index of $i_{q^\prime,j}$ into one of the disjoint sets of equi-spacing $g_{i_{q^\prime,j}}$. We then update $n_j= (n_j-1) p^{k_{i_{q^\prime,j}}-k_{i_{q,j}}}$ and $U_j=(n_j-1)\cdot p^{s-k_{i_{q^\prime},j}}$.
Since this process repeats until $U_j\neq 0$ for all $j$, we have the claim. \hfill$\blacksquare$

%\vspace{0.5em}
\begin{lemma}\label{lem:property_est_h_l}
During the $\ell$-th training period, the channel estimate $\hat{\hbf}_{u,\ell |\ell}$ based on Kalman filtering is characterized by
$E\{\hat{\hbf}_{u,\ell |\ell}\} = \mathbf{0}$ and $E\{\hat{\hbf}_{u,\ell |\ell}\hat{\hbf}_{u,\ell |\ell}^H\} = \Rbf_{\hbf_u} - \Pbf_{u,\ell | \ell}$.
\end{lemma}

%\vspace{0.5em}
{\em Proof:}
For notational simplicity, we omit the lower index $u$.
From \eqref{eq:initialCond} and \eqref{eq:measurementupdateH}, the channel estimate $\hat{\hbf}_{\ell |\ell}$ for $\ell =0$ is given by
\begin{equation}
\hat{\hbf}_{0|0} = \Pbf_{0|-1}\Sbf_0  (\Sbf_0^H\Pbf_{0|-1}\Sbf_0 + \Ibf_{M_p})^{-1} \ybf_{0,pilot},
\end{equation}
where recall that $\ybf_{\ell,pilot}= [y_{\ell M+1},\cdots,y_{\ell M+M_p}]^T$ denotes the $\ell$-th received training symbols 
and $\Sbf_{u,\ell}=[\sbf_{u,\ell M+1} \cdots \sbf_{u,\ell M+M_p}]$ denotes the $\ell$-th training symbols, as shown in  \eqref{eq:statespacemodel_y2}.
%and $\Sigmabf_\ell:=\text{diag}(\bar{\sigma}_{w,\ell M+1}^2,\cdots,\bar{\sigma}_{w,\ell M+M_p}^2)$, as shown in \eqref{eq:multiuser_eq_training_v2}.
%because of $\hbf_{0|-1}=\mathbf{0}$ as shown in \eqref{eq:initialCond}.
Since $E\{\ybf_{0,pilot}\ybf_{0,pilot}^H\}=\Sbf_0^H\Pbf_{0|-1}\Sbf_0+\Ibf_{M_p}$, we have
{%\small
\begin{align*}
E\{\hat{\hbf}_{0|0}\hat{\hbf}_{0|0}^H\} %\nonumber\\
%&= \Kbf_0\left(\Sbf_0^H\Pbf_{0|-1}\Sbf_0+\sigma_w^2\Ibf_{M_p}\right)\Kbf_0^H \\
%&= \left(\Pbf_{0|-1}\Sbf_0(\Sbf_0^H\Pbf_{0|-1}\Sbf_0+\sigma_w^2\Ibf_{M_p})^{-1}\right) \nonumber\\
%&~~~ \left(\Sbf_0^H\Pbf_{0|-1}\Sbf_0+\sigma_w^2\Ibf_{M_p}\right) \nonumber\\
%&~~~ \left((\Sbf_0^H\Pbf_{0|-1}\Sbf_0+\sigma_w^2\Ibf_{M_p})^{-1}\Sbf_0^H\Pbf_{0|-1}\right) \\
&= \Pbf_{0|-1}\Sbf_0(\Sbf_0^H\Pbf_{0|-1}\Sbf_0+\Ibf_{M_p})^{-1}\Sbf_0^H\Pbf_{0|-1} \nonumber\\
%&= \Rbf_\hbf - (\Rbf_\hbf - \Pbf_{0|-1}\Sbf_0(\Sbf_0^H\Pbf_{0|-1}\Sbf_0+\sigma_w^2\Ibf_{M_p})^{-1}\Sbf_0^H\Pbf_{0|-1})\\
&\stackrel{(a)}{=} \Rbf_\hbf - \bigl(\Pbf_{0|-1} - \Pbf_{0|-1}\Sbf_0  \nonumber \\
&~~~~ 
(\Sbf_0^H\Pbf_{0|-1}\Sbf_0+\Ibf_{M_p})^{-1}\Sbf_0^H\Pbf_{0|-1}\bigr) \\% \label{eq:cov_est_h_0}\\
&
= \Rbf_\hbf - \Pbf_{0|0}, 
\end{align*}}\noindent
where $(a)$ holds by $\Pbf_{0|-1}=\Rbf_\hbf$. % in \eqref{eq:initialCond}. %the initial parameter for the Kalman filter $\Pbf_{0|-1}=\Rbf_\hbf$ in \eqref{eq:initialCond}.
%For the given channel model of $\hbf_\ell\sim\mathcal{CN}(\mathbf{0},\Rbf_\hbf)$ in \eqref{eq:statespacemodel_h} and $\bar{\wbf}_{\ell}\sim\mathcal{CN}(\mathbf{0},\Sigmabf_\ell)$ in \tcr{\eqref{eq:multiuser_eq_training_v2}},
Here, $E\{\hat{\hbf}_{0|0}\}=\mathbf{0}$ from $E\{\ybf_{0,pilot}\}=\Sbf_0^HE\{\hbf_0\}+E\{\wbf_0\}=\mathbf{0}$.
%Note that $E\{\ybf_{0,pilot}\}=\Sbf_0^HE\{\hbf_0\}+E\{\wbf_0\}=\mathbf{0}$ and thereby $E\{\hat{\hbf}_{0|0}\}=\mathbf{0}$.

During the $\ell$-th training period, %for $\ell\in\mathbb{N}$, 
the channel estimate $\hat{\hbf}_{\ell |\ell}$ is given by from \eqref{eq:measurementupdateH} and \eqref{eq:timeupdateH}:
{%\small
\begin{align}
\hat{\hbf}_{\ell |\ell} 
&= a\hat{\hbf}_{\ell-1|\ell-1} +  \Pbf_{\ell |\ell-1}\Sbf_\ell(\Sbf_\ell^H\Pbf_{\ell |\ell-1}\Sbf_\ell+\Ibf_{M_p})^{-1} \nonumber\\
&~~~ 
(\ybf_{\ell,pilot}-\Sbf_\ell^H a\hat{\hbf}_{\ell-1|\ell-1}). \label{eq:hest_ll}
\end{align}}\noindent
Denote by $\ebf_\ell \in \mathbb{C}^{M_p}$ the innovation process of Kalman filter given by
%From \eqref{eq:decomh_k}, we obtain
{%\small
\begin{align}
\ebf_\ell 
&= \ybf_{\ell,pilot}-\Sbf_\ell^H(a\hat{\hbf}_{\ell-1|\ell-1}) \nonumber\\
&
= \left(\Sbf_\ell^H\hbf_\ell + \wbf_\ell\right) - \Sbf_\ell^H (a\hat{\hbf}_{\ell-1|\ell-1}) \\
&\stackrel{(a)}{=} \bigl(\Sbf_\ell^H\bigl(a(\hat{\hbf}_{\ell-1|\ell-1}+\tilde{\hbf}_{\ell-1})+\sqrt{1-a^2}\bbf_\ell\bigr) + \wbf_\ell\bigr) \nonumber \\
&~~~ 
- \Sbf_\ell^Ha\hat{\hbf}_{\ell-1|\ell-1} \nonumber\\ %\label{eq:innovationKalmanFilter}\\
&
= \Sbf_\ell^H\bigl(a\tilde{\hbf}_{\ell-1}+\sqrt{1-a^2}\bbf_\ell\bigr) + \wbf_\ell,
\end{align}}\noindent
where %the equality \eqref{eq:innovationKalmanFilter} 
$(a)$
holds by \eqref{eq:statespacemodel_h} and $\tilde{\hbf}_{\ell}:=\hbf_\ell - \hat{\hbf}_{\ell |\ell}$.
Note that $\ebf_\ell$ is independent of $ \hat{\hbf}_{\ell-1|\ell-1}$ due to the orthogonality property of the MMSE estimation and an independent process noise $\bbf_\ell$, then we have that $\ebf_\ell$ has zero mean and covariance matrix $E\{\ebf_\ell\ebf_\ell^H\} = \Sbf_\ell^H\Pbf_{\ell |\ell-1}\Sbf_\ell+\Ibf_{M_p}$. Thus, we obtain $E\{\hat{\hbf}_{\ell |\ell}\} = aE\{\hat{\hbf}_{\ell-1|\ell-1}\}+\Kbf_\ell E\{\ebf_\ell\}=\mathbf{0}$.
From \eqref{eq:hest_ll}, $E\{\hat{\hbf}_{\ell |\ell}\hat{\hbf}_{\ell |\ell}^H\}$ is given by
{%\small
\begin{align}
%\hspace{-0.85em}
&
E\{\hat{\hbf}_{\ell |\ell}\hat{\hbf}_{\ell |\ell}^H\} \nonumber\\
&= a^2E\{\hat{\hbf}_{\ell-1|\ell-1}\hat{\hbf}_{\ell-1|\ell-1}^H\} + \Kbf_\ell E\{\ebf_\ell\ebf_\ell^H\}\Kbf_\ell^H \\
&= a^2(\Rbf_\hbf - \Pbf_{\ell-1|\ell-1})  \nonumber\\
&~~~ 
+ \bigl( \Pbf_{\ell|\ell-1}\Sbf_\ell(\Sbf_\ell^H\Pbf_{\ell|\ell-1}\Sbf_\ell+\Ibf_{M_p})^{-1}\Sbf_\ell^H\Pbf_{\ell|\ell-1} \bigr) \\
&= a^2\Rbf_\hbf - \bigl( \Pbf_{\ell|\ell-1}-(1-a^2)\Rbf_\hbf\bigr) \nonumber\\
&~~~ 
+ \bigl( \Pbf_{\ell |\ell-1}\Sbf_\ell(\Sbf_\ell^H\Pbf_{\ell|\ell-1}\Sbf_\ell+\Ibf_{M_p})^{-1}\Sbf_\ell^H\Pbf_{\ell |\ell-1} \bigr) \label{eq:cov_est_h_l}\\
&= \Rbf_\hbf - \bigl( \Pbf_{\ell |\ell-1} - \Pbf_{\ell |\ell-1}\Sbf_\ell(\Sbf_\ell^H\Pbf_{\ell |\ell-1}\Sbf_\ell+\Ibf_{M_p})^{-1}  \nonumber\\
&~~~~ 
\Sbf_\ell^H\Pbf_{\ell |\ell-1} \bigr) \nonumber\\
&
= \Rbf_\hbf - \Pbf_{\ell |\ell}, 
\end{align}}\noindent
where the equality \eqref{eq:cov_est_h_l} follows \eqref{eq:timeupdateH}.
Since this Kalman recursion repeats, we have the claim. $\hfill \blacksquare$
\subsection{Proof of Proposition \ref{prop:deterministicSINR}}\label{subsec:deterministicSINR}
%\vspace{-0.4em}

To derive the deterministic quantity for $SINR_{u,\ell}$ in the limit of $N_t\rightarrow \infty$, we use the analysis technique \cite{Hoydis&Brink&Debbah:13JSAC}.
Applying Lemma \ref{lem:property_est_h_l}, we have
{%\small
\begin{equation}
%\hspace{-0.4em}
\frac{1}{N_t^2} |\hat{\hbf}_{u,\ell |\ell}^H\hat{\hbf}_{u,\ell |\ell}|^2  -  \frac{1}{N_t^2} |\text{tr}(\Rbf_{\hbf_u}-\Pbf_{u,\ell |\ell})|^2 
 ~\substack{a.s.\\ \overrightarrow{N_t\rightarrow \infty}} ~0,\label{eq:determinstic_v1}
%\hspace{-0.4em}\frac{1}{N_t^2} |\hat{\hbf}_{u,\ell |\ell}^H\hat{\hbf}_{u,\ell |\ell}|^2  -  \frac{1}{N_t^2} |\text{tr}(\Rbf_{\hbf_u}-\Pbf_{u,\ell |\ell})|^2 
%& ~\substack{a.s.\\ \overrightarrow{N_t\rightarrow \infty}} ~0,\label{eq:determinstic_v1}
\end{equation}}\noindent
where $\substack{a.s.\\ \longrightarrow}$ denotes the almost sure convergence.
If $u\neq u^\prime$, then $\hbf_{u,\ell}$ and $\hat{\hbf}_{u^\prime,\ell |\ell}$ are mutually independent, %in \eqref{eq:multiuser_eq_training}, 
thus we have using Lemma \ref{lem:property_est_h_l} as
{%\small
\begin{equation}
\frac{1}{N_t^2} |\hbf_{u,\ell}^H\hat{\hbf}_{u^\prime,\ell |\ell}|^2   -   \frac{1}{N_t^2} \text{tr}\left(\Rbf_{\hbf_u}(\Rbf_{\hbf_{u^\prime}}-\Pbf_{u^\prime,\ell |\ell})\right) %\nonumber\\
~\substack{a.s.\\ \overrightarrow{N_t\rightarrow \infty}} ~0,\label{eq:determinstic_v2}
\end{equation}}\noindent
%If $u\neq u^\prime$, then $\hat{\hbf}_{u,\ell |\ell}$ and $\hat{\hbf}_{u^\prime,\ell |\ell}$ are mutually independent, thus we have using Lemma \ref{lem:property_est_h_l} as
%\begin{align}
%&\frac{1}{N_t^2} |\hat{\hbf}_{u,\ell |\ell}^H\hat{\hbf}_{u^\prime,\ell |\ell}|^2  -  \frac{1}{N_t^2} \text{tr}\left((\Rbf_{\hbf_u}-\Pbf_{u,\ell |\ell})(\Rbf_{\hbf_{u^\prime}}-\Pbf_{u^\prime,\ell |\ell})\right) \nonumber\\
%& ~\substack{a.s.\\ \overrightarrow{N_t\rightarrow \infty}} ~0,\label{eq:determinstic_v2}
%\end{align}
Since $\tilde{\hbf}_{u,\ell}$ is independent of $\hat{\hbf}_{u,\ell |\ell}$ %for $u\neq u^\prime$ 
by the orthogonality property of the MMSE estimate, we obtain by using Lemma \ref{lem:property_est_h_l} and $\tilde{\hbf}_{u,\ell | \ell}\sim\mathcal{CN}(\mathbf{0},\Pbf_{u,\ell |\ell})$ as
{%\small
\begin{equation}
\frac{1}{N_t^2} |\tilde{\hbf}_{u,\ell}^H\hat{\hbf}_{u,\ell |\ell}|^2  -  \frac{1}{N_t^2} \text{tr}\left(\Pbf_{u,\ell |\ell}(\Rbf_{\hbf_{u}}-\Pbf_{u,\ell |\ell})\right) %\nonumber\\
 ~\substack{a.s.\\ \overrightarrow{N_t\rightarrow \infty}} ~0.\label{eq:determinstic_v3}
\end{equation}}\noindent
Substituting \eqref{eq:determinstic_v1}, \eqref{eq:determinstic_v2}, and \eqref{eq:determinstic_v3} into \eqref{eq:sinr_ul} with 
$\alpha_u^2=(\text{tr}(\Rbf_{\hbf_u}-\Pbf_{u,\ell|\ell}))^{-1}$ for $1\le u \le U$,
%$\alpha = (\sum_{u=1}^U \text{tr}(\Rbf_{\hbf_u}-\Pbf_{u,\ell |\ell}))^{-1}$, 
we have the deterministic equivalent SINR, given by% as shown in \eqref{eq:sinr_ul_v2}. 
{%\small
\begin{align}
%&
\overline{SINR}_{u,\ell} %\nonumber\\
&= 
|\text{tr}(\Rbf_{\hbf_u}-\Pbf_{u,\ell |\ell})|^2 \nonumber\\
&
\left(\frac{1}{\alpha_u^2\rho} + \text{tr}\left(\Pbf_{u,\ell |\ell}(\Rbf_{\hbf_{u}}-\Pbf_{u,\ell |\ell})\right)  \right.\nonumber\\
&\left. 
+ \sum_{u^\prime=1:u^\prime\neq u}^U  \frac{\alpha_{u^\prime}^2}{\alpha_u^2} \text{tr}\left(\Rbf_{\hbf_u}(\Rbf_{\hbf_{u^\prime}}-\Pbf_{u^\prime,\ell |\ell})\right)\right)^{-1} \label{eq:sinr_ul_v2}.
\end{align}}\noindent

Note that the estimation error covariance matrix $\Pbf_{u,\ell |\ell}$ has the same set of eigenvectors of $\Rbf_{\hbf_u}$ over all $\ell$ when we use its eigenvectors as the training signals \cite{Noh&Zoltowski&Sung&Love:14STSP}.
That is, given the ED of $\Rbf_{\hbf_u}=\Ubf_u\Lambdabf_u\Ubf_u^H$, % and $\Pbf_{u,0|-1}=\Ubf_u\Lambdabf_u^{(0)}\Ubf_u^H=\Rbf_{\hbf_u}$, 
$\Pbf_{u,\ell |\ell}$ is eigen-decomposed by $\Pbf_{u,\ell |\ell}=\Ubf_u\bar{\Lambdabf}_u^{(\ell)}\Ubf_u^H$.
From $\text{tr}(\Abf\Bbf\Cbf)=\text{tr}(\Bbf\Cbf\Abf)$, the terms in \eqref{eq:sinr_ul_v2} are then given by
{%\small 
\begin{align}
\text{tr}(\Rbf_{\hbf_u}-\Pbf_{u,\ell |\ell}) 
&= \text{tr}(\Lambdabf_{u} - \bar{\Lambdabf}_{u}^{(\ell)} ) \label{eq:determinstic_v4}\\
\text{tr}\left(\Rbf_{\hbf_u}(\Rbf_{\hbf_{u^\prime}}-\Pbf_{u^\prime,\ell |\ell})\right) 
&=  \text{tr}\bigl(\Lambdabf_u\Ubf_u^H\Ubf_{u^\prime} \nonumber\\
&
 (\Lambdabf_{u^\prime}-\bar{\Lambdabf}_{u^\prime}^{(\ell)} )  \Ubf_{u^\prime}^H\Ubf_u\bigr) \label{eq:determinstic_v5}\\
\text{tr}\left(\Pbf_{u,\ell |\ell}(\Rbf_{\hbf_{u^\prime}}-\Pbf_{u^\prime,\ell |\ell})\right) 
&= \text{tr}\bigl(  \bar{\Lambdabf}_{u}^{(\ell)} (\Lambdabf_{u} - \bar{\Lambdabf}_{u}^{(\ell)} )  \bigr) \label{eq:determinstic_v6}
\end{align}}\noindent
By substituting \eqref{eq:determinstic_v4}, \eqref{eq:determinstic_v5}, and \eqref{eq:determinstic_v6} into \eqref{eq:sinr_ul_v2}, the SINR expression is rewritten as \eqref{eq:sinr_ul_v3}.
$\hfill\blacksquare$

\subsection{Derivation of the lower bound in \eqref{eq:sinr_ul_steadystate_v2}}\label{subsec:lowerbound_steadystate_SINR}
%\subsection{Proof of Remark \ref{rem:lowerbound_steadystate_SINR}}\label{subsec:lowerbound_steadystate_SINR}
%\vspace{-0.2em}

By substituting $\gbf_u\in\mathbb{N}^{n_d}$ into \eqref{eq:minSSMSE_v2} and \eqref{eq:maxSSMSE_v2}, we can derive ${\lambdabf}_{\gbf_u}^{(\underline{\infty})}$ and ${\lambdabf}_{\gbf_u}^{(\overline{\infty})}$. From an inequality of \eqref{eq:boundSSMSE}, it follows that
%\begin{align} \label{eq:inequality_for_sinr_ul}
%\bigl\|\lambdabf_u - {\lambdabf}_{\gbf_u}^{(\overline{\infty})}\bigr\|_1  
%&\le \lim_{\ell\rightarrow\infty} \bigl\|\lambdabf_u - \bar{\lambdabf}_{\gbf_u}^{(\ell)}\bigr\|_1 \\
%%
%\lim_{\ell\rightarrow\infty} \bigl\|\bar{\lambdabf}_{\gbf_u}^{(\ell)}\odot(\lambdabf_{u} - \bar{\lambdabf}_{\gbf_u}^{(\ell)})\bigr\|_1
%&\le 
%\bigl\|{\lambdabf}_{\gbf_u}^{(\overline{\infty})}\odot(\lambdabf_{u} - {\lambdabf}_{\gbf_u}^{(\underline{\infty})})\bigr\|_1, \nonumber 
%\end{align}
%and 
%\begin{align} \label{eq:inequality_for_sinr_ul_v2}
%&\lim_{\ell\rightarrow\infty} \text{tr}\bigl(\Lambdabf_u\Ubf_u^H\Ubf_{u^\prime}(\Lambdabf_{u^\prime} - \bar{\Lambdabf}_{\gbf_{u^\prime}}^{(\ell)})\Ubf_{u^\prime}^H\Ubf_u\bigr) %\nonumber\\
%%&
%\le  
%\text{tr}\bigl( \Lambdabf_u\Ubf_{u}^H\Ubf_{u^\prime}(\Lambdabf_{u^\prime} - {\Lambdabf}_{\gbf_{u^\prime}}^{(\underline{\infty})}) \Ubf_{u^\prime}^H\Ubf_{u} \bigr),   %\\
%\end{align}
{%\small
\begin{align} \label{eq:inequality_for_sinr_ul}
\bigl\|\lambdabf_u - {\lambdabf}_{\gbf_u}^{(\overline{\infty})}\bigr\|_1  
&\le \lim_{\ell\rightarrow\infty} \bigl\|\lambdabf_u - \bar{\lambdabf}_{\gbf_u}^{(\ell)}\bigr\|_1 \nonumber \\
\lim_{\ell\rightarrow\infty} \bigl\|\bar{\lambdabf}_{\gbf_u}^{(\ell)}\odot(\lambdabf_{u} - \bar{\lambdabf}_{\gbf_u}^{(\ell)})\bigr\|_1
&\le 
\bigl\|{\lambdabf}_{\gbf_u}^{(\overline{\infty})}\odot(\lambdabf_{u} - {\lambdabf}_{\gbf_u}^{(\underline{\infty})})\bigr\|_1 \nonumber \\
%\end{align}
%and 
%\begin{align} \label{eq:inequality_for_sinr_ul_v2}
\lim_{\ell\rightarrow\infty} \text{tr}\bigl(\Lambdabf_u\Ubf_u^H\Ubf_{u^\prime}(\Lambdabf_{u^\prime} - &\bar{\Lambdabf}_{\gbf_{u^\prime}}^{(\ell)})\Ubf_{u^\prime}^H\Ubf_u\bigr) \nonumber\\
%&
\le  
\text{tr}\bigl( \Lambdabf_u\Ubf_{u}^H&\Ubf_{u^\prime}(\Lambdabf_{u^\prime} - {\Lambdabf}_{\gbf_{u^\prime}}^{(\underline{\infty})}) \Ubf_{u^\prime}^H\Ubf_{u} \bigr),  %\nonumber %\\
\end{align}}\noindent
%where $\Pbf_{u,0|-1}=\Ubf_u\Lambdabf_u^{(0)}\Ubf_u^H=\Rbf_{\hbf_u}$ with $\Lambdabf_u^{(0)}=\text{diag}(\lambdabf_u^{(0)})$ as in \eqref{eq:initialCond} and
%$\Pbf_{u,\ell |\ell} = \Ubf_u\bar{\Lambdabf}_{u}^{(\ell)}\Ubf_u^H$ with $\bar{\Lambdabf}_{u}^{(\ell)}=\text{diag}(\bar{\lambdabf}_{\gbf_u}^{(\ell)})$.
where $\Pbf_{u,\ell |\ell} = \Ubf_u\bar{\Lambdabf}_{u}^{(\ell)}\Ubf_u^H$ with $\bar{\Lambdabf}_{u}^{(\ell)}=\text{diag}(\bar{\lambdabf}_{\gbf_u}^{(\ell)})$, ${\Lambdabf}_{\gbf_{u^\prime}}^{(\underline{\infty})}=\text{diag}({\lambdabf}_{\gbf_{u^\prime}}^{(\underline{\infty})})$, and the initial conditions \eqref{eq:initialCond}. % and \eqref{eq:ToeplitzApproxTheorem_v2}.
By applying the inequalities \eqref{eq:inequality_for_sinr_ul} %and \eqref{eq:inequality_for_sinr_ul_v2}  
to \eqref{eq:sinr_ul_v3},
we obtain the closed-form lower bound on the steady-state SINR, as shown in \eqref{eq:sinr_ul_steadystate_v2}. % \eqref{eq:sinr_ul_steadystate}   
%{\small
%\begin{align}
%\overline{SINR}_{u} 
%&= \lim_{\ell\rightarrow\infty} \overline{SINR}_{u,\ell} \nonumber\\
%&\ge
%\bigl\|\lambdabf_u - {\lambdabf}_{\gbf_u}^{(\overline{\infty})}\bigr\|_1^2 %\nonumber\\
%%&~~~
%\left( \frac{1}{\alpha\rho} 
%+ \bigl\|{\lambdabf}_{\gbf_u}^{(\overline{\infty})}\odot(\lambdabf_{u} - {\lambdabf}_{\gbf_u}^{(\underline{\infty})})\bigr\|_1   \right.\nonumber\\
%&~~~\left. 
%+ \sum_{u^\prime=1: u^\prime \neq u}^U   \text{tr}\left(\Lambdabf_u\Ubf_u^H\Ubf_{u^\prime}(\Lambdabf_{u^\prime} - {\Lambdabf}_{\gbf_{u^\prime}}^{(\underline{\infty})}) \Ubf_{u^\prime}^H\Ubf_u\right) \right)^{-1}.  \label{eq:sinr_ul_steadystate}
%\end{align}}\noindent
%Note that the results of \eqref{eq:inequality_for_sinr_ul}, \eqref{eq:inequality_for_sinr_ul_v2}, and \eqref{eq:sinr_ul_steadystate} are obtained under the exact ED of $\Rbf_{\hbf_s}$, i.e., $\Rbf_{\hbf_s}=\Ubf_s\Lambdabf_s\Ubf_s^H$.
%By the assumption \eqref{eq:ToeplitzApproxTheorem_v2}, the lower bound of \eqref{eq:sinr_ul_steadystate} is approximated as shown in \eqref{eq:sinr_ul_steadystate_v2}.
$\hfill\blacksquare$

%\vspace{-0.5em}
%%%%%%%%%%%%%%%%%%%%%%%%%%%%%%%%%%%%%%%%%%%%%%%%%%%%%%%%%%%%%%%%%%%%%%%
% References
%%%%%%%%%%%%%%%%%%%%%%%%%%%%%%%%%%%%%%%%%%%%%%%%%%%%%%%%%%%%%%%%%%%%%%%
%\begin{spacing}{0.98}
%\bibliographystyle{ieeetr}
\bibliographystyle{IEEEbib}
%\bibliographystyle{IEEEtran}
%\bibliography{IEEEabrv,\HOME/bibs/referenceBibs}
\bibliography{IEEEabrv,referenceBibs}
%\end{spacing}

\end{document}

%% file: macros.tex
\def\psfancypar#1#2{\begingroup\def\par{\endgraf\endgroup\lineskiplimit=0pt}
               \setbox2=\hbox{\large\sc #2}
               \newdimen\tmpht \tmpht \ht2 \advance\tmpht by \baselineskip
               \font\hhuge=Times-Bold at \tmpht
               \setbox1=\hbox{{\hhuge #1}}
               \count7=\tmpht \count8=\ht1
               \divide\count8 by 1000 \divide\count7 by \count8 
               \tmpht=.001\tmpht\multiply\tmpht by \count7 
               \font\hhuge=Times-Bold at \tmpht
               \setbox1=\hbox{{\hhuge #1}}
               \noindent
                \hangindent1.05\wd1
               \hangafter=-2 {\hskip-\hangindent
               \lower1\ht1\hbox{\raise1.0\ht2\copy1}%
                \kern-0\wd1}\copy2\lineskiplimit=-1000pt}

\newcommand{\E}{\mbox{{\rm E}}}

\newcommand{\abf}{\mbox{${\bf a}$}}

\def\boxit#1{\vbox{\hrule\hbox{\vrule\kern3pt
        \vbox{\kern3pt#1\kern3pt}\kern3pt\vrule}\hrule}}

\def\reals{ { {\rm  I \kern-0.15em R }  } }
\def\complex{ {\,{{\rm C} \kern-0.50em \raise0.20ex {  |}}\, }}

\def\lambdabf{\hbox{\boldmath$\lambda$\unboldmath}}
\def\mubf{\hbox{\boldmath$\mu$\unboldmath}}

\def\Sigmabf{\hbox{$\bf \Sigma$}}

\def\Lambdabf{\mbox{$ \bf \Lambda $}}

\def\abf{{\bf a}}
\def\bbf{{\bf b}}

\def\dbf{{\bf d}}
\def\ebf{{\bf e}}
\def\fbf{{\bf f}}
\def\gbf{{\bf g}}
\def\hbf{{\bf h}}

\def\qbf{{\bf q}}

\def\sbf{{\bf s}}

\def\ubf{{\bf u}}
\def\vbf{{\bf v}}
\def\wbf{{\bf w}}
\def\xbf{{\bf x}}
\def\ybf{{\bf y}}

\def\xbf{{\bf x}}
\def\ybf{{\bf y}}
\def\Abf{{\bf A}}
\def\Bbf{{\bf B}}
\def\Cbf{{\bf C}}
\def\Dbf{{\bf D}}

\def\Fbf{{\bf F}}

\def\Hbf{{\bf H}}
\def\Ibf{{\bf I}}

\def\Kbf{{\bf K}}

\def\Pbf{{\bf P}}

\def\Rbf{{\bf R}}
\def\Sbf{{\bf S}}

\def\Ubf{{\bf U}}
\def\Vbf{{\bf V}}

\def\Ic{{\cal I}}

\def\Nc{{\cal N}}
\def\Oc{{\cal O}}

\def\Sc{{\cal S}}

\def\be{\vskip .3cm \begin{equation}}
\def\ee{\end{equation} \vskip .4cm \noindent}
\newcommand{\R}{\mbox{$\hat {\bf R}_{N}$}}

\def\Rxx{\Rbf_{\ssstyle X\kern-.1em X}}

\let\ssstyle=\scriptscriptstyle

\def\Kout{\setbox1=\hbox{\Huge\bf K}\hbox to
1.05\wd1{\hspace{.05\wd1}
}}
\def\Sout{\setbox1=\hbox{\Huge\bf S}\hbox to 1.05\wd1{\hspace{.05\wd1}
}}

%% file: labelfig.tex
  \ifx\LabelFigloaded\MYundefined\relax
  \else
   \fi

  \def\LabelFigloaded{\relax}% now loaded

  \chardef\LabelFigCatAt\the\catcode`\@
  \catcode`\@=11

 \let\LabelFigwlog@ld\wlog
 \def\wlog#1{\relax}

 \ifx\\\MYundefined@
    \let\\\relax
 \fi

  \def\ms@g{\immediate\write16}

 \def\N@wif{\csname newif\endcsname }
 \def\Temp@ {\N@wif\ifIN@}
 \ifx\INN@\MYundefined@
    \else \let\Temp@\relax
 \fi
 \Temp@

  \def\IN@{\expandafter\INN@\expandafter}
  \long\def\INN@0#1@#2@{\long\def\NI@##1#1##2##3\ENDNI@
    {\ifx\m@rker##2\IN@false\else\IN@true\fi}%
     \expandafter\NI@#2@@#1\m@rker\ENDNI@}
  \def\m@rker{\m@@rker}
 
  \newtoks\Initialtoks@  \newtoks\Terminaltoks@
  \def\SPLIT@{\expandafter\SPLITT@\expandafter}
  \def\SPLITT@0#1@#2@{\def\TTILPS@##1#1##2@{%
     \Initialtoks@{##1}\Terminaltoks@{##2}}\expandafter\TTILPS@#2@}

 \def\Shifted@@#1#2#3{\setbox0=\hbox{#3}%
   \raise -\dp0\vbox {\kern-#2%
       \hbox {\kern#1\unhbox0\kern-#1}%
           \kern#2}}

 \newcount\gridcount
 \newbox\auxGridbox@ \newbox\hGridbox@ \newbox\vGridbox@
 \newbox\Labelbox@ \newbox\auxLabelbox@
 \newbox\Coordinatebox@
 \newtoks\Labeltoks@
 \newdimen\Wdd@ \newdimen\Htt@
 \newdimen\Wddd@ \newdimen\Httt@
 
 \def\Wr@{\immediate\write16}

 \newdimen\GL@wd%% grid-line width
 \def\GridLineWidth#1{\GL@wd=#1}

 \def\gobble#1{}
 \def\EdgeErr@{\Wr@{}%
      \Wr@{\string\Edges\space argument
      1, 10, 100 or 1000 please\string!}%
      }

 \newcount\Edgect@

 \def\Sweepup#1\endSweepup{}

 \def\SetEdges@{%
    \edef\Zr@@s{\expandafter\gobble\number\Edgect@\empty}%
        \count255=0\Zr@@s\relax
        \ifnum\count255=\z@\else\EdgeErr@\show\tailtest\fi
        \count255=1\Zr@@s\relax%\showthe\count255
        \ifnum\count255=\Edgect@\relax\else\EdgeErr@\show\leadtest\fi
    \EdgGl@b\edef\Zr@s{\expandafter\gobble\Zr@@s\empty}%\show\Zr@s
    \ifnum\Edgect@>\@ne\relax\EdgGl@b\let\L@Dc\empty
        \else\EdgGl@b\edef\L@Dc{\string.}\fi
    \ifnum\Edgect@>\@ne\relax
        \EdgGl@b\edef\Edgescale@##1{\divide##1 by \Edgect@}%
        \else\EdgGl@b\edef\Edgescale@##1{}\fi
    }

 \def\Edges#1{\Edgect@=#1\relax
     \let\EdgGl@b\global \SetEdges@}

 \Edges{1}%% default

 \def\hhrule{\hrule height \GL@wd\vskip-.\GL@wd}

 \def\hRule@{%
   \advance\gridcount -2%
   \vfil\hhrule\vfil
   \llap{\smash{\raise -2.5pt
     \hbox{\L@Dc\number\gridcount\Zr@s\kern2pt}}}%
   \hhrule
   }

\def\vvrule{\vrule width \GL@wd \kern-\GL@wd}

 \def\vRule@{\advance\gridcount 2%
   \hfil\vvrule\hfil
   \setbox\auxGridbox@=\vbox to 0pt
      {\vskip \Htt@\vskip 2pt
        \hbox to 0pt{\hss\L@Dc\number\gridcount\Zr@s\hss}\vss}%
      \wd\auxGridbox@=0pt \box\auxGridbox@
   \vvrule
   }

 \def\PlaceGrid@@{\gridcount=10 
  \setbox\hGridbox@=\hbox{%
        \hbox{%
             \hskip-.4pt\vrule
             \vbox to \Htt@{%
               \offinterlineskip\parindent=\z@\relax
               \hbox to \Wdd@{\hfil}
               \hRule@\hRule@\hRule@\hRule@
               \vfil\hhrule\vfil}%
             \vrule\hskip-.4pt}
    }%
  \gridcount=0%
  \setbox\vGridbox@=\hbox{%
      \vbox{\offinterlineskip\parindent=0pt\hsize=0pt
         \vskip-.4pt\hrule%
         \hbox to \Wdd@{%
                 \vtop to \Htt@{\vfil}%
                 \vRule@\vRule@\vRule@\vRule@
                 \hfil\vvrule\hfil}%
         \hrule\vskip-.4pt}}%
  \wd\hGridbox@=0pt\ht\hGridbox@=0pt
  \wd\vGridbox@=0pt\ht\vGridbox@=0pt
  \hbox{\box\hGridbox@\box\vGridbox@}%
  }

 \def\LabelsGlobal{\def\LabGl@b{\global}}
 \def\LabelsLocal{\def\LabGl@b{}}
 \LabelsGlobal %% default

 \def\SetLabels#1\endSetLabels{%
   \LabGl@b\Labeltoks@={#1()\\}%
   }

 \LabGl@b\Labeltoks@={()\\}

 \def\ShowGrid{\LabGl@b\let\PlaceGrid@\PlaceGrid@@}
 \def\HideGrid{\LabGl@b\let\PlaceGrid@\relax}
 \def\Grids{\ShowGrid\LabGl@b\let\GridSwitch@\ShowGrid}
 \def\noGrids{\HideGrid\LabGl@b\let\GridSwitch@\HideGrid}

 \noGrids

 \def\bAdjust@@{%
     \setbox\auxLabelbox@=\hbox{\raise \dp\auxLabelbox@
            \box\auxLabelbox@}}
 \def\bAdjust@{\let\vAdjust@\bAdjust@@}

 \def\eAdjust@@{\dimen0=-.5\ht\auxLabelbox@
     \advance\dimen0 by .5\dp\auxLabelbox@
     \setbox\auxLabelbox@=
            \hbox{\raise\dimen0\box\auxLabelbox@}}
 \def\eAdjust@{\let\vAdjust@\eAdjust@@}

 \def\tAdjust@@{%
     \setbox\auxLabelbox@=\hbox{\raise-\ht\auxLabelbox@
            \box\auxLabelbox@}}
 \def\tAdjust@{\let\vAdjust@\tAdjust@@}

 \let\vAdjust@\relax

 \def\lAdjust@{\let\hAdjust@\rlap}
 \def\rAdjust@{\let\hAdjust@\llap}

 \let\hAdjust@\relax\let\vAdjust@\relax

 \def\FetchLabel@#1(#2)#3\\{%
     \IN@0#2@@\ifIN@
        \setbox0=\hbox{\ignorespaces#1#3\unskip}%
        \ifdim\wd0>0pt
           \ms@g{}%
           \ms@g{ !!! Bad label(s)? !!!}%
           \message{ #1(#2)#3}%
        \fi
        \def\LabelMole@##1\endFetchLabel@{%
            \IN@0()\\@##1@%
            \ifIN@\def\Temp@{\FetchLabel@##1\endFetchLabel@}%
            \else\def\Temp@{}%
            \fi
            \Temp@
           }%
     \else
       \ignorespaces#1\unskip
       \setbox\auxLabelbox@=%
         \hbox to 0pt{\hss\ignorespaces\hAdjust@
          {\ignorespaces#3\unskip}\hss}%
       \vAdjust@
       \let\hAdjust@\relax\let\vAdjust@\relax
       \AugmentLabelBox@@{#2}%
       \ht\Labelbox@=0pt\dp\Labelbox@=0pt
       \let\LabelMole@\FetchLabel@%
     \fi\LabelMole@}

 \newtoks\XYSep@ %\XYSep@{*}
 \def\SetXYSeparator#1{%
     \IN@0#1@@\ifIN@\XYSep@{*}%
     \else
     \XYSep@{#1}%
     \fi
     }

 \SetXYSeparator*

 \def\AugmentLabelBox@@#1{%
     \IN@0\the\XYSep@ @#1@\ifIN@
       \SPLIT@0\the\XYSep@ @#1@%
       \setbox\Labelbox@=\hbox to 0pt{%
         \unhbox\Labelbox@
         \Shifted@@{\the\Initialtoks@\Wddd@}%
         {\the\Terminaltoks@\Httt@}%
         {\box\auxLabelbox@}}%
     \else
         \ms@g{}%
         \ms@g{ !!! Bad insertion point. !!!}%
         \message{ (#1\ this point was rejected.)}%
     \fi
    }

 \def\FetchOption@#1[#2]#3\endFetchOption@{%
    \def\temp{#1}%\show\temp
    \ifx\temp\empty
       \Edgect@=#2\relax%\showthe\Edgect@
       \let\EdgGl@b\relax
       \SetEdges@%\def\Edgescale@##1{\divide##1 by \Edgect@\relax}%
       \Cleaner@#3%
    \fi}

 \def\Cleaner@#1[@]{\Labeltoks@{#1}}
     
 \def\PlaceLabels@@{\mathsurround=0pt%\bgroup
     \def\Cr@{\\}%
     \let\L\lAdjust@\let\R\rAdjust@
     \let\B\bAdjust@\let\E\eAdjust@\let\T\tAdjust@
     \expandafter\FetchOption@\the\Labeltoks@[@]\endFetchOption@
     \Wddd@=\Wdd@ \Edgescale@\Wddd@ %\showthe\Edgect@
     \Httt@=\Htt@ \Edgescale@\Httt@
     \expandafter\FetchLabel@\the\Labeltoks@\endFetchLabel@
     \box\Labelbox@%\egroup
     }%

 \let \PlaceLabels@\PlaceLabels@@

 \def\AffixLabels#1{\setbox\Coordinatebox@=\hbox{#1}%
      \Wdd@=\wd\Coordinatebox@ \Htt@=\ht\Coordinatebox@
      \advance\Htt@ \dp\Coordinatebox@
      \hbox{\copy\Coordinatebox@\kern-\Wdd@ 
           \Shifted@@{0pt}{-\dp\Coordinatebox@}%
           {\PlaceLabels@\PlaceGrid@}%
           \kern\Wdd@}%
      \GridSwitch@ %% next grid hidden
      \LabGl@b\Labeltoks@{()\\}%
      }
 
   \let\wlog\LabelFigwlog@ld   %%restore logging
   \catcode`\@=\LabelFigCatAt  %%12 or 13

                                By

              Raymond S\'eroul <A18645@FRCCSC21.BITNET>
                                and 
              Laurent Siebenmann <lcs@topo.math.u-psud.fr>
    
              VERSIONS: July 1991, Oct 1991, Jan 1992, July 1992

INTRODUCTION

      This labelling package is intended for TeX users who
rely on non-TeX sources for for their graphics inserts.  It
provides means for adding TeX labels to such inserts with a
minimum of fuss. 

       For most labels, TeX users have in the past found it
reasonably convenient to rely on non-TeX sources. Typical
occasions when an inescapable need for TeX labels seemed to
arise are

 (a) when the graphics program lacks certain exotic or complex
mathematical symbols

 (b) when the very highest typographical quality is wanted for the
labels

 (c) when labels included with the graphics fail to print, 
 and you cannot figure out why (cf. boxedeps.doc).  The labels
 provided by labelfig.tex are 100% portable.

       Since this package first appeared, many users, who in the
past scarcely dreamed of using TeX labels, have come to use
nothing but.  So it is now appropriate to add

Intoxication Warning:  TeX labels may be addictive and expensive. 

     If you have a fast preview you may disagree, and even find
that this package provides an agreeable paste-up environment; see
extra applications at end.

     Note to publishers: It is possible and convenient to ultimately
export the TeX labels produced by labelfig.tex to become an integral
part of the EPS file. This is often desired by a publisher who typically
uses an "upmarket" graphics or page layout program, with which the
staff is skilled in perfecting figures.  See Appendix I for
a recipe.

     The authors are grateful to Patrick Ion of Math Reviews for
helpful comments and encouragement.

BASIC INSTRUCTIONS

    After reading in the macro file using

\input labelfig.tex
preview or proof your figure with a coordinate grid printed on
top, by typing the following:

    \ShowGrid  % shows grid  for next figure only
    \AffixLabels{<the graphics insertion>}

Here <the graphics insertion> is what you would type to insert
the graphics object alone without the grid.  This must provide
for the space around it. For example <the graphics insertion>
might well be \BoxedEPSF{MyFigure scaled 700} using the
boxedeps.tex macro package (from same source); this provides a
TeX box containing the encapsulated PostScript insert specified by
the file MyFigure. \AffixLabels{...} provides the grid (supposing
\ShowGrid is present) and later, once you have specified labels
using the grid, it will "tack on" the labels.

     The grid is a sort of (usually elongated) checkerboard of
ten rows and ten columns and its (internal) partitions are by
default numbered  .1, ... ,.9  both horizontally (X-coordinate
running left to right) and vertically (Y-coordinate running bottom
to top).  Thus the points enclosed by the grid correspond to the
points of the unit square in the cartesian "X-Y" plane, the lower
left corner corresponding to the origin (0,0).  By extrapolation,
the full page corresponds to a larger rectangle in the plane.

     These coordinates serve to position labels as follows.
Before the \AffixLabels{...} command type label specifications:

  \SetLabels
   (<X-coordinate>*<Y-coordinate>) <first label> \\
   .
   .
   .
   (<X-coordinate>*<Y-coordinate>)  <last label> \\
  \endSetLabels

Each row specifies one label and is terminated by \\.  In each
row, the position indicator comes first; it is written as a
standard cartesian point except that the X- and Y- coordinates
are separated by * rather than a comma because TeX allows a
comma as decimal point. There are no dimension units to specify
as the unit is the grid itself.

     By default, this cartesian point specifies where the middle
of the baseline of the label will be located.  However if you precede
the point by \L [or \R] the left [or right] edge of the baseline will
be located there. Similarly you may also precede the point by \T, \E,
or \B to vertically align the top equator or bottom of the label box
at the specified point.  This gives nine standard positions of
the label with respect to the insertion point --- corresponding to
the eight principle points of the compas and the center

                     \L\T     \T      \R\T

                     \L\E     \E      \R\E

                     \L\B     \B      \R\B

But this neglects the default "baseline" level of TeX,
giving potentially three more positions

                     \L    <no tag>   \R

For text, the baseline level is often the preferred. Its relation to
the others is variable. It will often coincide with the bottom level,
as happens for "X".  But it is often distinct, as for "g", in which
case you have in all 12 distinct positions rather than 9.

     It is convenient to think of this specification of label
position as attaching the label by a thumb-tack to the coordinate
grid. There are up to twelve positions of the thumb-tack on the
label, while the position of the thumb-tack on the coordinate grid is
arbitrary.  Normally, one choses the position of the thumb-tack on
the label to be the one that is the closest to the item being
labeled.  There are good reasons for this "rule of thumb":

   (a)  It facilitates correct positioning at first try.

   (b)  If the scale of the figure must be altered after labels
have been affixed, the labels have a good chance of remaining well
positioned.

   (c)  The visible grid need not extend beyond the "bounding box"
for the figure, because the best preferred position is always
(at least almost) within the bounding box .

The second reason is particularly important. Indeed it often
happens that scale has to be altered after labelling begins, in
order to either provide space for the labels, or to adjust
proportions between the labels and the figure.  (The size of labels
is unaffected by scaling.)

     Here is an artificial but self-contained test which uses
TeX rules to make a graphics object.

TEST

    Do not skip this!

\input labelfig
 \def\FrameIt#1{\hbox{\vrule$\vcenter {\hrule\kern3pt%
             \hbox {\kern3pt #1\kern3pt}%
               \kern3pt\hrule}$\relax\vrule}}

 \def\Caption#1#2{\FrameIt{%
       \vtop {\hsize=#1\relax \parindent=0pt
         \leftskip=0pt \rightskip=0pt plus15pt
         \parfillskip=0pt
         \lineskip=1pt\baselineskip=0pt
         #2}}}

 \def\FirstQuadrant{\hbox to 100pt{\vrule\vbox to 100pt{%
        \hbox to 100pt{\hfil}\vfil\hrule}\hss}}

  \SetLabels
    \R(.5*.2) $\zeta\,\cdot$\\
    (.9*-.10) $\xi$\\
    \R(-.03*.9) $\eta$\\
    \T(.5*.9) \Caption{70pt}{%
          \it The norm of
          $g(\xi+i\eta)$ is indicated on
          contours of this invisible surface.}\\
  \endSetLabels

  \AffixLabels{\FirstQuadrant}

  \end

  Note that the coordinates to use for labels are indicated on the
edges of the grid (when visible) corresponding to the conventional
x- and y- axes of the Cartesian plane. By default the grid is
1-by-1. However, by the command \Edges{100}, you can change this
to 100-by-100 and many users find this alternative most
convenient. Place the command \Edges{...} in your style file (or
header) since its effect is is global. Other possible edge values
are 10 and 1000.

  If you use the command \Edges{...} at all, do so with care.  For
if you accidentally delete an \Edges{...} command your labels will
abruptly be badly misplaced and may logically but mysteriously
generate "dimension too big" errors under TeX and "off page" errors
under your driver.  

  You can dictate the edgescale for an individual figure by giving
the scale in brackets immediately after \SetLabels.  Thus, to
import into an article using say \Edge{100} a figure labelled using
another edgescale, say the original 1-by-1 default, you can use
\SetLabels[1]...\endSetLabels.

GETTING IT DOWN PAT

     Complicated labeling deserves the same respect as
complicated mathematics.  Do not expect it to come out perfect the
first time!  What is needed in either case is a mechanism to
repeatedly typeset troublesome pieces.

     One mechanism is always available.  One does complicated
labelling in a separate "test" file involving just the figure being
labelled;  a texpert will know how to \dump TeX's current state as
a temporary format that restarts rapidly at each retry.  Usually,
one then pastes the completed labelled figure back into the main
TeX file, but, of course, one can also \input it as an auxiliary
file.

     If you do not have a TeXpert at handy, here is a first
approximation to an efficient setup. By deletions reduce a copy
of your article to just a few lines before and after the figure.
Now label the figure, and finally, copy and paste the labelled
figure to the original article. Then copy the next figure to label
into this testbed and repeat. The TeXpert can improve the  speed
at which TeX starts up, by compiling a format specifically for
your article; just one caution: best NOT include in the format
ephemeral details of setup like \Set<mydriver>ArtSpecials (from
boxedeps.tex because this reads  figure dimensions which you may
change during your work session.

     An improved mechanism to repeatedly typeset troublesome
pieces is now available on the Macintosh; it is called LinoTeX;
see the same ftp sources.  It could be set up on many types
of computer.

     Before using labelfig.tex to attach labels to a graphics
object inserted using boxedeps.tex or BoxedArt.tex, make it a
firm rule to carefully adjust the bounding box using the trimming
commands of these packages, and also at least tentatively scale
and position the object. Beware of changing the grid inadvertently
after the labels have been positioned.  For example, correcting
the bounding box of a PostScript graphics object can foul up the
labels by changing the coordinate grid to which the labels are
attached. This is particularly true for the trimming  commands of
boxedeps.tex and BoxedArt.tex. However, as noted already, change
of scale is much less disruptive, and modest adjustments should be
well tolerated.

     Sometimes the labels protrude so far from the bounding box
of a figure that the figure has to be repositioned.  Best do this
by ad hoc spacing, say using \hglue and \vglue; altering the
bounding box would create a vicious circle.

     Remember that you are responsible for preventing labels
from overlapping. You are responsible for all label typography
including size and style. A label is really just about anything
that can be put in a TeX box. Note that spaces at the beginning
and end of labels will normally be suppressed; if you really want
them you must protect them with TeX braces.

     This package temporarily sets the \mathsurround parameter
of TeX to zero  while the labels are being affixed. This is done
because nonzero \mathsurround space would influence the position
of left and right aligned labels; then, when a texpert or printer
modifies mathsurround, diagram labeling might be disastrously
altered. There is a small price to pay involving labels that are
formatted as caption boxes including mathematics: you  may want or
need to specify an explicit mathsurround space within the caption
box; it will not influence anything outside.

     Those hostile to the use of * as separator between
the X and Y coordinates of label insertion points, are free to
impose another using \SetXYSeparator{<the new separator>}.  
Americans may prefer "," to "*" since they never use a 
comma as a decimal point; on the other hand, * may be more visible.

APPENDIX (I)  MERGING labelfig.tex LABELS INTO AN EPSF GRAPHICS OBJECT.

     As promised in the introduction, here is a recipe useful for
publishers. It works at least on Macintosh and at least for vectorized
graphics and Adobe type1 fonts.  (There is surely a similar recipe for
PCs under MSWindows.)

 (a)  Use boxedeps.tex utility to integrate the figure given by the eps
file, "x.eps" say, with a visible frame around it.  See
\ShowDisplacementBoxes command in boxedeps.tex.  To get precise results
automatically it is important to use the \Trim... commands of
boxedeps.tex making the "DisplacementBox" neatly fit the figure.

 (b)  Use the TeX printer driver and LaserWriter (versions >= 8.1.1) to
export to an EPSF the DVI page containing the integrated, labelled
figure. You now have an EPS file  "xx.eps"  that contains too much, and at
the wrong scale, and at wrong position.

 (c)  Convert the EPSF to an Adode Illustrator format EPSF using
the shareware utility called epsConvert by Sam Weiss
1993-- (currently $25).

 (d)  In Illustrator (or a compatible program), group the labels and the
"DisplacementBox"; copy them to the clipboard and paste them into "x.ps".
This step requires that all the label fonts be "visible to the Macintosh.

 (e)  Translate and scale the pasted group consisting of the labels plus
the "DisplacementBox" so as to make the "DisplacementBox" the bounding
box of (labelless) figure represented by "x.eps".  At this point the
labels will be correctly placed on the figure "x.eps".

 (f)  Ungroup and delete the "DisplacementBox".  The result is the
desired single EPS file, "x+.eps" say, It contains the original figure
plus its labels.  

     Using grouping and ungrouping appropriately in "x+.eps", a
publisher's staff can very efficiently improve label positions etc.

APPENDIX II)  SOME EXOTIC APPLICATIONS

     The grid of labelfig.tex is analogous to a light-table in
classical page makeup with wax or latex glue.  In principle, you
can use it to compose any page from its indivisible parts.  This
even has some of the artisanal charm of classical paste-up
provided you have a fast screen preview to make the process
"interactive".

     In practice labelfig.tex is a tool for nonstandard jobs.
Here are a few going beyond the labelling already discussed.

(I)  GRAPHICS INTEGRATION.

     This is accomplished by treating the imported graphics
objects as labels.  The underlying graphics object is then
typically an empty  \vbox to <dimension>{\vfill} in a TeX
\midinsert...\endinsert construction.  A label line
might be of the form

   (.1*.1) \\

The exact form of the special command varies from driver to
driver.  However, in the case of encapsulated PostScript graphics
(EPSF norm), by relying on boxedeps.tex, one can have the
following standard syntax (independant of driver  (see
boxedeps.doc for details.
  
  (.1*.1) \BoxedEPSF{MyFigure scaled <scale in mils>}\\

This may be slow since it requires TeX to read the PostScript
file to read bounding box using many complex macros.  So you
may want to try

  (.1*.1) \EPSFSpecial{MyFigure}{<scale in mils>}\\

which is fast and driver independant, but it squashes the
bounding box, normally to its lower left corner.

     Similarly for graphics of the Macintosh PICT norm ---
using BoxedArt.tex (same sources) in place of boxedeps.tex.

     This approach to integration is to be recommended when
one is assembling a composite graphics object.

 (II)  COMMUTATIVE DIAGRAM ENHANCEMENT

     Commutative diagrams or arrays of mathematical objects
connected by arrows of various sorts are common in mathematics.
The mathematical objects require the use of TeX.  Recently TeX
acquired a good collection of arrows of all slopes --- that of
LamSTeX --- plus pwerful macros to build the diagrams.

     However, even the LamSTeX collection is often
inadequate; it lacks for example double shafted arrows, dotted
arrows and curved arrows. Fortunately it is possible to produce
such arrows on an individual basis using sophisticated graphics
programs such as Illustrator and AldusFreehand (both serving
the EPSF norm) or using Metafont (with its public domain norm).
Since the creation of each new arrow is a work of love, you
probably want to limit the number of arrows by using LamSTeX
for most arrows. The 40K commutative diagram module of LamSTeX
has been adapted to work with AmSTeX and a copy may be posted
with LabelFig and related files. Unfortunately no one has yet
offered a version that works with Plain TeX or LaTeX.

       Suffice it here to say that when the exotic arrow has
been somehow imported into TeX, labelfig.tex treats it as a
label that one affixes to the commutative diagram.  Two other
steps will be treated in separate notes, namely the matter of
extracting the dimension specifications for the arrow and the
construction of the arrow --- for these steps are far from
unique and often depend intimately on your computer environment. 
Notes for the Macintosh-Textures-Illustrator combination are
found in the file ExoticArrows.doc.

 (III) NESTING 

Ingenuity pays off in exploiting labelfig.tex. One can
mix graphics and typography quite freely.  labelfig.tex is good
for freeform or overlapping arrangements, while boxedeps.tex (or
BoxedArt.tex) is best for regimented non-overlapping
arrangements --- and the two can be combined.

     The default behavior of labelfig.tex is not ideal 
for nesting objects, because to prevent trouble for beginners
the register for labels is globally cleared when \AffixLabels
concludes.  But there are switches available

      \LabelsGlobal      \LabelsLocal

which change this.  To understand this, extend the above test 
by something like:

 \LabelsLocal
 \SetLabels
    (.5*.5) AAA\\
 \endSetLabels

 {%%% Watch for influence of braces!!
 \SetLabels
    (.5*.5) ZZZ\\
 \endSetLabels
   \AffixLabels{\FirstQuadrant}
 }

   \AffixLabels{\FirstQuadrant}

     There are however potential pitfalls.  Neither
labelfig.tex nor boxedeps.tex has been tested under extreme
conditions. Problems may occur if their procedures are
indiscriminately nested. For boxedeps.tex (not labelfig.tex)
there is a precise cause for worry, namely many of its
variables are "global", which means that TeX braces will not
provide the protection one might expect.

COMMAND SUMMARY FOR labelfig.tex

  Here [...] means optional (one or zero)
       [...]* means any number of such constructs

  \SetLabels
    [[<P>](<X><Sep><Y>) <label> \\]*
  \endSetLabels
  \ShowGrid  % this makes the grid visible (once)
  \AffixLabels{<the figure>}

   --- <P> is tack position, one of eleven or empty
              order irrelevant

                   \L\T      \T      \R\T

                   \L\E      \E      \R\E

                     \L               \R

                   \L\B      \B      \R\B

   --- (<X><Sep><Y>) insertion point;
  <Sep> is separator, = * by default;
  \SetXYSeparator{<Sep>} changes it.
   <X> and <Y> are real numbers

  --- <label> a label to attach 

  --- <the figure> the figure to label 

  \GlobalLabels (default)     
  \LocalLabels  setting for nested constructs.

 \Grids makes ALL grids appear; \HideGrid then makes just next disappear.
 \noGrids returns to default.  The commands are always global.

 \GridLineWidth{<dimension>} adjusts width of grid lines. Default is very
small, to give "hairline" effect. If your grid lines are missing try
setting \GridLineWidth{1pt}.

 \Edges#1 globally changes the edge size of all grids to the numerical 
value #1, which must be 1, 10, 100, or 1000.  The default is 1.

VERSION HISTORY.
 --- Jan 1993: \Edges#1 and [??] option after \SetLabels
 --- July 1992: \Grids, \noGrids, \HideGrid;
       Gridlines become hairlines; \GridLineWidth{<dimension>}.
 --- Oct 1991, Jan 1992: \SetXYSeparator{<Sep>},  \LabelsGlobal,
       \LabelsLocal.
 --- July 1991: first release

Address for bugs and other feedback:

        Raymond S\'eroul
        IREM and Lab. de Typographie Informatise
        Univ. Rene Descartes
        Strasbourg

    Tel 33-88-41-63-45
    Email:  A18645@FRCCSC21.BITNET

        Laurent Siebenmann
        Mathematique, Bat. 425,
        Univ de Paris-Sud,
        91405-Orsay,
        France

    Tel 33-1-6941-7949; 
    Email: lcs@topo.math.u-psud.fr